# Resolving the Quantum Origin of Anisotropic Raman Response in Quasi-1D $Ta_2Ni_3Te_5$

*Fei Wang[1], Qiaohui Zhou[1], Hong Tang[1]*, Fan Zhang[2], Yanxing Li[2], Steven A. Hindmarsh[3], Chengkun Xing[4], Keyuan Bai[1], Sidra Younus[1], Ana M. Sanchez[3], Haidong Zhou[4], Chih-Kang Shih[2], Adrienn Ruzsinszky[1], Xin Lu[1]*, and Jiang Wei[1]**

[1]Department of Physics and Engineering Physics, Tulane University, New Orleans, Louisiana 70118, United States.

[2]Department of Physics, The University of Texas at Austin, Austin, Texas 78712, United States.

[3]Department of Physics, University of Warwick, Coventry, CV4 7AL, United Kingdom.

[4]Department of Physics and Astronomy, University of Tennessee, Knoxville, Tennessee 37996, United States.

*Emails: htang5@tulane.edu, xlu5@tulane.edu, jwei1@tulane.edu

**ABSTRACT:** Probing anisotropic electron-phonon interactions (EPIs) is fundamental to understanding low-dimensional quantum materials, yet isolating their intrinsic signatures remains a challenge due to the confounding influence of extrinsic effects. Here, we resolve this ambiguity in quasi-one-dimensional $Ta_2Ni_3Te_5$ through a combination of angle-resolved polarized Raman spectroscopy and density functional perturbation theory (DFPT). Anomalous four-lobe angular dependences were observed in $A_g$ phonon modes, which deviate from the real-valued Raman tensor prediction. Further, we explicitly highlight that, despite their distinct physical origins, light absorption and birefringence yield mathematically identical angular dependencies in $Ta_2Ni_3Te_5$, rendering phenomenological fitting alone insufficient to identify the true physical mechanism. By employing a full quantum treatment incorporating detailed interaction mechanisms, we capture the qualitative characteristics and mode-to-mode diversity of the anomalous angular responses observed experimentally. The DFPT calculations further reveal mode-specific and momentum-resolved EPIs, providing microscopic support for anisotropic EPIs as the mechanism of anomalous Raman responses.

Electron-phonon interactions (EPIs) constitute a fundamental pillar of condensed matter physics, governing the electrical resistivity, carrier mobility, thermal conductivity, and superconductivity.[1, 2] In low-dimensional systems, the reduced dimensionality and weakened dielectric screening enhance the interactions, fostering emergent quantum phenomena.[3-11] In highly anisotropic quantum materials, EPIs can further manifest in unconventional and strongly direction-dependent manners, profoundly altering the physical properties. Consequently, investigating EPIs in such systems is not only fundamentally intriguing but also technologically crucial, promising to unveil unprecedented phenomena and providing deeper insights into the physics.

The $Ta_2M_3Te_5$ (M = Pd, Ni) will be an ideal system for this study. This family has recently garnered significant attention due to its extraordinary electronic and topological properties, including quantum spin Hall effect,[12, 13] pressure-induced superconductivity in $Ta_2Pd_3Te_5$[14] and $Ta_2Ni_3Te_5$ coexisting with $Z_2$ band topology,[15] and double-band inversion induced second-order topology.[16] Structurally, $Ta_2M_3Te_5$ hosts pronounced in-plane anisotropy, forming one-dimensional (1D) atomic chains within the layers. The reduction in crystalline symmetry gives rise to strong anisotropy,[17, 18] leading to remarkable direction-dependent electrical,[19-21] optical,[21, 22] thermal,[23, 24] and phononic[25-27] properties. The inherent structural anisotropy endows $Ta_2M_3Te_5$ with a critical degree of freedom for tailoring emergent phenomena, making it an excellent platform for investigating the interplay between topology and anisotropic interactions in reduced dimensions. While the in-plane anisotropy has been recently reported in this family,[28] the physical origin of the experimental ambiguity remains obscured. Current models often conflate intrinsic electronic anisotropy with extrinsic effects like light absorption or birefringence,[25, 26, 28-32] failing to distinguish between the two and leading to inconsistent and ambiguous interpretations. Anisotropic EPIs could be a possible origin of the anomalous Raman response, following the pioneering inference of anisotropic EPIs in black Phosphorus.[33] However, a direct verification of the mode-specific and momentum-resolved EPIs is still lacking in this system. Elucidating the true nature of EPIs in low-symmetry quantum materials is essential for understanding interaction-driven phenomena and achieving broad quantum device applications.

Here, we resolve this ambiguity by uncovering mode-specific and momentum-resolved anisotropic EPIs in quasi-1D $Ta_2Ni_3Te_5$ nanoflakes. The angle-resolved polarized Raman measurements reveal pronounced anomalies in $A_g$ phonon modes. Moving beyond a phenomenological description, we explicitly identify a mathematical degeneracy: the light absorption and birefringence effects yield identical angular Raman dependencies. This renders conventional phenomenological fitting insufficient and necessitates a full quantum mechanical treatment. By employing a full quantum framework combining density functional perturbation theory (DFPT) calculations, we demonstrate that these anomalies arise as macroscopic manifestation of intrinsically anisotropic EPIs. Furthermore, mode-specific and momentum-resolved EPIs are directly mapped across the first Brillouin zone (BZ), establishing anisotropic EPIs as an intrinsic characteristic of $Ta_2Ni_3Te_5$. These findings reveal the underlying quantum-interaction mechanism governing the optoelectronic properties, providing a general paradigm for exploring EPIs in low-dimensional quantum materials.

Layered van der Waals $Ta_2Ni_3Te_5$ crystallizes in the orthorhombic $Pnma$ space group, exhibiting distinctive quasi-1D confinement and atomic-scale anisotropy within a two-dimensional structure composed of monolayers stacked along the $a$-axis (Figure 1a). Within each layer, pronounced in-plane anisotropy arises from well-defined atomic chains where Ta, Ni, and Te atoms align along the $b$-axis (Figure 1b), establishing a strict quasi-one-dimensional structural and consequently electronic confinement. This intrinsic anisotropy is macroscopically evident in the crystal morphology; as-grown crystals (see Supporting Information S1) develop as lustrous needles, and exfoliated flakes retain this ribbon-like geometry with remarkably straight edges aligned to the chains (Figure 1c, d). To validate the atomic-scale integrity of this 1D confinement, we performed high-resolution transmission electron microscopy (HRTEM) and scanning tunneling microscopy (STM). HRTEM imaging (Figure 1e) reveals pristine lattice fringes that perfectly match the $bc$-plane crystal structure, confirming that the ribbon edges are naturally cleaved along the $b$-axis. While the material exhibits surface sensitivity (Figure 1f, g), cross-sectional high-angle annular dark-field scanning transmission electron microscopy (HAADF-STEM) confirms a self-limiting degradation mechanism where amorphous surface layers protect the crystalline interior. Furthermore, *in-situ* STM topography of freshly cleaved monolayer surfaces (Figure 1h, i) explicitly visualizes the atomic chains, showing defect-free 1D chained structures essential for studying intrinsic quantum responses.

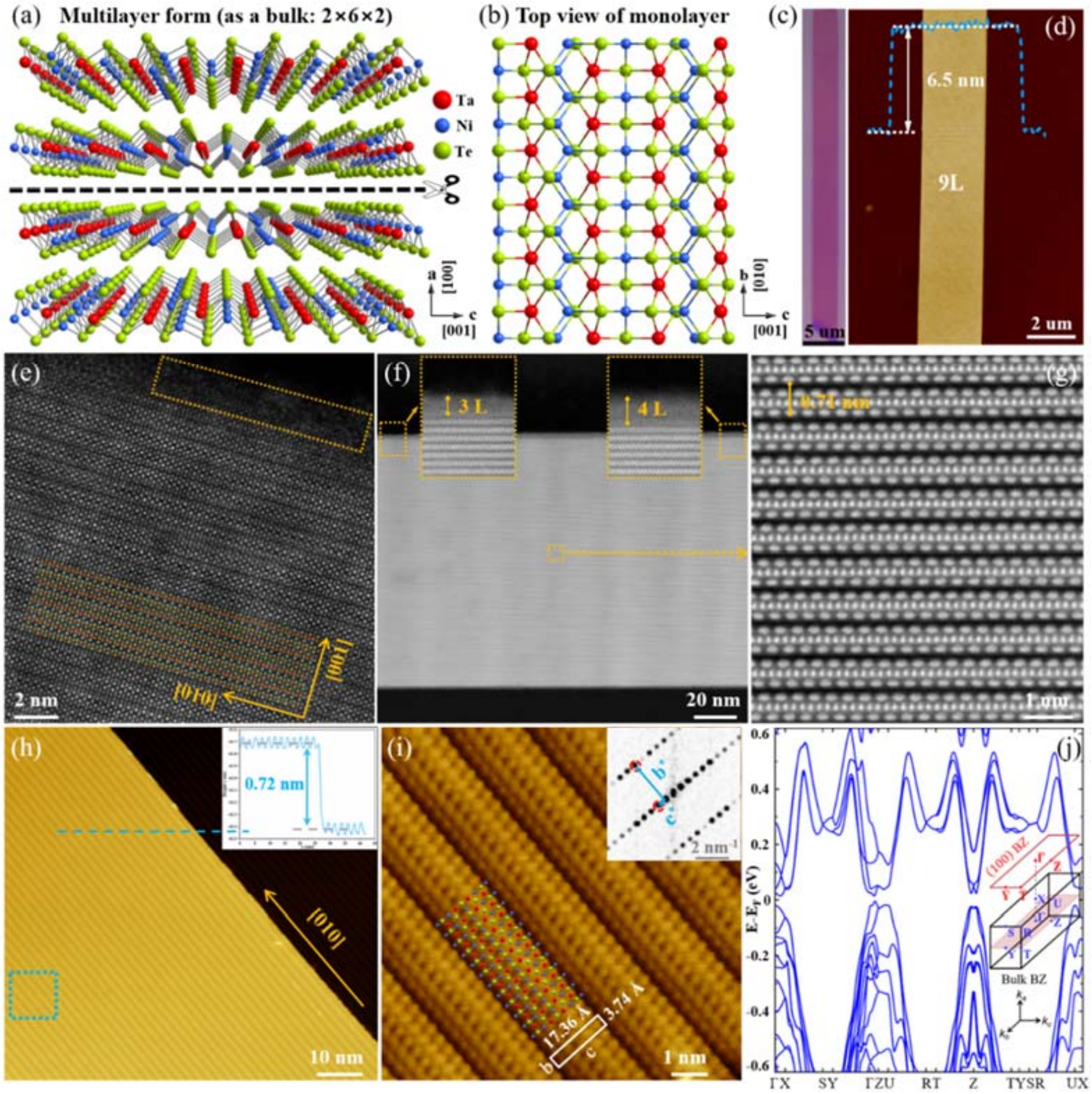

**Figure 1.** Structural and electronic anisotropy of quasi-1D $Ta_2Ni_3Te_5$. (a) Crystal structure viewed along the *b*-axis. (b) Top view of a monolayer. (c, d) Optical and atomic force microscopy images of exfoliated 9-layer $Ta_2Ni_3Te_5$ nanoflake, showing well-defined ribbon-like edges aligned with the atomic chains. Inset is the height profile. (e) HRTEM image confirming the high crystalline quality and the ribbon edge aligned along the *b*-axis. Inset shows that the HRTEM image matches well with the crystal structure in the *bc* plane. (f, g) Cross-sectional HAADF-STEM image showing the atomic stacking order; surface oxidation is observed but the bulk crystallinity remains intact. Insets show zoomed-in areas on the top surface, indicating 3-4 layers of degradation. (h, i) STM topography of the single crystal surface cleaved *in-situ* under high vacuum, directly visualizing the defect-free 1D atomic chains. A monolayer step

edge along $b$-axis is also identified. (j) Calculated electronic band structure, revealing strong dispersion anisotropy near the Fermi level, characteristic of the 1D confinement. Inset shows the bulk and (100) surface BZ.

This structural confinement manifests directly as pronounced anisotropy in the electronic and vibrational manifolds. Our calculated electronic band structure (Figure 1j) reveals significant dispersion anisotropy around the Fermi level, implying strongly direction-dependent electrical, optical and phononic properties as well as their anisotropic interactions. To probe the vibrational consequences of this anisotropy, we characterized the phonon modes using Raman spectroscopy (Figure 2a), resolving twenty-three peaks in excellent agreement with our DFT calculations (Supporting Information S2). Crucially, the vibrational eigenvectors exhibit distinct spatial symmetries: as illustrated in the insets of Figure 2a, atoms in the $A_g$ modes vibrate predominantly in the $ac$-plane (perpendicular to the atomic chains), whereas atoms in the $B_{3g}$ modes vibrate primarily along the $b$-axis (parallel to the chains). This directional orthogonality suggests that $A_g$ and $B_{3g}$ phonons will probe different components of the electron-phonon interaction field. The calculated phonon dispersion (Figure 2b) shows no imaginary frequencies, confirming the lattice stability.

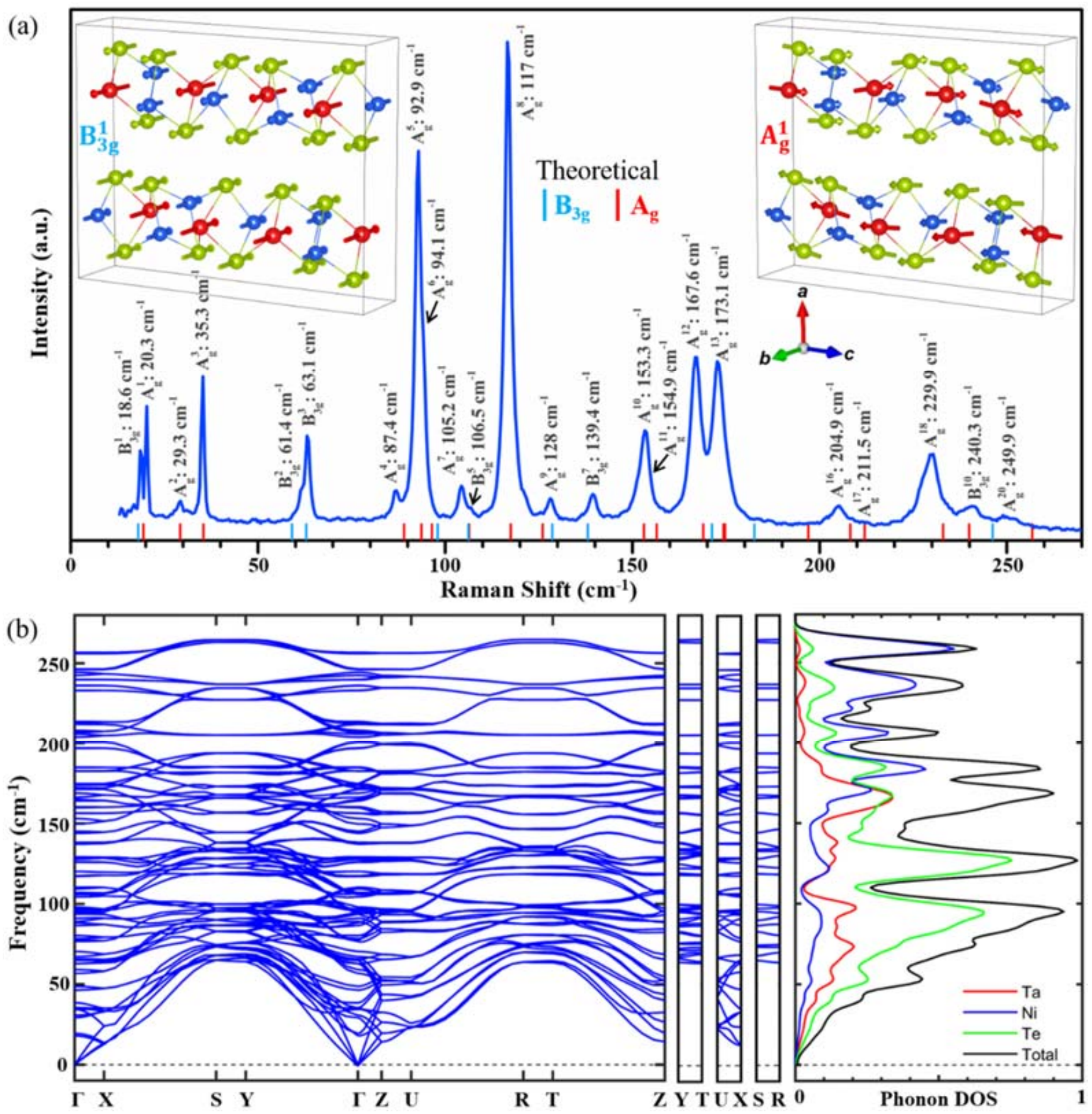

**Figure 2.** Phonon modes of $Ta_2Ni_3Te_5$. (a) Representative unpolarized Raman spectrum of $Ta_2Ni_3Te_5$ nanoflake, with calculated Raman modes indicated by red ($A_g$) and blue ($B_{3g}$) vertical lines. Insets: Schematic representations of the vibrational eigenvectors for the $B_{3g}^1$ and $A_g^1$ phonon modes. $A_g$ mode involves atomic displacements primarily in the ac-plane (perpendicular to the chains), while $B_{3g}$ mode involves displacements along the b-axis (parallel to the chains). (b) Calculated phonon dispersion along high-symmetry paths (left) and corresponding phonon partial-DOS (right).

To systematically characterize the anisotropic properties, we performed angle-resolved polarized Raman measurements under parallel polarization on a 12L (~8.7 nm). $Ta_2Ni_3Te_5$ nanoflake (Figure 3a shows the measurement setup and details are described in Supporting Information S3). Figure 3b presents the comprehensive intensity map of angular dependent Raman spectra. Representative polar plots are shown in Figure 3c-h, with detailed polar plots of Raman modes provided in Figures S8 and S9. Considering Raman tensor as real numbers, it is expected that $B_{3g}$ and $A_g$ should exhibit four-fold and two-fold symmetry, as represented in Figure 3c and 3d, respectively (see Supporting Information S3.1 for detailed analysis).

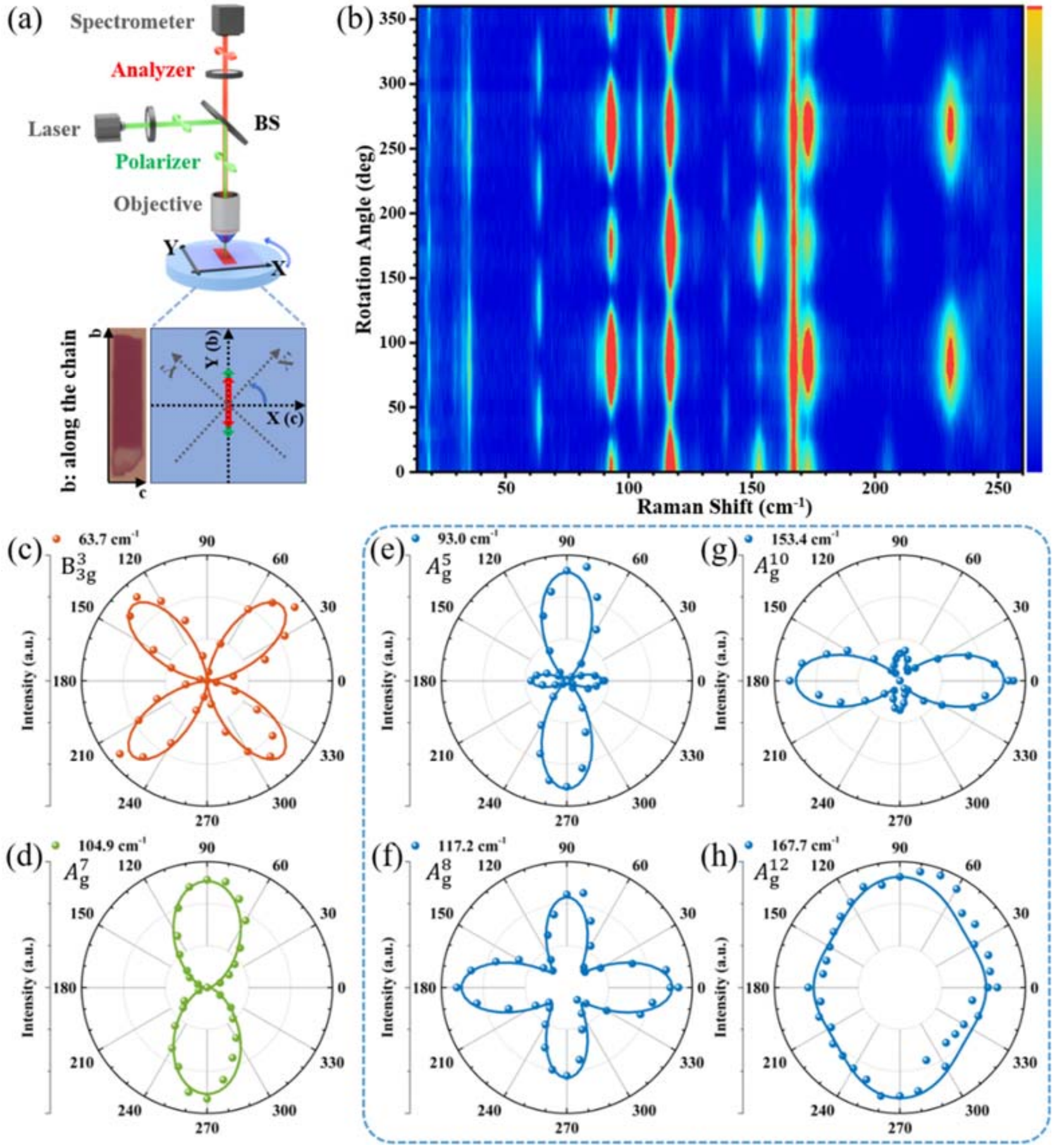

**Figure 3.** Anomalous anisotropy in polarized Raman response. (a) Schematic of the angle-resolved polarized Raman measurement setup in a backscattering geometry. The analyzer is set parallel to the polarizer (parallel configuration). The inset illustrates the sample alignment: the quasi-1D atomic chains (b-axis) of the $Ta_2Ni_3Te_5$ nanoflake are initially aligned with the laser polarization (Y-axis) at a rotation angle of 0°. (b) False-

color map of the Raman intensity evolution as a function of sample rotation angle, highlighting the distinct periodic modulations of different phonon modes. (c–h) Representative polar plots of normalized Raman intensity versus polarization angle θ (dots) overlaid with theoretical fits (solid lines). (c) The $B_{3g}$ mode (63.7 $cm^{-1}$) exhibits the expected four-fold cloverleaf symmetry. (d) The $A_g$ mode 104.9 $cm^{-1}$ displays the standard two-fold dipolar symmetry. (e–h) Anomalous $A_g$ modes (93.0, 117.2, 153.4, and 167.7 $cm^{-1}$) that deviate significantly from the simple two-fold prediction shown in (d), motivating the microscopic analysis of mode-dependent anisotropic electron-phonon interactions below.

However, as shown in Figure 3e-h, there are many $A_g$ modes exhibiting unexpected four-lobe angular patterns. This is anomalous for the $D_{2h}$ point group, where a simple two-fold response is expected from the real-valued Raman tensor (Figure S10a). To elucidate the origin of these anomalies, we first extend beyond the real-valued Raman tensor formalism by introducing complex Raman tensors. In highly anisotropic crystals, both light absorption[25, 26, 29, 30] and birefringence[31, 32] can introduce complex components (detailed analyses can be found in Supporting Information S3.2 and S3.3). Crucially, we identify a fundamental mathematical degeneracy for the $D_{2h}$ point group: although light absorption and birefringence introduce phases through entirely different physical mechanisms, they lead to mathematically identical angular dependencies for the $A_g$ modes, a result not recognized in previous studies.[28, 31, 32] Importantly, this is not a material-specific artifact but a conclusion applicable to diverse 2D crystals ranging from orthorhombic to monoclinic. (Table 1, see Supporting Information S3.4 for comprehensive analyses). Consequently, the phenomenological fitting is inherently insufficient to identify the true physical mechanism, leading to ambiguous interpretations. Crucially, both light absorption and birefringence are macroscopic optical effects. Birefringence introduces a phase retardation dependent on sample thickness, imposing a uniform shift across all $A_g$ modes equally. For the $A_g$ modes, the symmetry of the optical transitions is the same, and hence light absorption would also affect Raman modes of the same symmetry simultaneously. However, the experimental results reveal dramatic variations in phase differences among different anomalous $A_g$ modes (Figure 3d-h and Figure S8). This phonon mode-specificity rules out macroscopic optical artifacts as the primary origin and points to an intrinsic, mode-specific quantum origin. Moreover, the anomalous profiles remain consistent between 12- and 17-layer flakes (Figures S23-S25), despite the thickness dependence expected for the birefringence-induced $\delta(t)$ (Equation S12, Supporting Information S3.3). The observed consistency further highlights the microscopic mechanism underlying the anomalous polarization dependence and provides evidence against birefringence being the primary origin. The excellent fits obtained using the complex Raman tensors (solid line in Figure 3c-h) are therefore necessary but insufficient evidence of the intrinsic physical mechanism.

**Table 1.** Comparison of light absorption and birefringence effect in the Raman measurement of low-symmetry systems.[a]

| System | Point group | Raman mode | Representation of angle-dependent Raman intensity | |
|---|---|---|---|---|
| Orthorhombic<br>**$Ta_2Ni_3Te_5$**,<br>BP, bulk $WTe_2$,<br>Td-$MoTe_2$,<br>… | $D_2$<br>$C_{2v}$<br>$D_{2h}$<br><br>Tensor<br>(classical) | A<br>$A_1$<br>$A_g$<br><br>$\begin{pmatrix} a & 0 & 0 \\ 0 & b & 0 \\ 0 & 0 & c \end{pmatrix}$ | Light absorption:<br>$I \propto b^2 cos^4 \theta + c^2 sin^4 \theta$<br>$+2bc\, cos^2 \theta\, sin^2 \theta\, cos\, \phi_{bc}$ | Birefringence:<br>$I \propto b^2 cos^4 \theta + c^2 sin^4 \theta$<br>$+2bc\, cos^2 \theta\, sin^2 \theta\, cos\, 2\delta(t)$ |
| Monoclinic<br>N-layer $WTe_2$,<br>1T′ $MoTe_2$,<br>GaTe<br>… | $C_s$<br>$C_2$<br>$C_{2h}$<br><br>Tensor<br>(classical) | A′<br>A<br>$A_g$<br><br>$\begin{pmatrix} a & 0 & d \\ 0 & b & 0 \\ d & 0 & c \end{pmatrix}$ | Light absorption:<br>$I \propto a^2 cos^4 \theta + b^2 sin^4 \theta$<br>$+2ab\, cos^2 \theta\, sin^2 \theta\, cos\, \phi_{ab}$ | Birefringence:<br>$I \propto a^2 cos^4 \theta + b^2 sin^4 \theta$<br>$+2ab\, cos^2 \theta\, sin^2 \theta\, cos\, 2\delta(t)$ |
| Triclinic<br>$ReS_2$,<br>$ReSe_2$,<br>… | $C_i$<br><br>Tensor<br>(classical) | $A_g$<br><br>$\begin{pmatrix} a & d & e \\ d & b & f \\ e & f & c \end{pmatrix}$ | Light absorption:<br>$I \propto b^2 cos^4 \theta + a^2 sin^4 \theta + d^2 sin^2 2\theta + 4ad\, sin^3 \theta\, cos\, \theta\, cos\, \phi_{ad}$<br>$+4bd\, cos^3 \theta\, sin\, \theta\, cos\, \phi_{bd} + 2ab\, sin^2 \theta\, cos^2 \theta\, cos\, \phi_{ab}$<br>Birefringence:<br>$I \propto b^2 cos^4 \theta + a^2 sin^4 \theta + d^2 sin^2 2\theta + 4ad\, sin^3 \theta\, cos\, \theta\, cos\, \delta(t)$<br>$+4bd\, cos^3 \theta\, sin\, \theta\, cos\, \delta(t) + 2ab\, sin^2 \theta\, cos^2 \theta\, cos\, 2\delta(t)$ | |

[a]Note: $\phi_{bc}$ is the phase difference $(\phi_b - \phi_c)$ of the tensor element complex phase $\phi_b$ and $\phi_c$ (same for $\phi_{ab}$, $\phi_{ad}$, and $\phi_{bd}$); $\delta(t)$ is the phase difference induced by birefringence. In a specific measurement, the incident light may initially polarize along *b* or *a* direction. Then, the element *b* could be replaced by *a* in the above representations. The conclusion will not be changed.

To pinpoint the microscopic origin of the anomaly, we investigated the scattering processes using a full quantum mechanical Raman model, implemented via density functional perturbation theory. Unlike the classical approaches that rely on macroscopic Raman tensors, the quantum framework explicitly evaluates the scattering probability using third-order time-dependent perturbation theory, where the Raman intensity $I_\nu$ for a phonon mode $\nu$ is governed by: [33-36]

$$I_\nu(E_L) = \left| \sum_{i,m,m'} \frac{\langle i|H_{op}|m'\rangle\langle m'|H_{ep}^{\nu}|m\rangle\langle m|H_{op}|i\rangle}{(E_L - \Delta E_{mi})(E_L - \hbar\omega_\nu - \Delta E_{m'i})} \right|^2 \quad (1)$$

Here, $E_L$ is the incident laser photon energy, and $\hbar\omega_\nu$ is the phonon energy. The indices $i$, $m$, and $m'$ denote the initial/final, first intermediate, and second intermediate electronic states, respectively. $E_{i,m,m'}$ is the corresponding energy at the states. The terms $\Delta E_{mi}$ $(\Delta E_{m'i})$=$E_m$ $(E_{m'})-E_i-i\gamma$ represent the energy differences between the intermediate and initial states, including a phenomenological damping factor $\gamma$. The matrix elements in the numerator describe the sequential quantum pathway: first, the term $\langle m|H_{op}|i\rangle$ represents the photon absorption process where an electron is excited from the initial valence band $|i\rangle$ to a virtual intermediate band $|m\rangle$. Subsequently, the term $\langle m'|H_{ep}^{\nu}|m\rangle$ describes the phonon emission, wherein the electron scatters from band $|m\rangle$ to a second intermediate band $|m'\rangle$ via the electron-phonon interaction. Finally, the term $\langle i|H_{op}|m'\rangle$ corresponds to photon emission, as the electron relaxes from band $|m'\rangle$ back to the original electronic state $|i\rangle$, releasing a Raman photon. Figure 4a depicts this process for the $A_g$ mode. Within the dipole approximation,[37] the electron-photon interactions can be expressed as $\langle m|H_{op}|i\rangle \propto \vec{P} \cdot \vec{D}$, where $\vec{P}$ is the light polarization vector and $\vec{D} = \langle m|\nabla|i\rangle$ is the dipole vector of the transition. For a nonvanishing $\langle m|H_{op}|i\rangle$, $\vec{D}$ must possess a nonzero component parallel to $\vec{P}$, which dictates specific electronic bands participating in the Raman scattering process (symmetry-allowed optical transitions are summarized in Table S4).

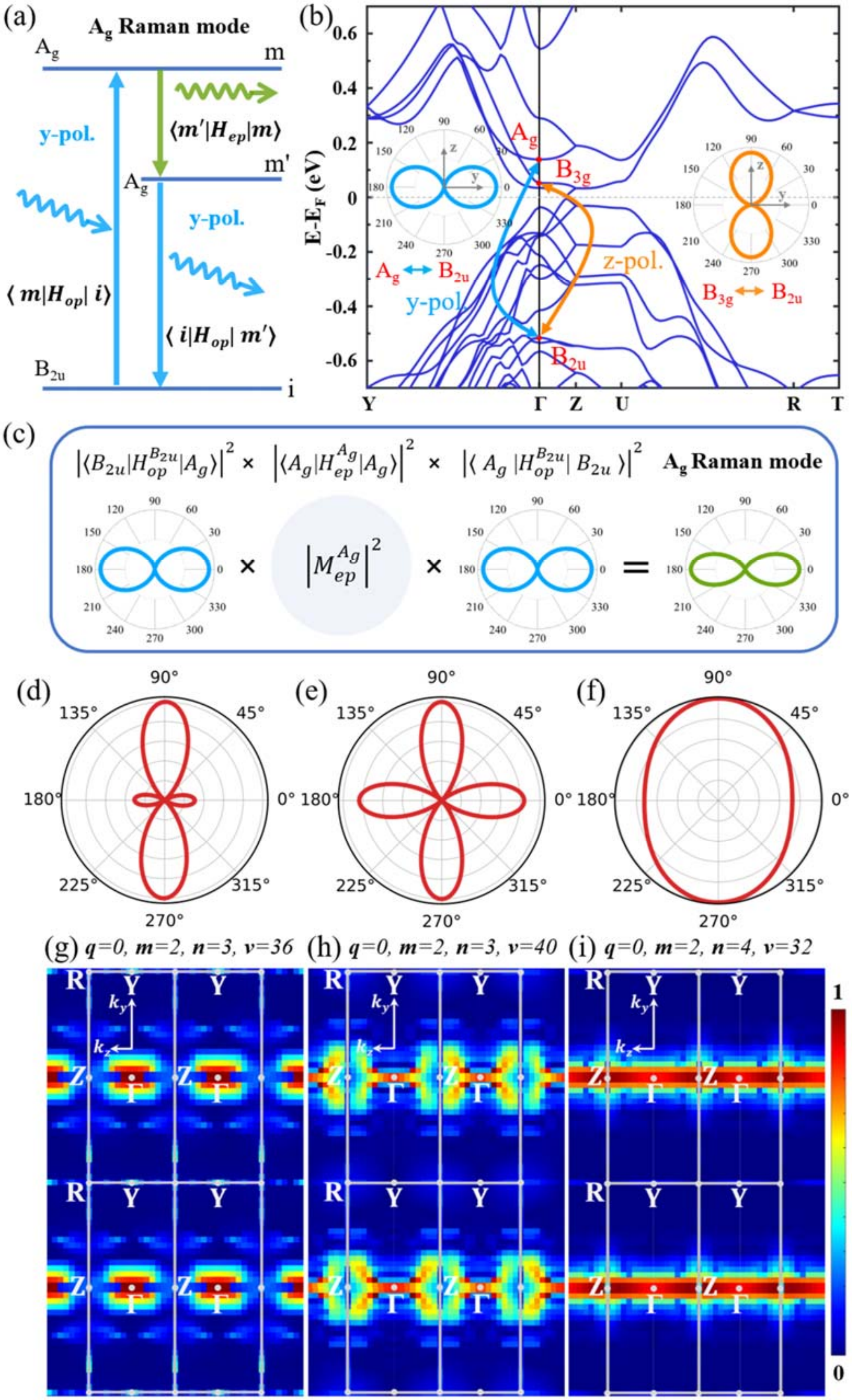
(a)
$A_g$ Raman mode
$A_g$
m
y-pol.
$\langle m'|H_{ep}|m\rangle$
m'
$A_g$
y-pol.
$\langle m|H_{op}|i\rangle$
$\langle i|H_{op}|m'\rangle$
$B_{2u}$
i
(b)
E-E$_F$ (eV)
$A_g$
$B_{3g}$
$A_g \leftrightarrow B_{2u}$
y-pol.
z-pol.
$B_{3g} \leftrightarrow B_{2u}$
$B_{2u}$
Y
Γ
Z
U
R
T
(c)
$\left|\langle B_{2u}|H_{op}^{B_{2u}}|A_g\rangle\right|^2 \times \left|\langle A_g|H_{ep}^{A_g}|A_g\rangle\right|^2 \times \left|\langle A_g|H_{op}^{B_{2u}}|B_{2u}\rangle\right|^2$
$A_g$ Raman mode
$\left|M_{ep}^{A_g}\right|^2$
(d)
(e)
(f)
(g) q=0, m=2, n=3, v=36
(h) q=0, m=2, n=3, v=40
(i) q=0, m=2, n=4, v=32
R
Y
Z
Γ
$k_y$
$k_z$
1
0

**Figure 4.** Anisotropic electron-phonon interactions in $Ta_2Ni_3Te_5$. (a) Schematic representation of the Raman scattering process for the bulk $A_g$ modes, including symmetry-allowed selection rules. (b) Symmetry-allowed optical transitions represented in the band structure of bulk $Ta_2Ni_3Te_5$, with irreducible representations marked for bands at Γ point. The blue (orange) arrow indicates the transition from $B_{2u}$ ($B_{2u}$) valence band to $A_g$ ($B_{3g}$) conduction band for the $y$-axis ($z$-axis) polarized light. Their anisotropic absorptions are shown in the insets. (c) Calculated polar plots of the electron-photon interactions and corresponding bulk $A_g$ phonon modes in the Raman scattering, where the anisotropy of EPIs is neglected. (d-f) Three representative calculated polar plots (168.1, 175.1, and 250.8 $cm^{-1}$) for monolayer $A_1$ modes, selected from Figure S16, corresponding to qualitatively distinct types of anomalous angular patterns observed in the 12 and 17 layers: (d) dominant lobes near the c-axis with weak perpendicular secondary lobes; (e) four lobes of comparable intensity along both axes; (f) dominant lobes perpendicular to the chain axis. The full quantum calculation incorporating anisotropic EPIs faithfully capture the qualitative characteristics and mode-to-mode diversity of the anomalous angular responses observed experimentally. (g-i) Anisotropic electron-phonon interaction maps in the first BZ of monolayer $Ta_2Ni_3Te_5$. The band symmetries at the Γ point are labeled in Figure S15a. (g, h) EPIs maps of the $v$=36 and $v$=40 $A_1$ phonon modes with vertical electronic transition between 3rd and 2nd conduction bands (two bands of the same $A_1$ symmetry). (i) is included to demonstrate that $\boldsymbol{k}$-space anisotropy also occurs for a different phonon symmetry: EPIs map of the $v$=32 $A_2$ phonon mode with vertical electronic transition between 4th ($A_2$ symmetry) and 2nd ($A_1$ symmetry) conduction bands. The high symmetry points in the BZ have been labelled. The vertical direction ($k_y$) is along the chains and the horizontal direction ($k_z$) is perpendicular to the chains.

We map the specific optical transitions onto the electronic band structure in Figure 4b to identify the resonant scattering in the bulk $Ta_2Ni_3Te_5$ (detailed calculations are provided in Supporting Information S4). The valence band at the $\Gamma$ point exhibits $B_{2u}$ symmetry. The dipole selection rules impose the optical transitions: when the crystal is excited with light polarized along the atomic chains ($y$-axis, $E \parallel b$), symmetry mandates a transition to the $A_g$ conduction band, indicated by the blue arrow together with its anisotropic absorption. Conversely, polarization perpendicular to the chains ($z$-axis, $E \parallel c$) couples exclusively to the $B_{3g}$ band, marked by the orange arrow.

The symmetry-allowed Raman processes can be analyzed through direct products of relevant interaction matrices. For the bulk $A_g$ mode, the intermediate states $|m\rangle$ and $|m'\rangle$ share the same symmetry (selection rules are summarized in Table S5). To understand the specific role of the electron-phonon interaction, we first consider the conventional expectation where this interaction is assumed to be isotropic. Under this assumption, the Raman response of bulk $Ta_2Ni_3Te_5$ should follow predictions driven by the optical selection rules, where the two electron-photon interaction matrices possess identical polarization dependence and the symmetry of which are the same for all $A_g$ modes, producing a simple two-fold symmetry in the $A_g$ mode (Figure 4c).

However, the experimental reality for the bulk $A_g$ mode exhibits dramatic mode-specificity and stands in sharp contrast to this simple two-fold prediction. This discrepancy indicates that if the electron-phonon interaction is treated as isotropic, even within the quantum framework, the dramatic variations among these anomalous $A_g$ modes still cannot be explained. The mode-specific anomaly is therefore unlikely to be a simple optical artifact. Instead, the electron-phonon interactions vary for each specific phonon mode, serving as the fundamental microscopic origin of the mode-specific anomaly. Herein, the mode-dependent anisotropic electron-phonon interactions mix the intermediate electronic states and impose a profound angular modulation on the scattering amplitude.

It is important to clarify that Figure 4c illustrates the result of bulk $A_g$ mode under the isotropic EPI assumption for the dominant scattering pathway, where the electron-photon matrix elements are zero at 90° by symmetry; the anisotropic EPI does not modify this single pathway but instead activates inter-band scattering channels whose coherent sum, weighted by their mode-specific EPI matrix elements, produces the finite intensity at 90° observed in Figures 4d–f.

The complete physical picture is revealed only when the anisotropic EPIs are fully activated within the quantum model. Due to the immense computational cost associated with ab initio calculations of these interactions in bulk $Ta_2Ni_3Te_5$, which exceeds our current resources, we performed the calculations for monolayer $Ta_2Ni_3Te_5$. Crucially, our DFPT implementation goes beyond simple symmetry arguments by rigorously calculating the EPI matrix elements for each scattering channel (Supporting Information S5.2). This explicitly accounts for how the specific orbital textures of the initial ($m$) and final ($m'$) electronic states couple to the lattice vibrations. Notably, although the point group symmetry reduces from $D_{2h}$ (bulk) to $C_{2v}$ (monolayer), the monolayer $A_1$ mode retains the same Raman tensor form as the bulk $A_g$ mode. Furthermore, the calculation results (Figure S15) demonstrate that the key aspects of above analysis persist in the monolayer and are consistent with the bulk findings (Figure S13). Figure 4d–f displays three representative calculated polar plots incorporating both anisotropic electron-photon interactions and rigorous anisotropic EPI matrix elements. The overall calculated angular responses and, in particular, the systematic mode dependence of the Raman patterns remain consistent with the symmetry-allowed framework and capture the qualitative characteristics and diversity of the anomalous angular responses observed experimentally in the 12- and 17-layer samples. It is important to emphasize that while anisotropic optical absorption is present, it alone cannot produce the mode-specific anomaly. It is therefore the highly anisotropic and mode-specific EPI that is the critical component required to replicate the experimentally observed mode-specific anomalies.

To visualize the deep origin of this directionality, we calculated the momentum-resolved EPI matrix in monolayer $Ta_2Ni_3Te_5$ to directly verify that such anisotropic EPIs are indeed critical features of this system (Supporting Information S5.3). The EPI matrix is defined as $g_{m,n,\nu}(\boldsymbol{k},\boldsymbol{q}) = \langle \psi_{n,\boldsymbol{k}+\boldsymbol{q}} | \partial_{\boldsymbol{q}\nu} V | \psi_{m,\boldsymbol{k}} \rangle$, where $\psi_{m,\boldsymbol{k}}$ is the electronic

wavefunction for band $m$ with wavevector $\boldsymbol{k}$, and $\partial_{\boldsymbol{q}\nu}V$ is the derivative of the self-consistent potential associated with a phonon mode $\omega_{\boldsymbol{q}\nu}$ of wavevector $\boldsymbol{q}$ and branch index $\nu$. Representative EPI maps over the entire first BZ of monolayer $Ta_2Ni_3Te_5$ are shown in Figure 4g-i (band symmetries at the Γ point are labeled in Figure S15a). This k-space mapping serves as the microscopic counterpart to the macroscopic Raman anomalies. As a representative example, Figure 4g depicts the EPI map for the monolayer $A_1$ phonon mode with branch $\nu$ =36, corresponding to the vertical electronic transition between the 3rd and 2nd conduction bands (two bands of the same $A_1$ symmetry), providing supportive evidence of the anisotropic EPIs. We note that the strict symmetry classification applies rigorously at the Γ point, where the bands can be labeled by the irreducible representations. Away from the Γ point, at generic wavevectors $\boldsymbol{k}$, the little group of $\boldsymbol{k}$ is lower than the full point group, so finite inter-band coupling develops continuously across the zone, which is precisely what the maps visualize. These maps therefore reveal the momentum-space anisotropy of the EPIs as an intrinsic property of $Ta_2Ni_3Te_5$. Although the detailed EPIs distributions vary among different phonon modes (Figure 4i depicts the EPI map for the monolayer $A_2$ phonon mode with branch $\nu$ =32), they illustrate that the pronounced $\boldsymbol{k}$-space anisotropic electron–phonon interactions emerge as a universal feature in $Ta_2Ni_3Te_5$. This indicates that phonon modes interact selectively and intensely with electronic states residing in specific regions of the BZ, rather than uniformly. Notably, for most phonon modes, the interaction is highly localized and enhanced along the *Γ-Z* direction (inter-chain), while remaining substantially weaker along the *Γ-Y* direction (parallel to the chain).

This result implies that while the electronic states are spatially confined along the quasi-1D Ta-Ni chains ($b$-axis), their energy levels are most critically perturbed by inter-chain lattice distortions. Phonons propagating along the *Γ-Z* direction ($c$-axis) effectively modulate the distance and hybridization between adjacent atomic chains. Given the sensitivity of the electronic band structure to the inter-chain coupling in this low-dimensional system, even small displacements perpendicular to the chains result in large deformation potentials.

Within the dipole approximation, the quantum Raman scattering intensity in Equation 1 can be reformulated as $I_v(E_L) = \left|\hat{P}_s \cdot \hat{R}_v \cdot \hat{P}_i\right|^2$, where $\hat{P}_i$ and $\hat{P}_s$ are the polarization vectors of the incident and scattered light respectively, and $R_v$ is the Raman tensor expressed within the quantum mechanical framework: (complete derivation is provided in Supporting Information S5.4)

$$\hat{R}_v = \sum_{i,m,m'} \frac{\langle i|\nabla|m'\rangle\langle m'|H_{ep}^{v}|m\rangle\langle m|\nabla|i\rangle}{(E_L - \Delta E_{mi})(E_L - \hbar\omega_v - \Delta E_{m'i})} \quad (2)$$

The complex Raman tensor, which can effectively fit the experimental data, originates from the anisotropic electron-photon and electron-phonon interactions. The quantum and tensor formalisms are therefore fully consistent and complementary: the tensor formalism characterizes the macroscopic scattering symmetry, while the quantum framework reveals the microscopic physical origin. Moreover, because the EPI matrix elements $\langle m' \mid H_{ep}^{v} \mid m \rangle$ are intrinsically mode-specific and momentum-dependent, the interference among the different inter-band scattering channels in the coherent sum differs from mode to mode. This is indeed the microscopic origin of the mode-specific complex phases of the Raman tensor elements: the anisotropic EPI matrix elements reweight the complex-valued contributions of the different inter-band channels, producing the mode-dependent phase differences observed experimentally (Figures 3e–h and S8). This quantum framework explains not only the anomalous phonon mode observed in the angle-dependent Raman spectroscopy but also the dramatic variations in the phase differences observed among different modes. Our results therefore provide microscopic support for interpreting the anomalous phonon response as a macroscopic manifestation of intrinsically anisotropic, mode-specific EPIs in $Ta_2Ni_3Te_5$. More broadly, the framework established here—specifically the identification of the mathematical degeneracy between optical artifacts and the use of mode-specific DFPT

calculations—provides a pathway for investigating anomalous phonon responses in other low-symmetry 2D materials, as also exemplified in triclinic $ReSe_2$.[38, 39]

In summary, we investigated quasi-1D $Ta_2Ni_3Te_5$ nanoflakes and resolved the origin of pronounced phonon anomalies by rigorous quantum treatment and DFPT implementation. By mapping the mode-specific and momentum-resolved EPIs landscape, we established a microscopic foundation for the experimental observations, revealing that the anomalies arise from complex, highly anisotropic quantum interactions inherent to the quasi-1D structure. This validation of strong, directionally selective EPIs highlights its potential as a pristine platform for manipulating phonon-mediated quantum phases, including excitonic insulating phase transition and topological superconductivity. More broadly, this work highlights a protocol for probing anisotropic interactions in van der Waals heterostructures, providing a framework for the design of next-generation direction-dependent quantum devices where electron-phonon interaction is the governing design parameter.

**ASSOCIATED CONTENT**

Supporting Information

The Supporting Information is available free of charge to readers.

Crystal growth and characterizations; mechanical exfoliations and optical/AFM images; TEM and STEM studies. First-principles calculations of phonon modes. Polar plots and corresponding fitted Raman tensor components. Derivations of the polarization dependent equations for Raman modes with real and complex Raman tensor elements. First-principles calculations of anisotropic electron-photon and electron-phonon interactions in bulk and monolayer $Ta_2Ni_3Te_5$. Complex Raman tensors calculation of $Ta_2Ni_3Te_5$. Anisotropic electron-phonon interactions mapping of $Ta_2Ni_3Te_5$ and graphene. Temperature-dependent XRD, Raman spectra and phonon decay process. Thickness-dependent polarized Raman measurements. Graphite encapsulation control experiment for surface degradation.

## AUTHOR INFORMATION

Corresponding Authors

Hong Tang - Department of Physics and Engineering Physics, Tulane University, New Orleans, Louisiana 70118, United States

Xin Lu - Department of Physics and Engineering Physics, Tulane University, New Orleans, Louisiana 70118, United States

Jiang Wei - Department of Physics and Engineering Physics, Tulane University, New Orleans, Louisiana 70118, United States

Authors

Fei Wang - Department of Physics and Engineering Physics, Tulane University, New Orleans, Louisiana 70118, United States

Qiaohui Zhou - Department of Physics and Engineering Physics, Tulane University, New Orleans, Louisiana 70118, United States

Fan Zhang - Department of Physics, The University of Texas at Austin, Austin, Texas 78712, United States

Yanxing Li - Department of Physics, The University of Texas at Austin, Austin, Texas 78712, United States

Steven A. Hindmarsh - Department of Physics, University of Warwick, Coventry, CV4 7AL, United Kingdom

Chengkun Xing - Department of Physics and Astronomy, University of Tennessee, Knoxville, Tennessee 37996, United States

Keyuan Bai - Department of Physics and Engineering Physics, Tulane University, New Orleans, Louisiana 70118, United States

Sidra Younus - Department of Physics and Engineering Physics, Tulane University, New Orleans, Louisiana 70118, United States

Ana M. Sanchez - Department of Physics, University of Warwick, Coventry, CV4 7AL, United Kingdom

Haidong Zhou - Department of Physics and Astronomy, University of Tennessee, Knoxville, Tennessee 37996, United States

Chih-Kang Shih - Department of Physics, The University of Texas at Austin, Austin, Texas 78712, United States

Adrienn Ruzsinszky - Department of Physics and Engineering Physics, Tulane University, New Orleans, Louisiana 70118, United States

Author Contributions

F.W. and J.W. conceived the project. F.W. did the single crystal growth, mechanical exfoliation, and corresponding characterizations. F.W. and Q.Z. carried out the spectroscopic measurements. H.T. and A.R. performed the first-principles calculations. F.Z., Y.L., and C.-K.S. carried out the STM measurements. F.W., S.A.H., A.M.S., K.B., and S.Y. prepared TEM samples and performed the characterizations. C.X. and H.Z. measured the temperature-dependent XRD. J.W., X.L., and A.R. supervised the project. F.W. wrote the manuscript with contributions from all authors.

Notes

The authors declare no competing financial interest.

**ACKNOWLEDGEMENTS**

This work is supported by the National Science Foundation under Grant 1752997 and the Louisiana Board of Regents under Grant 082ENH-22. We acknowledge the support from Tulane Instrumentation for Nanoscience & Innovation (TINI). We also acknowledge support from Prof. Julie Albert for atomic force microscopy. X.L. and Q.Z. acknowledge support from Tulane University startup fund and the Louisiana Board of Regents Support Fund (BoRSF) under award # LEQSF(2022-25)-RD-A-23. H.T. and A.R. acknowledge the support from Tulane University startup fund and the donors of ACS Petroleum Research Fund under New Directions Grant 65973-ND10 for the work on Raman spectroscopy. For the work on momentum-resolved electron-phonon interactions, A.R. acknowledges the U.S. Department of Energy, Office of Science, Office of Basic Energy Sciences, under Award Number DE-SC0026293. F.Z.,

Y.L. and C.-K.S. acknowledge the National Science Foundation under Grant DMR-2219610. The work at the University of Tennessee (C.X. and H.Z.) is supported by the U.S. Department of Energy under Grant No. DE-SC0020254.

**Supporting Information:**

# Resolving the Quantum Origin of Anisotropic Raman Response in Quasi-1D $Ta_2Ni_3Te_5$

*Fei Wang[1], Qiaohui Zhou[1], Hong Tang[1]*, Fan Zhang[2], Yanxing Li[2], Steven A. Hindmarsh[3], Chengkun Xing[4], Keyuan Bai[1], Sidra Younus[1], Ana M. Sanchez[3], Haidong Zhou[4], Chih-Kang Shih[2], Adrienn Ruzsinszky[1], Xin Lu[1]*, and Jiang Wei[1]**

[1]Department of Physics and Engineering Physics, Tulane University, New Orleans, Louisiana 70118, United States.

[2]Department of Physics, The University of Texas at Austin, Austin, Texas 78712, United States.

[3]Department of Physics, University of Warwick, Coventry, CV4 7AL, United Kingdom.

[4]Department of Physics and Astronomy, University of Tennessee, Knoxville, Tennessee 37996, United States.

*Emails: htang5@tulane.edu, xlu5@tulane.edu, jwei1@tulane.edu

**Navigational Summary of Supporting Information**

| **Key Topic** | **Section** | **Figures and Tables** |
|---|---|---|
| Crystal growth, EDS, and powder XRD characterization | S1.1 | Figure S1 |
| Mechanical exfoliation and AFM characterization of nanoflakes | S1.2 | Figure S2 |
| TEM and SAED characterization of nanoflake structure | S1.3 | Figures S3, S4 |
| Raman modes identification and DFT phonon mode calculations | S2 | Figures S5, S6, S7, Tables S1, S2 |
| Experimental polar plots of all observed Ag and $B_{3g}$ modes | S3 | Figures S8, S9 |
| Real Raman tensor derivations for $A_g$, $B_{3g}$, $B_{1g}$, $B_{2g}$ modes | S3.1 | Figure S10 |
| Complex Raman tensor derivations including light absorption | S3.2 | Figure S11 |
| Complex Raman tensor derivations including birefringence | S3.3 | — |

| Key Topic | Section | Figures and Tables |
| --- | --- | --- |
| Generalization of mathematical degeneracy to other low symmetry 2D materials ($WTe_2$, $MoTe_2$, $ReS_2$, $ReSe_2$) | S3.4 | — |
| Symmetry allowed optical transitions and selection rules in bulk $Ta_2Ni_3Te_5$ | S4 | Figures S12, S13, Tables S3, S4, S5, S6 |
| Symmetry allowed optical transitions and selection rules in monolayer $Ta_2Ni_3Te_5$ | S5.1 | Figures S14, S15, Tables S7, S8, S9, S10 |
| DFPT complex Raman tensor calculations incorporating anisotropic EPI (monolayer) | S5.2 | Figures S16, S17, Tables S11, S12 |
| Anisotropic EPI maps across the first Brillouin zone (monolayer $Ta_2Ni_3Te_5$ and graphene) | S5.3 | Figures S18, S19 |
| Quantum mechanical complex Raman tensor derivations | S5.4 | — |
| Temperature dependent XRD confirming structural stability | S6 | Figure S20 |
| Temperature dependent Raman spectra and phonon decay process | S6 | Figure S21, S22, Table S13 |
| Thickness dependent angle resolved Raman spectroscopy confirming intrinsic anomalous patterns | S7 | Figure S23, S24, S25 |
| Graphite encapsulation control experiment confirming surface degradation has no impact on Raman signal | S8 | Figure S26 |

## S1. Crystal growth and characterizations

### S1.1. Bulk crystal synthesis and characterizations

Large-scale and high-quality $Ta_2Ni_3Te_5$ single crystals were grown using vapor phase transport, as illustrated in Figure S1a. Stoichiometric elemental powders of tantalum (Alfa Aesar, 99.97% purity), nickel (Alfa Aesar, 99.99% purity), and tellurium (Alfa Aesar, 99.99% purity) were mixed with iodine as transport agent (9mg/cm$^3$) and placed in a quartz tube (13 × 1.6 cm). The ampule was evacuated and sealed at a pressure of around 5x10$^{-5}$ torr. Then, the ampule was placed in a double zone furnace with the source positioned in the hot end. The hot end was then heated to 950 °C while the cold end was kept at 850 °C. Then, the furnace was maintained at that temperature difference for two weeks. Finally, the furnace was naturally cooled to room temperature. $Ta_2Ni_3Te_5$ bulk crystals were obtained in the cold end.

Elemental composition of the crystal was examined by energy dispersive X-ray spectroscopy (EDS) (Figure S1b). The ratio of Ta:Ni:Te is 2:3:5.05. To verify the crystal structure, powder XRD was conducted on $Ta_2Ni_3Te_5$ bulk crystals, and the result was plotted in Figure S1c. Very strong diffraction peaks (upper side) have been observed and matched with the standard *(l00)* diffraction planes of $Ta_2Ni_3Te_5$, resulting from the 2D nature of the bulk crystals and indicating the high quality of as-grown crystals. As mentioned, $Ta_2Ni_3Te_5$ monolayers are stacked along the [100] direction, since $Ta_2Ni_3Te_5$ bulk crystals were ground into powders, there is high probability that the powder flakes are still stacked along the [100] direction, persisting the 2D nature of the crystal, as indicated by the strong *(l00)* diffraction peaks. The zoom-in pattern shows weaker diffraction peaks corresponding to different diffraction planes of $Ta_2Ni_3Te_5$, which have been labeled in the downside of Figure S1c. Our characterizations prove successful growth of high-quality $Ta_2Ni_3Te_5$ bulk crystals and support the 2D nature and in-plane anisotropic character in the crystal structure.

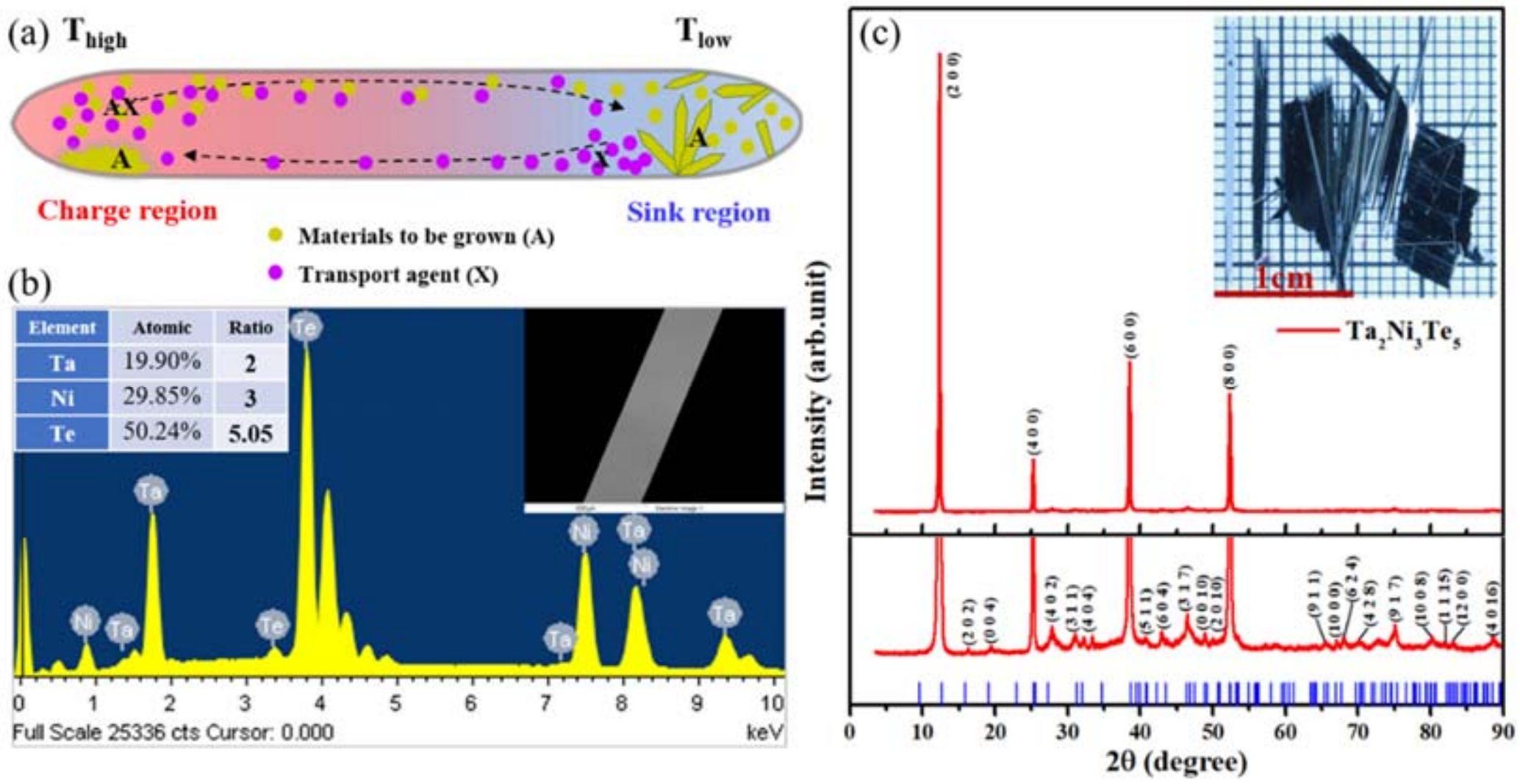

**Figure S1.** Synthesis and characterization of $Ta_2Ni_3Te_5$ bulk crystal. (a) Schematic of $Ta_2Ni_3Te_5$ bulk crystal growth via iodine vapor transport. (b) EDS spectrum of $Ta_2Ni_3Te_5$. The composition obtained from the spectrum data is shown in the table of the top left inset. The SEM image, top right inset, shows a typical $Ta_2Ni_3Te_5$ bulk crystal in a ribbon-like shape with parallel edges. (c) Powder XRD pattern of $Ta_2Ni_3Te_5$ crystals. Red and blue lines denote the measured data and reference of $Ta_2Ni_3Te_5$. The downside shows the zoom-in powder XRD diffraction peaks of $Ta_2Ni_3Te_5$. Inset presents the optical image of $Ta_2Ni_3Te_5$ bulk crystals after taking out the crystals from the quartz tube.

### S1.2. Mechanical exfoliation and nano-flake morphology

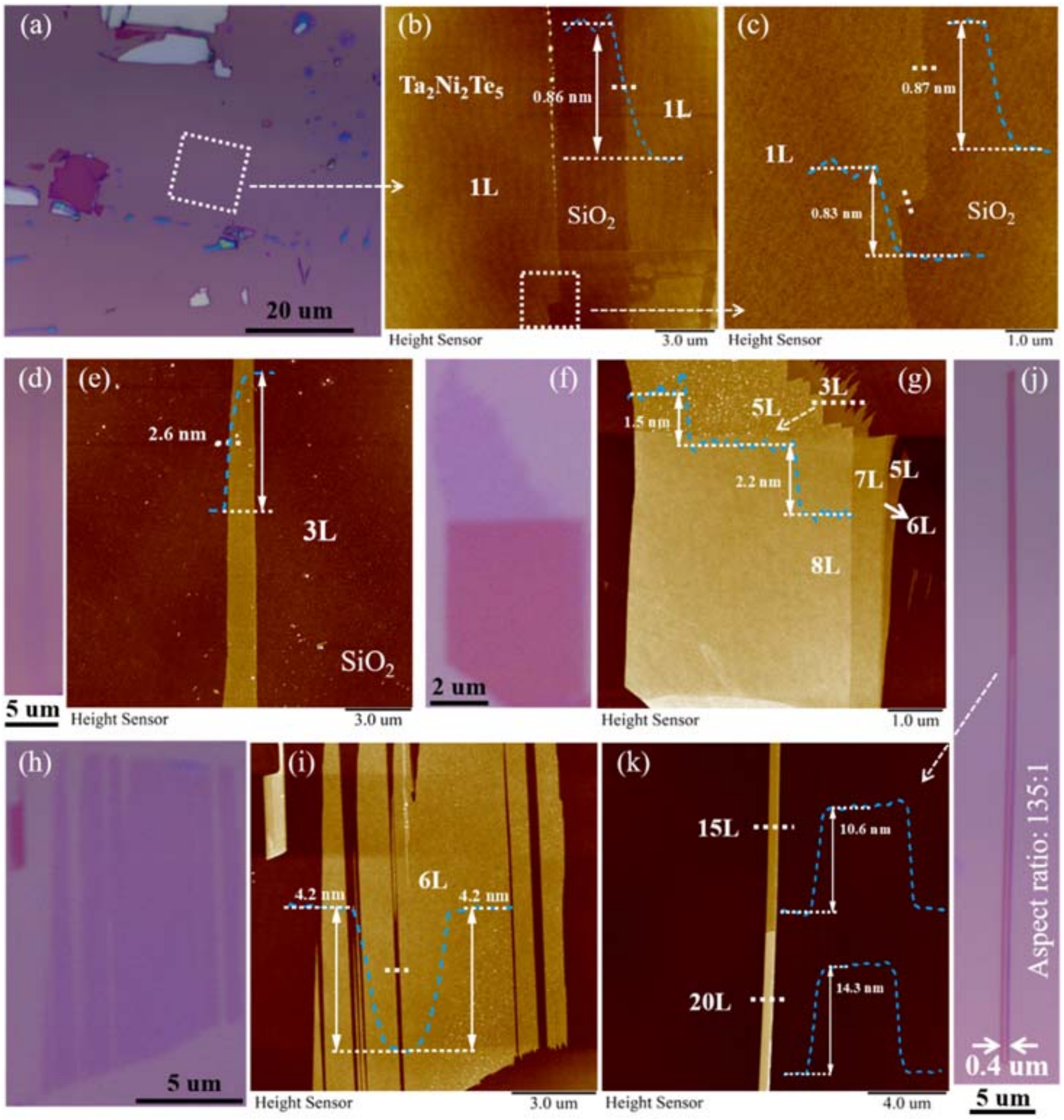

**Figure S2.** The exfoliation and corresponding AFM characterizations of monolayer and few-layer $Ta_2Ni_3Te_5$ nano-flakes.

Leveraging the layered structure and weak interlayer vdW interactions, mechanical exfoliation was subsequently employed to obtain $Ta_2Ni_3Te_5$ nano-flakes for comprehensive characterization. Figure S2 documents the exfoliation and atomic force microscopy (AFM) characterizations of nano-flakes with varying thicknesses. Exfoliated thin $Ta_2Ni_3Te_5$ flakes became significantly transparent at reduced thicknesses, exhibiting a subtle red or pink appearance on Si substrate with 280 nm $SiO_2$. Consequently, AFM was employed for accurate characterization of the morphology and thicknesses of the exfoliated nano-flakes. As shown in Figure S2a and d, 1L and 3L $Ta_2Ni_3Te_5$ flakes appear extremely transparent, making them nearly indistinguishable from the substrate optically. Notably, Figure S2j illustrates exfoliated flakes displaying an elongated nanoribbon-like shape with an exceptionally high aspect ratio up to 135:1. AFM images confirm that the exfoliated flakes possess smooth and clean surfaces with remarkably straight edges along a specific orientation, which will be discussed in the TEM and STM studies. These results provide supporting evidence for the high quality of exfoliated nano-flakes. The successful exfoliation down to few-layer and even monolayer $Ta_2Ni_3Te_5$ flakes not only facilitates the investigation of intriguing low-dimensional and anisotropic physical properties but also pave the way for further studies on high-order topology in $Ta_2Ni_3Te_5$. Moreover, the straight, pristine long edges aligned with the quasi-1D chain in these exfoliated elongated nanoribbon-like thin flakes, where the edges are dominant, will provide a unique platform for exploring fundamental topological edge states and 1D physics, potentially advancing edge-state-based device applications.

## S1.3. TEM studies of the anisotropic crystal structure on the nano-flakes

To investigate the in-plane anisotropy, we developed a "pick-up-and-transfer" technique to conduct transmission electron microscopy (TEM) studies on the nanoflakes (Figure S3a).

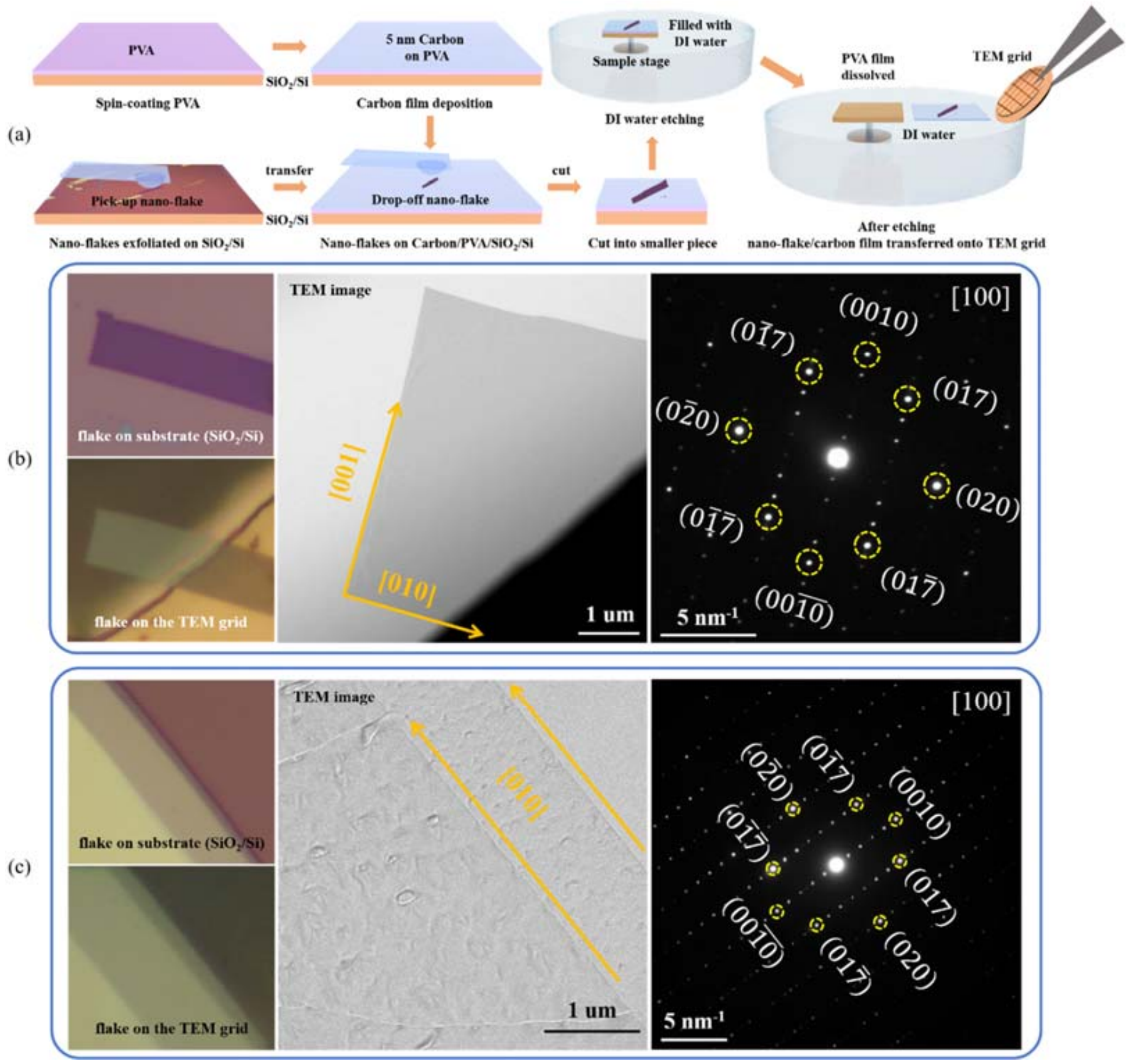

**Figure S3.** (a) Schematic representation of the "pick-up-and-transfer" technique developed for the TEM study on the exfoliated $Ta_2Ni_3Te_5$ nano-flakes. (b) and (c) show the optical images of the nano-flakes exfoliated on the substrate and transferred onto TEM grid, together with the corresponding TEM images and [100] zone (perpendicular to layers) selected area electron diffraction (SAED) pattern.

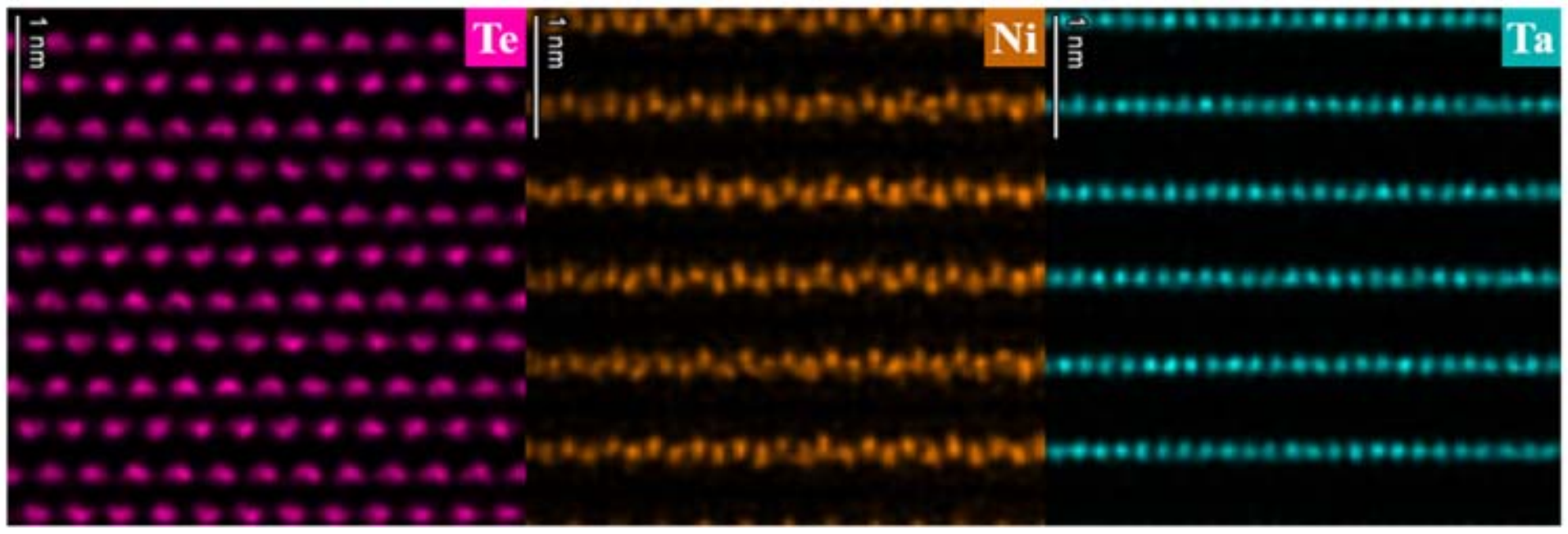

**Figure S4.** Elemental mappings of the side surface represent the layered structure, which is consistent with the description in the main text.

**S2. Raman spectroscopy and phonon properties of $Ta_2Ni_3Te_5$**

The micro-Raman spectra were measured using a home-built setup. Measurements were conducted in backscattering geometry with a 532 nm laser. The Raman response in $Ta_2Ni_3Te_5$ can be understood through the Raman tensor of the point group. The Raman tensor elements of this material are given by the derivative of the dielectric tensor. Depending on the point group of the material, the tensor's elements have different values and hence different optical and electronic properties along those directions. Unlike most layered 2D materials, $Ta_2Ni_3Te_5$ hosts a large and rather complex unit cell, each unit cell contains 40 atoms with two monolayers of 20 atoms each, this could make the phonon modes very complex, which will be discussed in detail in the following parts. Based on the symmetry of the $D_{2h}$ point group, shown in Table S1,[1] $Ta_2Ni_3Te_5$ has 120 eigenmodes at the zone center Γ points predicted through group theory, including 3 acoustic and 117 optical modes. The irreducible representation of the zone center phonon modes is $\Gamma = 20A_g + 10B_{1g} + 20B_{2g} + 10B_{3g} + 10A_u + 19B_{1u} + 9B_{2u} + 19B_{3u} + (B_{1u} + B_{2u} + B_{3u})_{acoustic}$,[1] where $A_g$, $B_{1g}$, $B_{2g}$, and $B_{3g}$ are Raman-active modes. Of the 60 Raman-active modes, only the 30 modes from $A_g$ and $B_{3g}$ modes can be detected in the backscattering geometry according to the selection rules, where the light travels along the *a*-axis and the polarization of the light is in the *bc* plane. The complementary measurement on a 12L flake is presented in Fig. S5. To assign each observed Raman peak to its corresponding phonon mode, we performed DFT calculations employing multiple functionals.

**Table S1.** Raman tensors of $Ta_2Ni_3Te_5$

| **Mode** | $\mathbf{A_g}$ | $\mathbf{B_{1g}}$ | $\mathbf{B_{2g}}$ | $\mathbf{B_{3g}}$ |
|---|---|---|---|---|
| **Tensor** | $\begin{pmatrix} a & 0 & 0 \\ 0 & b & 0 \\ 0 & 0 & c \end{pmatrix}$ | $\begin{pmatrix} 0 & d & 0 \\ d & 0 & 0 \\ 0 & 0 & 0 \end{pmatrix}$ | $\begin{pmatrix} 0 & 0 & e \\ 0 & 0 & 0 \\ e & 0 & 0 \end{pmatrix}$ | $\begin{pmatrix} 0 & 0 & 0 \\ 0 & 0 & f \\ 0 & f & 0 \end{pmatrix}$ |

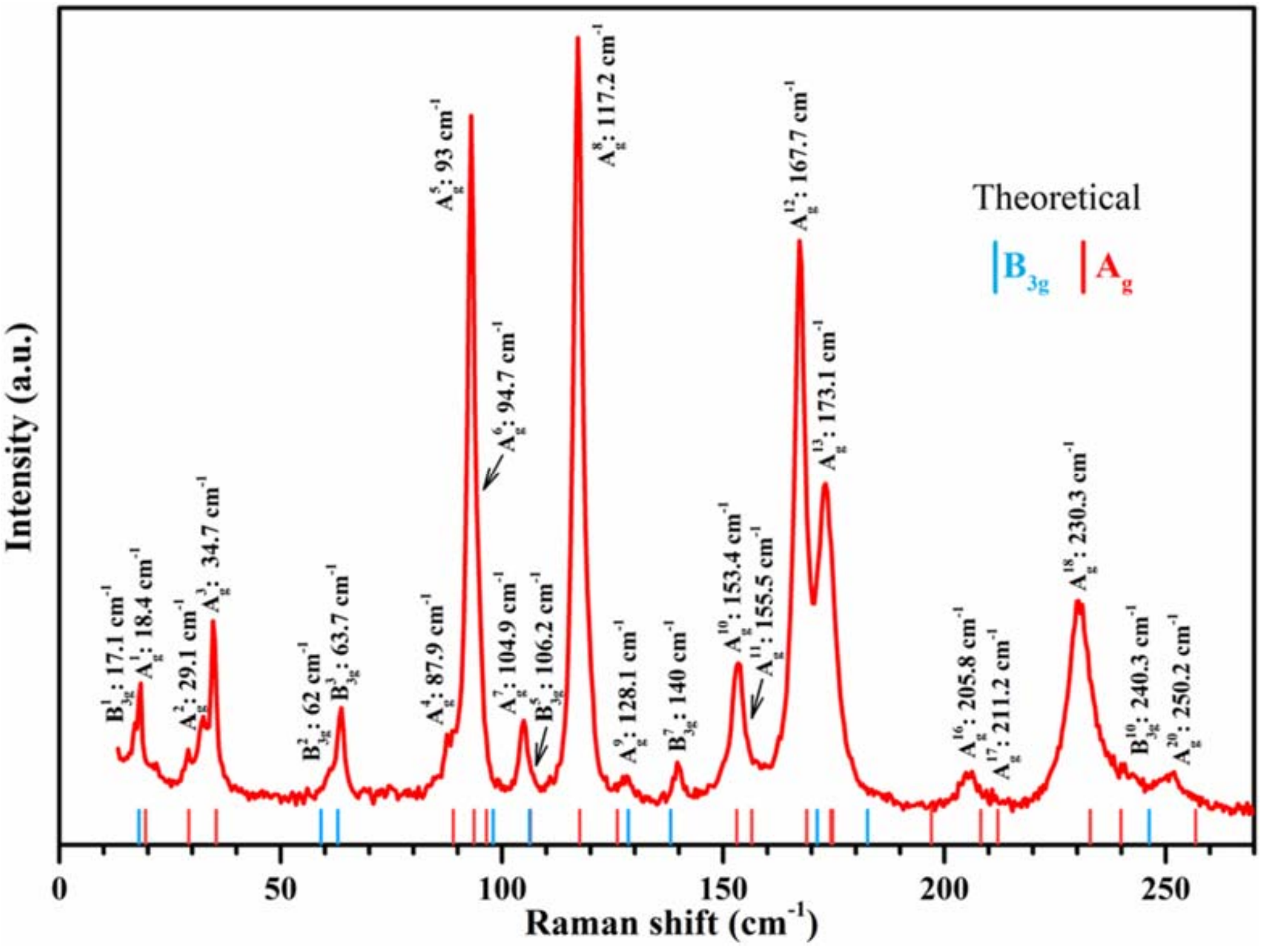

**Figure S5.** Raman spectrum of the 12L $Ta_2Ni_3Te_5$ flake, with calculated Raman peaks indicated by red ($A_g$ modes) and blue ($B_{3g}$ modes) vertical lines.

**Table S2. All Raman modes in $Ta_2Ni_3Te_5$: first-principles calculations and experimental observations.**

| Irreducible representation | Raman shift ($cm^{-1}$) PBE-D3 | Raman shift ($cm^{-1}$) PBE | Experiment (9L) | Experiment (12L) |
|---|---|---|---|---|
| $B_{3g}^{1}$ | 21.09283 | 18.00459 | 18.6 | 17.1 |
| $A_{g}^{1}$ | 22.38517 | 19.43626 | 20.3 | 18.4 |
| $A_{g}^{2}$ | 30.40324 | 29.22605 | 29.3 | 29.1 |
| $A_{g}^{3}$ | 37.90397 | 35.42893 | 35.3 | 34.7 |
| $B_{3g}^{2}$ | 58.77421 | 59.07899 | 61.4 | 62 |
| $B_{3g}^{3}$ | 61.4584 | 62.8975 | 63.1 | 63.7 |
| $A_{g}^{4}$ | 89.57149 | 89.01606 | 87.4 | 87.9 |
| $A_{g}^{5}$ | 92.87634 | 93.66553 | 92.9 | 93 |
| $A_{g}^{6}$ | 95.3375 | 96.47436 | 94.1 | 94.7 |
| $B_{3g}^{4}$ | 97.51518 | 97.99251 | -- | -- |
| $B_{3g}^{5}$ | 101.64915 | 106.21419 | 106.5 | 106.2 |
| $A_{g}^{7}$ | 110.25045 | 106.39322 | 105.2 | 104.9 |
| $A_{g}^{8}$ | 117.92629 | 117.51431 | 117 | 117.2 |
| $A_{g}^{9}$ | 124.24839 | 126.02662 | 128 | 128.1 |
| $B_{3g}^{6}$ | 126.06078 | 128.57525 | -- | -- |
| $B_{3g}^{7}$ | 136.93867 | 138.10371 | 139.4 | 140 |
| $A_{g}^{10}$ | 154.78786 | 153.09349 | 153.3 | 153.4 |
| $A_{g}^{11}$ | 157.70509 | 156.46669 | 154.9 | 155.5 |
| $A_{g}^{12}$ | 167.54466 | 168.90786 | 167.6 | 167.7 |
| $B_{3g}^{8}$ | 167.06888 | 171.26311 | -- | -- |
| $A_{g}^{13}$ | 175.42073 | 174.3293 | 173.1 | 173.1 |
| $A_{g}^{14}$ | 176.42754 | 174.74196 | -- | -- |
| $B_{3g}^{9}$ | 180.71787 | 182.61547 | -- | -- |
| $A_{g}^{15}$ | 197.91217 | 197.01753 | -- | -- |
| $A_{g}^{16}$ | 200.80478 | 208.21559 | 204.9 | 205.8 |
| $A_{g}^{17}$ | 209.6008 | 212.04475 | 211.5 | 211.2 |
| $A_{g}^{18}$ | 231.02397 | 232.979 | 229.9 | 230.3 |
| $A_{g}^{19}$ | 240.48242 | 239.90888 | -- | -- |
| $B_{3g}^{10}$ | 243.4254 | 246.23045 | 240.3 | 240.3 |
| $A_{g}^{20}$ | 256.35367 | 256.77104 | 249.9 | 250.2 |

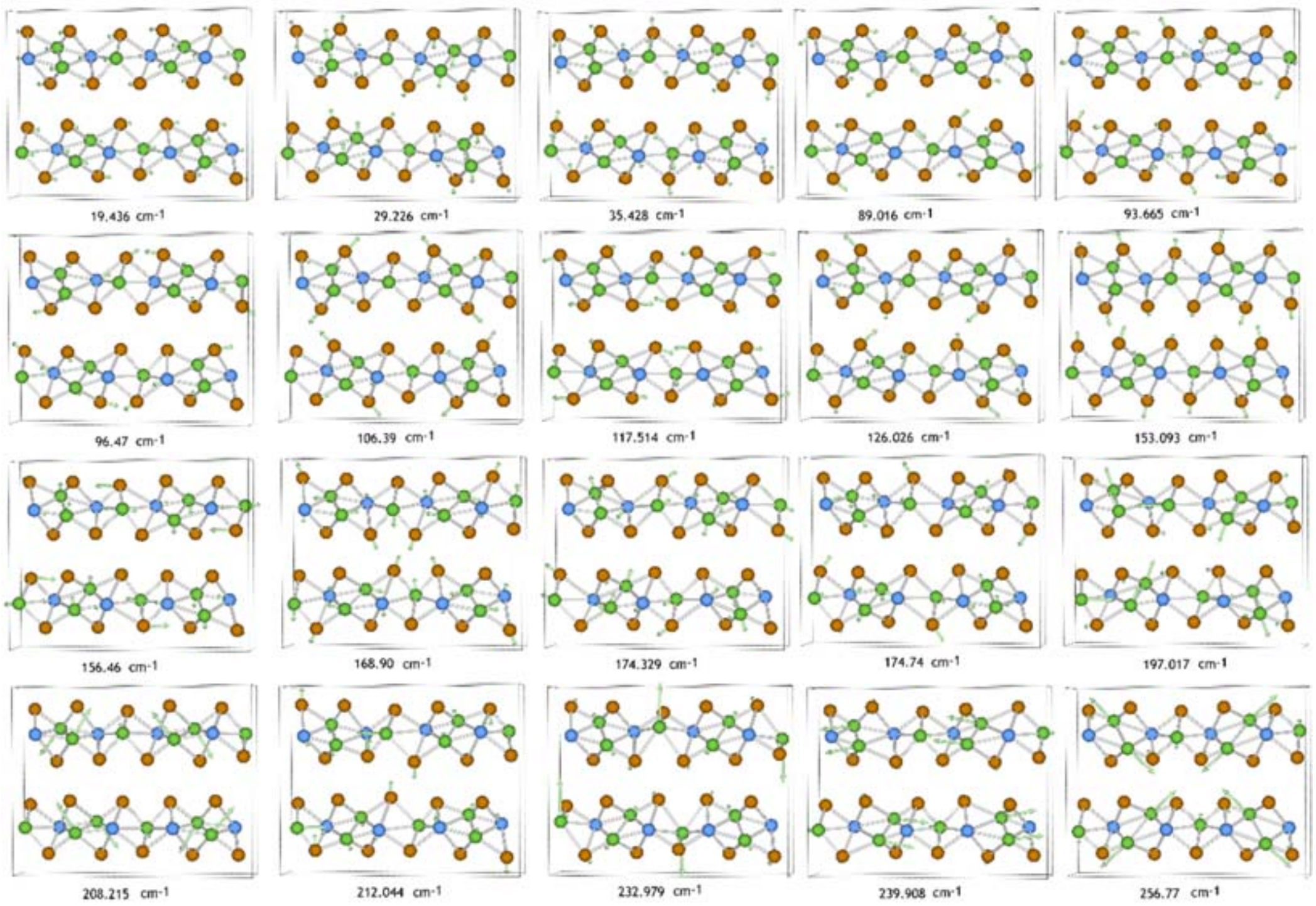

**Figure S6.** The vibrations of atoms in the 20 $A_g$ Raman modes of bulk $Ta_2Ni_3Te_5$. The blue sphere is Ta, green sphere is Ni, and red sphere is Te. The atom color code is different from that in Figure 1, due to different generating software. The vibrations of the atoms in the $A_g$ modes are mainly in the *ac* plane, perpendicular to the atomic chains (the *b*-axis). The number at each plot is the frequency of the mode in $cm^{-1}$.

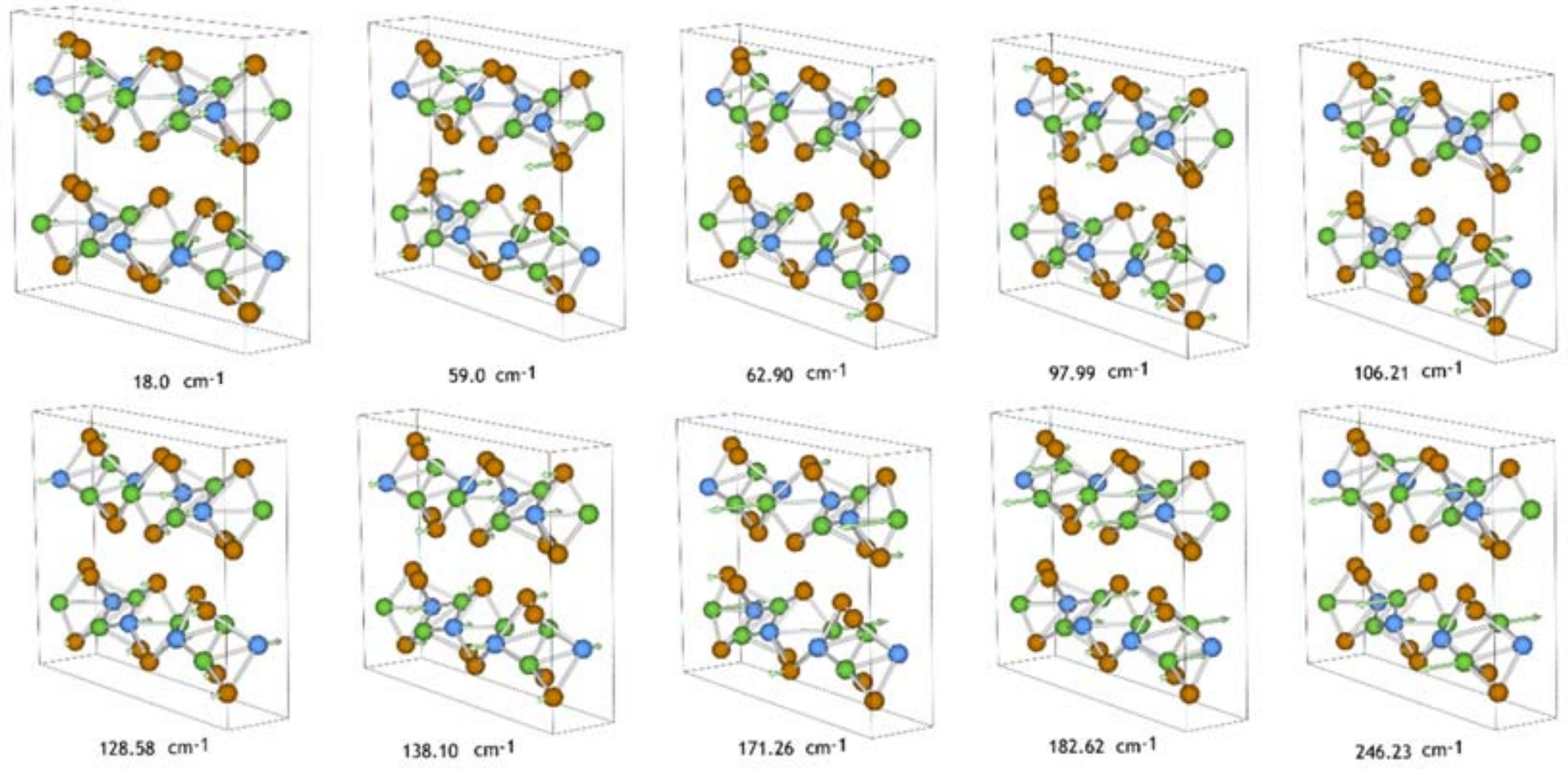

**Figure S7.** The vibrations of atoms in the 10 $B_{3g}$ Raman modes of bulk $Ta_2Ni_3Te_5$. The blue sphere is Ta, green sphere is Ni, and red sphere is Te. The atom color code is different from that in Figure 1, due to different generating software. The vibrations of the atoms in the $B_{3g}$ modes are mainly along the atomic chain direction (the *b*-axis). The number at each plot is the frequency of the mode in $cm^{-1}$.

**S3. Angle-resolved polarized Raman spectroscopy and unusual phonon response of the $A_g$ Raman modes in $Ta_2Ni_3Te_5$**

In this experimental section, angle-resolved polarized Raman spectroscopy measurements were conducted by rotating the sample in the *XY* plane as a function of varying rotational angles from 0° to 360° with a step of 10°. The Cartesian coordinates adopted here were defined as the laboratory coordinates of the *XYZ* stage, as illustrated in Figure 3a. The laser beam propagates along the *Z* direction (the *a*-axis of the crystal structure) with the incident light polarized always along the vertical (*Y*) direction, and an analyzer, placed before the entrance of the spectrometer, allowed for the analysis of the scattered light polarized parallel to the incident light polarization (parallel configuration). In the very beginning, the exfoliated nanoribbon-like $Ta_2Ni_3Te_5$ thin flakes (optical image is shown in Figure 3a) was put on the sample stage with the long edge (*b*-axis or atomic chain direction in the crystal structure) carefully aligned with the *Y* direction, which means that the incident light is polarized along the chain direction at zero degree. Since we have used a backscattering geometry and the laser propagates along the *a*-axis of the crystal structure, we can only analyze the polarized light in the *bc* plane. From the Raman tensor elements in Table S1, the spectra in this configuration will only show the $A_g$ and $B_{3g}$ modes because the $B_{1g}$ and $B_{2g}$ modes only have *ab* and *ac* non-null components, respectively.

The corresponding polar plots of each observed Raman modes are presented in Figure S8 and S9. The Raman peaks at 29.1, 34.7, 104.9, and 230.3 $cm^{-1}$ have 2-fold symmetry (two-lobed shape) with maximum intensities at 90° and 270°. On the other hand, the peaks at 87.9, 93.0, 94.7, and 173.1 $cm^{-1}$ show maximum intensities at 90° and 270° with a secondary local maximum at 0° and 180°, which is opposite for the peak at 153.4 $cm^{-1}$. There are peaks show clear four-lobe angular patterns. However, while the peaks at 17.1, 62, 63.7, 140.0, and 240.3 $cm^{-1}$ have the four lobes rotated 45° from the vertical and horizontal axis, the peaks at 18.4, 117.2, and 128.1 $cm^{-1}$ have the four lobes oriented closer to the axis. Meanwhile, the Raman peaks at 167.7 $cm^{-1}$ displays angle-dependent intensities that the symmetry cannot be easily concluded in the present situation. The peak at 250.2 $cm^{-1}$ looks to be angle-independent. The Raman mode symmetry will be discussed below based on the Raman tensor and experimental configuration. The absolute values of the obtained Raman tensor components for each peak are also presented here.

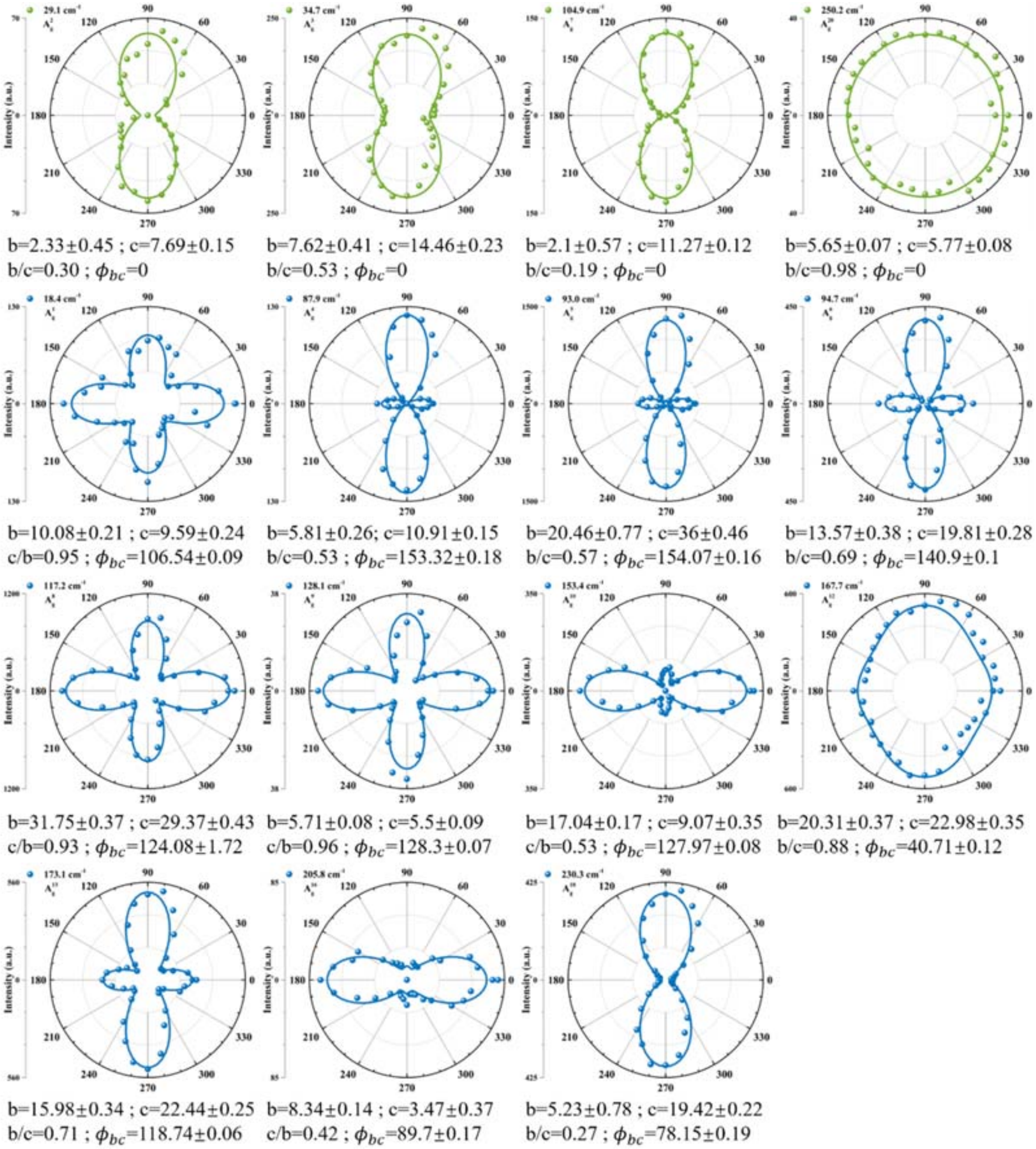

**Figure S8.** The polar plots and corresponding fitted Raman tensor components of each observed $A_g$ Raman modes. Experimental data are represented as dots, while numerical fittings are plotted in solid lines. The solid lines represent the fitted results obtained with considering complex Raman tensor.

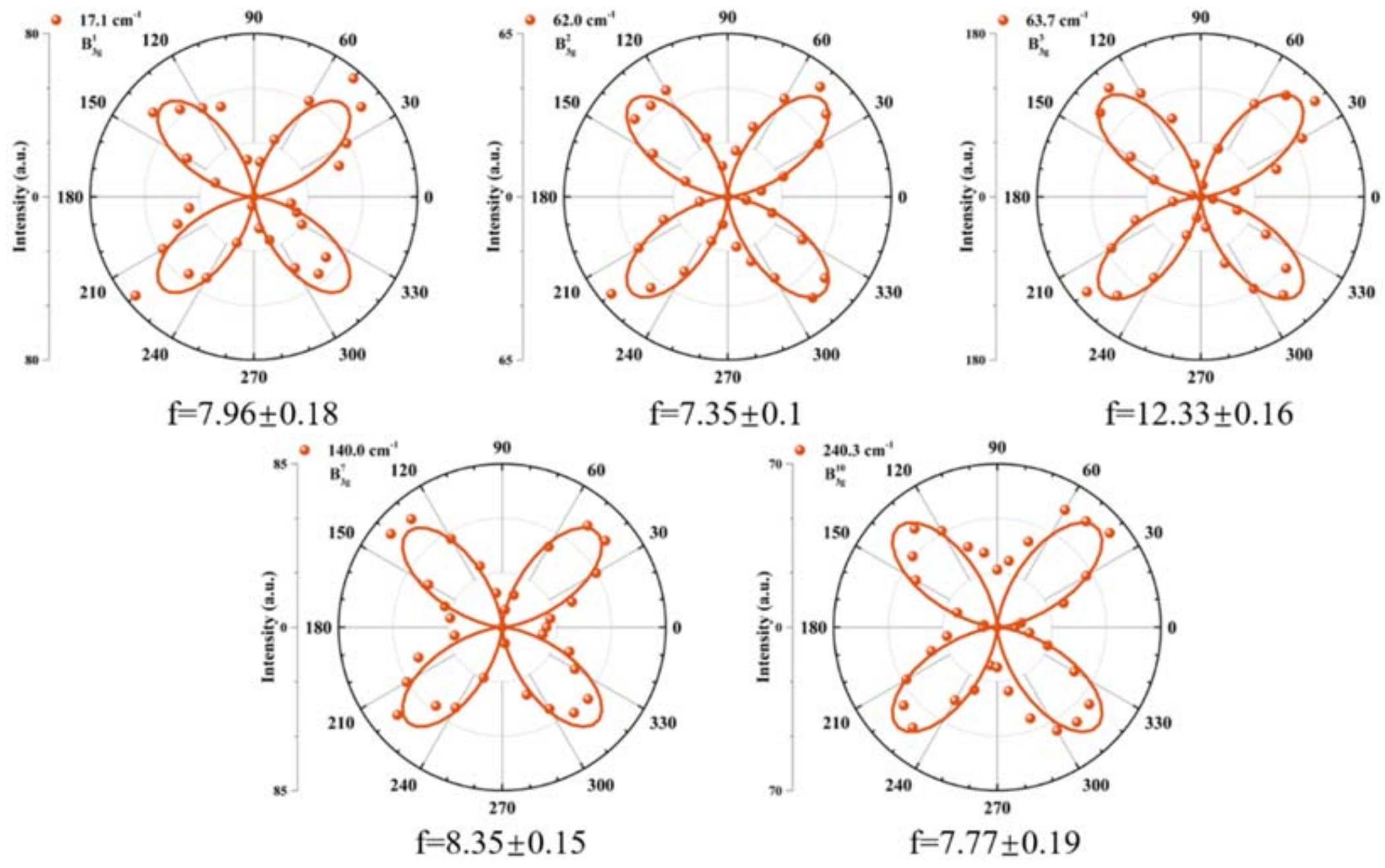

**Figure S9.** The polar plots and corresponding fitted Raman tensor components of each observed $B_{3g}$ Raman modes. Experimental data are represented as dots, while numerical fittings are plotted in solid lines. The solid lines represent the fitted results obtained with considering complex Raman tensor.

## S3.1. Derivation of the light-polarization dependent equations for the Raman modes with real Raman tensor elements

In the parallel configuration where the sample is rotated while keeping the polarization and analyzer fixed, we can write the unitary vectors as a function of angle $\theta$ with respect to the b crystal axis as follows: for the incident beam, $\hat{e}_i = (0 \cos\theta \sin\theta)$, while for the scattered light $\hat{e}_s = (0 \cos\theta \sin\theta)$.

The Raman cross section of a phonon mode that gives the scattering light intensity (*I*), can be written as: [1]

$$I \propto \left|\hat{e}_i \cdot \hat{R}_{ij} \cdot \hat{e}_s\right|^2 \tag{S1}$$

*S3.1.1. The $A_g$ modes in the parallel polarization configuration*

Utilizing the Raman tensor for the $A_g$ modes given in Table S1, we can obtain the angular dependence of the intensities of the $A_g$ modes:

$$\hat{R}_{A_g} = \begin{pmatrix} a & 0 & 0 \\ 0 & b & 0 \\ 0 & 0 & c \end{pmatrix}$$

$$I_{A_g} \propto \left| (0\; cos\,\theta\; sin\,\theta) \begin{pmatrix} a & 0 & 0 \\ 0 & b & 0 \\ 0 & 0 & c \end{pmatrix} \begin{pmatrix} 0 \\ cos\,\theta \\ sin\,\theta \end{pmatrix} \right|^2 \propto \left| (0\; bcos\,\theta\; csin\,\theta) \begin{pmatrix} 0 \\ cos\,\theta \\ sin\,\theta \end{pmatrix} \right|^2$$

$$\boldsymbol{I_{A_g} \propto \left| b * cos^2\,\theta + c * sin^2\,\theta \right|^2 \propto b^2\,cos^4\,\theta + \; c^2\,sin^4\,\theta + 2bc\,cos^2\,\theta\;sin^2\,\theta}$$

(S2)

In this configuration, the Raman intensity of $A_g$ mode shows a 2-fold symmetry that the maximum of two lobes varies with the ratio of the Raman tensor elements *b/c*. When $b<c$ ($b>c$), the maximum intensities will be at 90°/270° (0°/180°), especially when $b<<c$ ($b>>c$), the Raman intensity of the mode will be close to zero at 0°/180° (90°/270°), which means the corresponding Raman peak cannot be observed in the spectrum at 0°/180° (peaks at 29.1 and 104.9 cm$^{-1}$). Moreover, when $b=c$, the mode will show no angular dependence (peak at 250.2 cm$^{-1}$). Therefore, the peaks at 29.1, 34.7, 104.9, and 250.2 cm$^{-1}$ are $A_g$ modes. The simulation of the Raman intensity of $A_g$ modes with different *b/c* ratio presented in the polar plot is shown in Figure S10a.

*S3.1.2. The $B_{3g}$ modes in the parallel polarization configuration*

Utilizing the Raman tensor for the $B_{3g}$ modes given in Table S1, we can obtain the angular dependence of the intensities of the $B_{3g}$ modes:

$$\hat{R}_{B_{3g}} = \begin{pmatrix} 0 & 0 & 0 \\ 0 & 0 & f \\ 0 & f & 0 \end{pmatrix}$$

$$I_{B_{3g}} \propto \left| (0\; cos\,\theta\; sin\,\theta) \begin{pmatrix} 0 & 0 & 0 \\ 0 & 0 & f \\ 0 & f & 0 \end{pmatrix} \begin{pmatrix} 0 \\ cos\,\theta \\ sin\,\theta \end{pmatrix} \right|^2 \propto \left| (0\; fsin\theta\; fcos\theta) \begin{pmatrix} 0 \\ cos\,\theta \\ sin\,\theta \end{pmatrix} \right|^2$$

$$\boldsymbol{I_{B_{3g}} \propto |fsin\,\theta * cos\,\theta + fcos\,\theta * sin\,\theta|^2 \propto |2\,f\,(sin\,\theta * cos\,\theta)|^2 \propto |f\,sin\,2\theta|^2}$$

(S3)

In this configuration, the Raman intensity of $B_{3g}$ mode shows a four-lobe angular patterns that the intensities reach the maximum at 45°/135°/225°/315° with 90° variation period. The simulation of the Raman intensity of $B_{3g}$ modes presented in the polar plot is shown in Figure S10b.

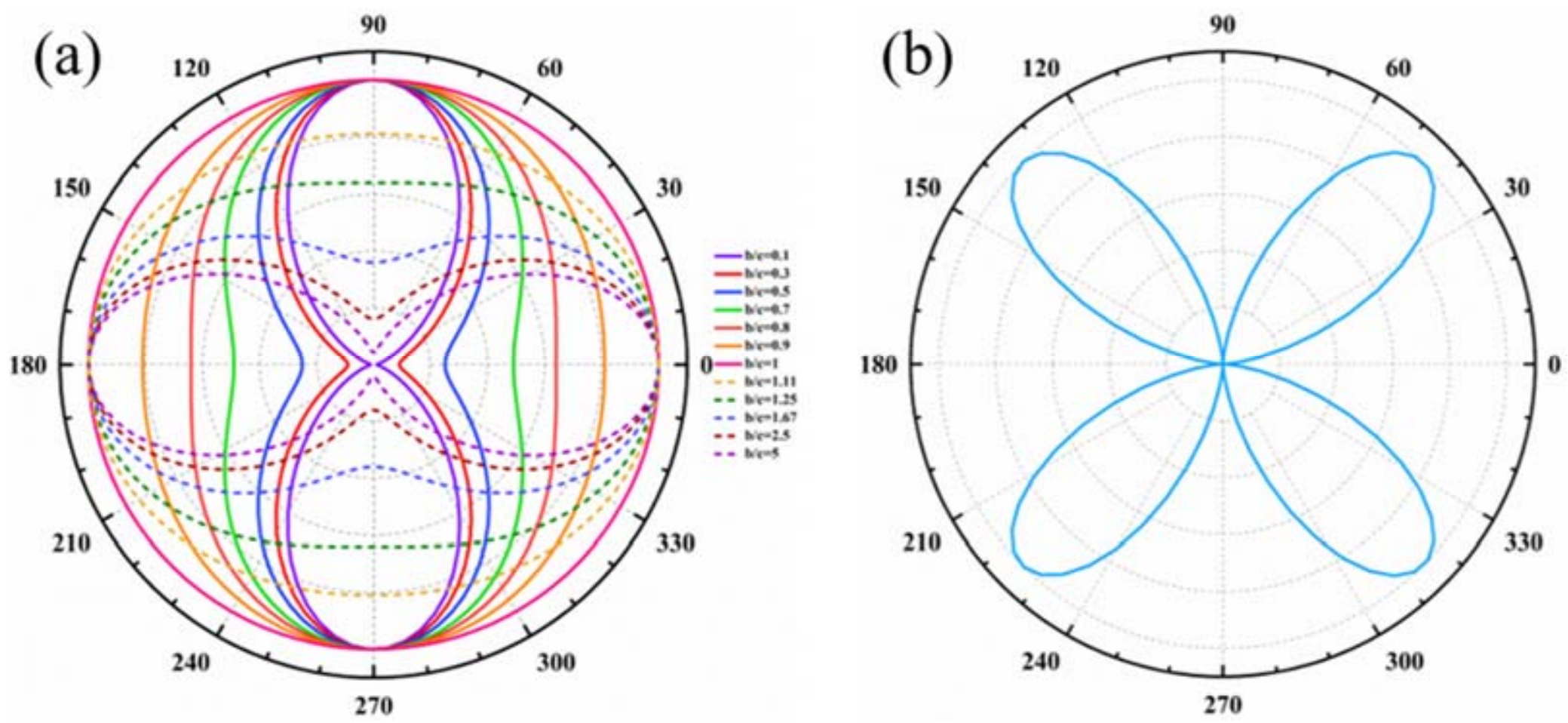

**Figure S10.** (a) Simulations of the Raman intensity of $A_g$ modes with different *b/c* ratio. (b) Simulation of the Raman intensity of $B_{3g}$ modes presented in the polar plot.

*S3.1.3. The $B_{1g}$ modes in the parallel polarization configuration*

Utilizing the Raman tensor for the $B_{1g}$ modes given in Table S1, we can obtain the angular dependence of the intensities of the $B_{1g}$ modes:

$$\hat{R}_{B_{1g}} = \begin{pmatrix} 0 & d & 0 \\ d & 0 & 0 \\ 0 & 0 & 0 \end{pmatrix}$$

$$I_{B_{1g}} \propto \left| (0\, cos\,\theta\; sin\,\theta) \begin{pmatrix} 0 & d & 0 \\ d & 0 & 0 \\ 0 & 0 & 0 \end{pmatrix} \begin{pmatrix} 0 \\ cos\,\theta \\ sin\,\theta \end{pmatrix} \right|^2 \propto |0|^2 \qquad \text{(S4)}$$

The intensities of $B_{1g}$ modes are zero, they cannot be observed for this scattering configuration.

*S3.1.4. The $B_{2g}$ modes in the parallel polarization configuration*

Utilizing the Raman tensor for the $B_{2g}$ modes given in Table S1, we can obtain the angular dependence of the intensities of the $B_{2g}$ modes:

$$\hat{R}_{B_{2g}} = \begin{pmatrix} 0 & 0 & e \\ 0 & 0 & 0 \\ e & 0 & 0 \end{pmatrix}$$

$$I_{B_{2g}} \propto \left| (0\, cos\,\theta\; sin\,\theta) \begin{pmatrix} 0 & 0 & e \\ 0 & 0 & 0 \\ e & 0 & 0 \end{pmatrix} \begin{pmatrix} 0 \\ cos\,\theta \\ sin\,\theta \end{pmatrix} \right|^2 \propto |0|^2 \qquad \text{(S5)}$$

The intensities of $B_{2g}$ modes are zero, they cannot be observed for this scattering configuration.

## S3.2. Raman modes with complex Raman tensor elements when considering light absorption

In the absorptive materials, each component $\epsilon_{ij}$ of the dielectric function tensor has real and imaginary parts and can be written as $\epsilon_{ij} = \epsilon'_{ij} + i\epsilon''_{ij}$. The element $R^k_{ij}$ of the Raman tensor is given by the derivative of the dielectric function element $\epsilon_{ij}$ with respect to the normal coordinate $q^k$: [1, 2]

$$R^k_{ij} = \frac{\partial \epsilon_{ij}}{\partial q^k} = \frac{\partial \epsilon'_{ij}}{\partial q^k} + i\frac{\partial \epsilon''_{ij}}{\partial q^k} \tag{S6}$$

Consequently, when there is absorption of light in the material, the Raman tensor elements also present complex values, with real and imaginary parts. The Raman tensor elements relevant to this work (*a*, *b*, *c*, and *f*) can thus be written as:

$$a = |a|e^{i\phi_a};\ b = |b|e^{i\phi_b};\ c = |c|e^{i\phi_c};\ f = |f|e^{i\phi_f}; \tag{S7}$$

where $\phi = arctg\ \frac{R^{Imaginary}_{ij}}{R^{Real}_{ij}} = arctg\ \frac{\frac{\partial \epsilon''_{ij}}{\partial q^k}}{\frac{\partial \epsilon'_{ij}}{\partial q^k}}$. Thus, the phases of the Raman tensor elements are given by:

$$\phi_a = arctg\ \frac{\frac{\partial \epsilon''_{aa}}{\partial q^{A_g}}}{\frac{\partial \epsilon'_{aa}}{\partial q^{A_g}}};\ \phi_b = arctg\ \frac{\frac{\partial \epsilon''_{bb}}{\partial q^{A_g}}}{\frac{\partial \epsilon'_{bb}}{\partial q^{A_g}}};\ \phi_c = arctg\ \frac{\frac{\partial \epsilon''_{cc}}{\partial q^{A_g}}}{\frac{\partial \epsilon'_{cc}}{\partial q^{A_g}}};\ \phi_f = arctg\ \frac{\frac{\partial \epsilon''_{cb}}{\partial q^{B_{3g}}}}{\frac{\partial \epsilon'_{cb}}{\partial q^{B_{3g}}}};$$

*S3.2.1. The $A_g$ modes in the parallel polarization configuration with considering light absorption:*

Substituting the real Raman tensor elements by the complex tensor elements, given by Equation S7, the angular dependencies for the $A_g$ modes are now given by:

$$\hat{R}_{A_g} = \begin{pmatrix} |a|e^{i\phi_a} & 0 & 0 \\ 0 & |b|e^{i\phi_b} & 0 \\ 0 & 0 & |c|e^{i\phi_c} \end{pmatrix}$$

$$I_{A_g} \propto \left| (0\ cos\,\theta\ sin\,\theta) \begin{pmatrix} |a|e^{i\phi_a} & 0 & 0 \\ 0 & |b|e^{i\phi_b} & 0 \\ 0 & 0 & |c|e^{i\phi_c} \end{pmatrix} \begin{pmatrix} 0 \\ cos\,\theta \\ sin\,\theta \end{pmatrix} \right|^2$$

$$\propto \left| \left(0\ |b|e^{i\phi_b} cos\,\theta\ |c|e^{i\phi_c} sin\,\theta\right) \begin{pmatrix} 0 \\ cos\,\theta \\ sin\,\theta \end{pmatrix} \right|^2 \propto \left| |b|e^{i\phi_b} * cos^2\,\theta + |c|e^{i\phi_c} * sin^2\,\theta \right|^2$$

$$\propto |b|^2 \cos^4 \theta + |c|^2 \sin^4 \theta + \left(|b|e^{i\phi_b} * cos^2\,\theta\right) * \left(|c|e^{-i\phi_c} * sin^2\,\theta\right) +$$

$$\left(|b|e^{-i\phi_b} * cos^2\,\theta\right) * \left(|c|e^{i\phi_c} * sin^2\,\theta\right) \propto |b|^2 \cos^4 \theta + |c|^2 \sin^4 \theta +$$

$$|b||c| \cos^2 \theta \, sin^2\,\theta * (e^{i(\phi_b-\phi_c)} + e^{-i(\phi_b-\phi_c)})$$

$$\boldsymbol{I_{A_g}} \propto |\boldsymbol{b}|^2 \mathbf{cos^4\,\theta} + |\boldsymbol{c}|^2 \mathbf{sin^4\,\theta} + \mathbf{2}|\boldsymbol{b}||\boldsymbol{c}| \mathbf{cos^2\,\theta\,sin^2\,\theta\,cos}\,\boldsymbol{\phi_{bc}} \tag{S8}$$

*S3.2.2. The $B_{3g}$ modes in the parallel polarization configuration with considering light absorption*

Substituting the real Raman tensor elements by the complex tensor elements, given by Equation S7, the angular dependencies for the $B_{3g}$ modes are now given by:

$$\hat{R}_{B_{3g}} = \begin{pmatrix} 0 & 0 & 0 \\ 0 & 0 & |f|e^{i\phi_f} \\ 0 & |f|e^{i\phi_f} & 0 \end{pmatrix}$$

$$I_{B_{3g}} \propto \left| (0\,cos\,\theta\,sin\,\theta) \begin{pmatrix} 0 & 0 & 0 \\ 0 & 0 & |f|e^{i\phi_f} \\ 0 & |f|e^{i\phi_f} & 0 \end{pmatrix} \begin{pmatrix} 0 \\ cos\,\theta \\ sin\,\theta \end{pmatrix} \right|^2$$

$$\propto \left| \left(0\,|f|e^{i\phi_f} sin\theta |f|e^{i\phi_f} cos\theta\right) \begin{pmatrix} 0 \\ cos\,\theta \\ sin\,\theta \end{pmatrix} \right|^2$$

$$I_{B_{3g}} \propto \left||f|e^{i\phi_f} sin\,\theta * cos\,\theta + |f|e^{i\phi_f} cos\,\theta * sin\,\theta\right|^2 \propto \left|2\,|f|e^{i\phi_f}\,(sin\,\theta * cos\,\theta)\right|^2$$

$$\boldsymbol{I_{B_{3g}}} \propto \left|\boldsymbol{|f|e^{i\phi_f}\,sin\,2\theta}\right|^2 \propto \left|\boldsymbol{|f|}\,\mathbf{sin}\,\boldsymbol{2\theta}\right|^2 \tag{S9}$$

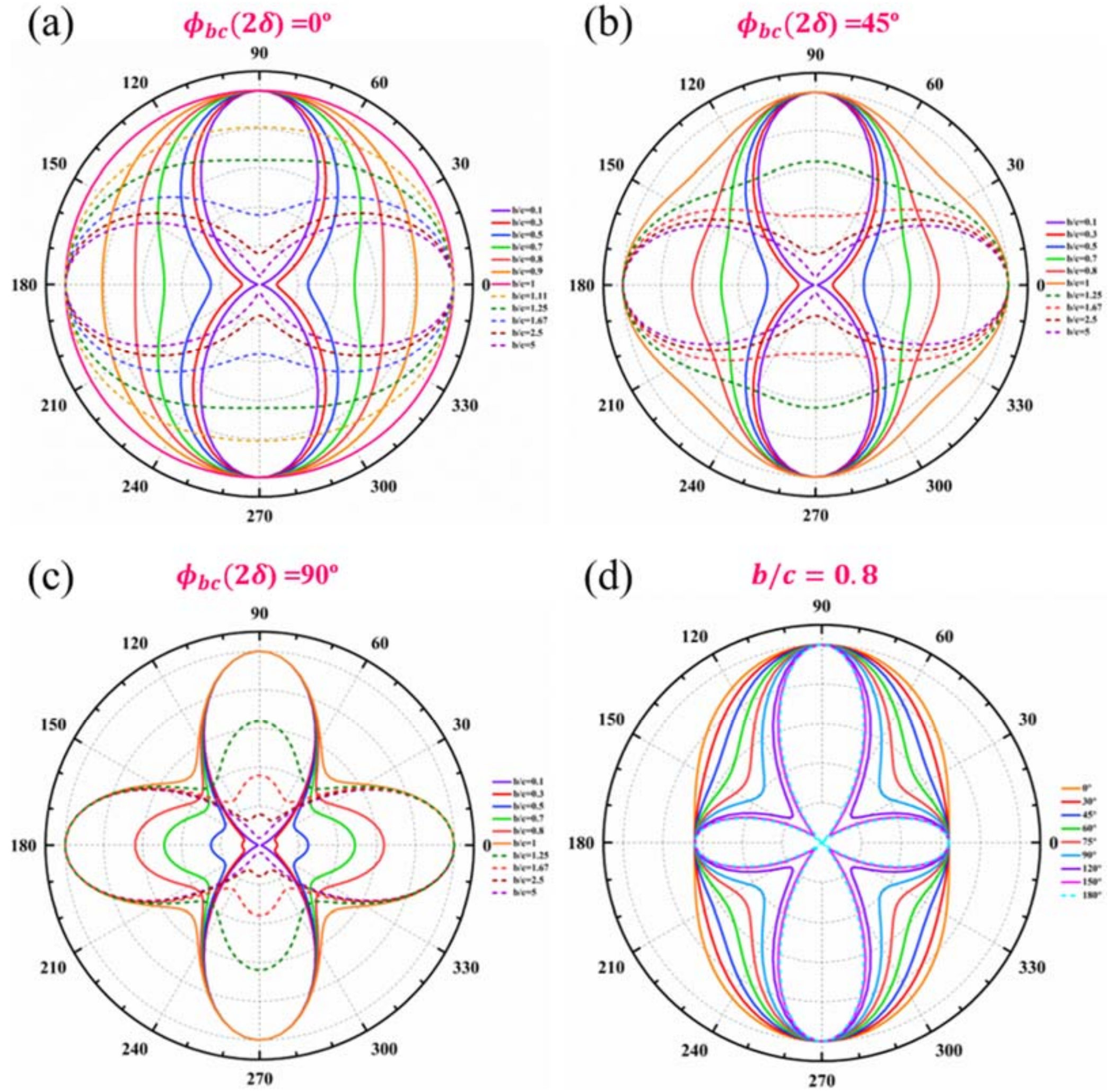

**Figure S11.** Simulations of the Raman intensity of $A_g$ modes with different *b/c* ratio and phase difference.

### S3.3. Raman modes with complex Raman tensor elements when considering birefringence effect

As for birefringence effect, the polarization vector of the incident light can be decomposed into two components along the two crystalline orientations, the chain direction (*b*-axis) and cross-chain direction (*c*-axis) respectively. The two components propagate in $Ta_2Ni_3Te_5$ with different phase velocities, resulting in a phase difference $\delta$. Therefore, the actual $\hat{e}_i$ interacting with the material could be slightly different from the original polarization vector of the incident light. Similarly, the scattered light also has the birefringence effect when exiting from the crystal. This effect on the polarization of the Raman scattering can be described applying a Jones matrix $J(t)$,[3] which depends on the sample thickness $t$. Since the scattered light also has to pass through the sample as the incident light, the Raman scattering intensity ($I$) expressed by Equation S1 becomes:[3]

$$I \propto \left|\hat{e}_i \cdot J(t) \cdot \hat{R}_{ij} \cdot J(t) \cdot \hat{e}_s\right|^2 \quad \text{(S10)}$$

Here, in our experimental configuration, the crystalline direction has been well aligned with our laboratory coordinates in the way as we mentioned before, which means that the two systems are equivalent. Therefore, the rotation matrix $S(\theta)$,[3] which transforms the crystal system to the laboratory system, does not need to be included in Equation S10. The Jones matrix $J(t)$ takes the simple diagonal form[3] in the basis of the allowed polarizations, with being normalized to the first entry:

$$J(t) = \begin{pmatrix} 1 & 0 & 0 \\ 0 & 1 & 0 \\ 0 & 0 & e^{i\cdot\delta(\mathrm{t})} \end{pmatrix} \quad \text{(S11)}$$

The phase shift $\delta(\mathrm{t})$ between the light with polarization parallel to the chain direction and cross-chain direction, with difference in refractive indices $\Delta n = (n_b - n_c)$, is:[3]

$$\delta(\mathrm{t}) = \frac{2\pi}{\lambda} \cdot \Delta n \cdot t \quad \text{(S12)}$$

Here, $n_b$ and $n_c$ are the refractive indices along the $b$ and $c$ directions, $\lambda$ is the wavelength, and $t$ is sample thickness. Since the incident and scattered light wavelength in the Raman scattering are very close to each other, the difference can be neglected in Equation S12. Therefore, $\delta(\mathrm{t})$ is identical in both Jones matrices.[3]

*S3.3.1. The $A_g$ modes in the parallel polarization configuration with considering birefringence effect*

Substituting in Equation S10 the unitary vectors $\hat{e}_i = (0 \cos\theta \sin\theta)$ and $\hat{e}_s = (0 \cos\theta \sin\theta)$ and the Raman tensor elements, the angular dependencies for the $A_g$ modes intensities are now given by:

$$I_{A_g} \propto \left|(0\, cos\,\theta\, sin\,\theta) \begin{pmatrix} 1 & 0 & 0 \\ 0 & 1 & 0 \\ 0 & 0 & e^{i\cdot\delta(\mathrm{t})} \end{pmatrix} \begin{pmatrix} a & 0 & 0 \\ 0 & b & 0 \\ 0 & 0 & c \end{pmatrix} \begin{pmatrix} 1 & 0 & 0 \\ 0 & 1 & 0 \\ 0 & 0 & e^{i\cdot\delta(\mathrm{t})} \end{pmatrix} \begin{pmatrix} 0 \\ cos\,\theta \\ sin\,\theta \end{pmatrix}\right|^2$$

$$\propto \left|(0\, cos\,\theta\, sin\,\theta) \begin{pmatrix} a & 0 & 0 \\ 0 & b & 0 \\ 0 & 0 & ce^{i\cdot 2\delta(\mathrm{t})} \end{pmatrix} \begin{pmatrix} 0 \\ cos\,\theta \\ sin\,\theta \end{pmatrix}\right|^2 \propto$$

$$\left|\left(0\; b\cos\theta\; ce^{i\cdot 2\delta(\mathrm{t})}\sin\theta\right)\begin{pmatrix}0\\ \cos\theta\\ \sin\theta\end{pmatrix}\right|^2$$

$$\propto \left|b * \cos^2\theta + ce^{i\cdot 2\delta(\mathrm{t})} * \sin^2\theta\right|^2$$

$$\propto \left(b^2\cos^4\theta + c^2\sin^4\theta + bce^{-i\cdot 2\delta(\mathrm{t})}\cos^2\theta\sin^2\theta\right.$$

$$\left. + cbe^{i\cdot 2\delta(\mathrm{t})}\sin^2\theta\cos^2\theta\right)$$

$$\boldsymbol{I_{A_g} \propto b^2\cos^4\theta + c^2\sin^4\theta + 2bc\cos^2\theta\sin^2\theta\cos 2\delta(t)} \qquad \text{(S13)}$$

*S3.3.2. The $B_{3g}$ modes in the parallel polarization configuration with considering birefringence effect*

Substituting in Equation S10 the unitary vectors $\hat{e}_i = (0\cos\theta\sin\theta)$ and $\hat{e}_s = (0\cos\theta\sin\theta)$ and the Raman tensor elements, the angular dependencies for the $B_{3g}$ modes intensities are now given by:

$$I_{B_{3g}} \propto$$

$$\left|\left(0\cos\theta\sin\theta\right)\begin{pmatrix}1 & 0 & 0\\ 0 & 1 & 0\\ 0 & 0 & e^{i\cdot\delta(\mathrm{t})}\end{pmatrix}\begin{pmatrix}0 & 0 & 0\\ 0 & 0 & f\\ 0 & f & 0\end{pmatrix}\begin{pmatrix}1 & 0 & 0\\ 0 & 1 & 0\\ 0 & 0 & e^{i\cdot\delta(\mathrm{t})}\end{pmatrix}\begin{pmatrix}0\\ \cos\theta\\ \sin\theta\end{pmatrix}\right|^2$$

$$\propto \left|\left(0\cos\theta\sin\theta\right)\begin{pmatrix}0 & 0 & 0\\ 0 & 0 & fe^{i\cdot\delta(\mathrm{t})}\\ 0 & fe^{i\cdot\delta(\mathrm{t})} & 0\end{pmatrix}\begin{pmatrix}1 & 0 & 0\\ 0 & 1 & 0\\ 0 & 0 & e^{i\cdot\delta(\mathrm{t})}\end{pmatrix}\begin{pmatrix}0\\ \cos\theta\\ \sin\theta\end{pmatrix}\right|^2$$

$$\propto \left|\left(0\; fe^{i\cdot\delta(\mathrm{t})}\sin\theta\; fe^{i\cdot\delta(\mathrm{t})}\cos\theta\right)\begin{pmatrix}0\\ \cos\theta\\ \sin\theta\end{pmatrix}\right|^2$$

$$\propto \left|fe^{i\cdot\delta(\mathrm{t})}\sin\theta * \cos\theta + fe^{i\cdot\delta(\mathrm{t})}\cos\theta * \sin\theta\right|^2 \propto \left|2\, fe^{i\cdot\delta(\mathrm{t})}\,(\sin\theta * \cos\theta)\right|^2$$

$$\boldsymbol{I_{B_{3g}} \propto \left|fe^{i\cdot\delta(\mathrm{t})}\sin 2\theta\right|^2 \propto |f\sin 2\theta|^2} \qquad \text{(S14)}$$

The angle-dependent responses of peaks assigned to the $A_g$ symmetry (18.4, 87.9, 93.0, 94.7,117.2, 128.1, 153.4, 167.7, 173.1, 205.8, and 230.3 $cm^{-1}$) deviate markedly from the classical predictions. Similar anomalies were also observed in other low-symmetry materials; however, the physical origins were variably attributed to light absorption[4-7] or birefringence,[8, 9] primarily through fitting-based analyses, leading to inconsistent and ambiguous interpretations.

Our comprehensive analyses reveal a significant finding: both light absorption and birefringence lead to mathematically identical angular dependencies for the $A_g$ modes. Consequently, simply fitting experimental angle-dependent Raman intensity to corresponding mathematical representation proves inherently insufficient to unambiguously determine the intrinsic mechanism responsible for these anomalies. This arises for the following reason: $J(t) \cdot \hat{R}_{ij} \cdot J(t)$ can be recast as an effective Raman tensor,[3] $\hat{R}_{ij(\mathrm{eff})} = \begin{pmatrix} a & 0 & 0 \\ 0 & b & 0 \\ 0 & 0 & c \cdot e^{i \cdot 2\delta(\mathrm{t})} \end{pmatrix}$, which also introduces complex components.

All $A_g$ and $B_{3g}$ modes can be accurately fitted using complex Raman tensors, as shown in Figure S8 and S9. The experimental data of the Raman mode symmetries exhibit well-defined clarity, with no angular deviations. In contrast, previous analyses[10] required artificially introduced angular shifts for unknow reason, some peaks cannot be properly fitted even in this way. The fundamental mechanism underlying complex Raman tensors has been elucidated with considering anisotropic electron-phonon interactions in the full quantum framework.

**S3.4. Derivation of the light-polarization dependent equations for the Raman modes in other 2D materials with low symmetry**

*S3.4.1. Tungsten Ditelluride ($WTe_2$)*

$WTe_2$ is a layered transition metal dichalcogenide (TMD) with sandwich structure composed of the Te-W-Te atomic layers stacking along the *c*-axis through van der Waals interactions between the Te layers. $WTe_2$ crystallizes in a distorted structure with octahedral coordination around the metal, known as Td-polytype (Td-$WTe_2$).[6] Unlike other TMDs such as $MoS_2$, $MoSe_2$, $WS_2$, and $WSe_2$, Td-$WTe_2$ is strongly distorted from the ideal hexagonal network due to the off-centering W atoms, which form the slightly buckled W-W zigzag chains along the *a*-axis of the orthorhombic unit cell. [6]

Bulk Td-$WTe_2$ belongs to the space group Pmn$2_1$ and point group $C_{2v}^7$, the irreducible representation of the phonon modes at $\Gamma$ point is: $\Gamma_{\text{bulk}} = 12A_1 + 7A_2 + 6B_1 + 11B_2$. The bilayer and thicker few-layer $WTe_2$ ($N \geq 2$) belong to space group *Pm* and point group $C_s^1$, the irreducible representation of its phonon modes at $\Gamma$ point is: $\Gamma_{N\text{-layer}} = 6NA' + 12NA''$. Monolayer (1L) $WTe_2$ belongs to space group $P2_1/m$ and point group $C_{2h}^2$, the irreducible representation of its phonon modes at $\Gamma$ point is: $\Gamma_{1\text{-layer}} = 6A_g + 3A_u + 3B_g + 6B_u$.[6, 11]

According to the table,[1] the Raman tensors in **bulk $WTe_2$** are:

$$\hat{R}_{A_1} = \begin{pmatrix} a & 0 & 0 \\ 0 & b & 0 \\ 0 & 0 & c \end{pmatrix}; \hat{R}_{A_2} = \begin{pmatrix} 0 & e & 0 \\ e & 0 & 0 \\ 0 & 0 & 0 \end{pmatrix}; \hat{R}_{B_1} = \begin{pmatrix} 0 & 0 & d \\ 0 & 0 & 0 \\ d & 0 & 0 \end{pmatrix}; \hat{R}_{B_2} = \begin{pmatrix} 0 & 0 & 0 \\ 0 & 0 & f \\ 0 & f & 0 \end{pmatrix};$$

The Raman tensors of the Raman active modes in **1L (or NL) $WTe_2$** are:

$$\hat{R}_{A_g(\text{orA}')} = \begin{pmatrix} a & 0 & d \\ 0 & b & 0 \\ d & 0 & c \end{pmatrix}; \hat{R}_{B_g(\text{orA}'')} = \begin{pmatrix} 0 & e & 0 \\ e & 0 & f \\ 0 & f & 0 \end{pmatrix};$$

In a back-scattering set-up with parallel configuration where the sample is rotated while keeping the polarizer and analyzer fixed, the polarization vectors lie in the *ab*-plane with the initial incident laser polarized along the *a*-axis (the atomic chain direction), and the unitary vectors can be written as a function of angle $\theta$ with respect to the *a*-axis as follows: for the incident beam, $\hat{e}_i = (\cos\theta \sin\theta\ 0)$, while for the scattered light $\hat{e}_s = (\cos\theta \sin\theta\ 0)$. Thus, the intensity of these Raman modes is:

$$I_{A_1} \propto \left| (\cos\theta \sin\theta\ 0) \begin{pmatrix} a & 0 & 0 \\ 0 & b & 0 \\ 0 & 0 & c \end{pmatrix} \begin{pmatrix} cos\,\theta \\ sin\,\theta \\ 0 \end{pmatrix} \right|^2 \propto |a * cos^2\theta + b * sin^2\theta|^2$$

$$\boldsymbol{I_{A_1} \propto a^2\, cos^4\theta + \ b^2\, sin^4\theta + 2ab\, cos^2\theta\, sin^2\theta} \quad \text{(S15)}$$

$$I_{A_2} \propto \left| (\cos\theta \sin\theta\ 0) \begin{pmatrix} 0 & e & 0 \\ e & 0 & 0 \\ 0 & 0 & 0 \end{pmatrix} \begin{pmatrix} cos\,\theta \\ sin\,\theta \\ 0 \end{pmatrix} \right|^2 \propto |e\, sin\,\theta * cos\,\theta + e\, cos\,\theta * sin\,\theta|^2$$

$$\boldsymbol{I_{A_2} \propto |2\, e\, (sin\,\theta * cos\,\theta)|^2 \propto |e\, sin\, 2\theta|^2} \quad \text{(S16)}$$

$$I_{B_1} \propto \left| (\cos\theta \sin\theta\ 0) \begin{pmatrix} 0 & 0 & d \\ 0 & 0 & 0 \\ d & 0 & 0 \end{pmatrix} \begin{pmatrix} cos\,\theta \\ sin\,\theta \\ 0 \end{pmatrix} \right|^2 \propto |0|^2 \quad \text{(S17)}$$

$$I_{B_2} \propto \left| (\cos\theta \sin\theta\ 0) \begin{pmatrix} 0 & 0 & 0 \\ 0 & 0 & f \\ 0 & f & 0 \end{pmatrix} \begin{pmatrix} cos\,\theta \\ sin\,\theta \\ 0 \end{pmatrix} \right|^2 \propto |0|^2 \quad \text{(S18)}$$

Here, the intensities of $B_1$ and $B_2$ modes are zero, they cannot be observed in this configuration.

$$I_{A_g(\text{orA}')} \propto \left| (\cos\theta \sin\theta\ 0) \begin{pmatrix} a & 0 & d \\ 0 & b & 0 \\ d & 0 & c \end{pmatrix} \begin{pmatrix} cos\,\theta \\ sin\,\theta \\ 0 \end{pmatrix} \right|^2 \propto |a * cos^2\theta + b * sin^2\theta|^2$$

$$\boldsymbol{I_{A_g(\text{orA}')} \propto a^2\, cos^4\theta + \ b^2\, sin^4\theta + 2ab\, cos^2\theta\, sin^2\theta} \quad \text{(S19)}$$

$$I_{B_g(\text{orA}'')} \propto \left| (\cos\theta \sin\theta\ 0) \begin{pmatrix} 0 & e & 0 \\ e & 0 & f \\ 0 & f & 0 \end{pmatrix} \begin{pmatrix} cos\,\theta \\ sin\,\theta \\ 0 \end{pmatrix} \right|^2 \propto |e\, sin\,\theta * cos\,\theta + e\, cos\,\theta * sin\,\theta|^2$$

$$\boldsymbol{I_{B_g(\text{orA}'')} \propto |2\, e\, (sin\,\theta * cos\,\theta)|^2 \propto |e\, sin\, 2\theta|^2} \quad \text{(S20)}$$

Here, the intensity of the $A_g$(orA′) Raman modes have the same angular dependence as $A_1$, and the $B_g$(orA′′) Raman modes have the same angular dependence as $A_2$.

When considering **light absorption** in the material, the Raman tensor elements also present complex values, with real and imaginary parts. The Raman tensor elements relevant to this work (*a*, *b*, *c*, *d*, *e* and *f*) can thus be written as:

$$a = |a|e^{i\phi_a};\ b = |b|e^{i\phi_b};\ c = |c|e^{i\phi_c};\ d = |d|e^{i\phi_d};\ e = |e|e^{i\phi_e};\ f = |f|e^{i\phi_f};$$

$$I_{A_1} \propto \left| (\cos\theta \sin\theta\ 0) \begin{pmatrix} |a|e^{i\phi_a} & 0 & 0 \\ 0 & |b|e^{i\phi_b} & 0 \\ 0 & 0 & |c|e^{i\phi_c} \end{pmatrix} \begin{pmatrix} cos\,\theta \\ sin\,\theta \\ 0 \end{pmatrix} \right|^2$$

$$\boldsymbol{I_{A_1} \propto |a|^2 \cos^4 \theta + |b|^2 \sin^4 \theta + 2|a||b| \cos^2 \theta \sin^2 \theta \cos \phi_{ab}} \tag{S21}$$

Here, $\phi_{ab}$ is the phase difference of Raman tensor element complex phase ($\phi_a - \phi_b$).

$$I_{A_2} \propto \left| (\cos\theta \sin\theta\ 0) \begin{pmatrix} 0 & |e|e^{i\phi_e} & 0 \\ |e|e^{i\phi_e} & 0 & 0 \\ 0 & 0 & 0 \end{pmatrix} \begin{pmatrix} \cos\theta \\ \sin\theta \\ 0 \end{pmatrix} \right|^2$$

$$\boldsymbol{I_{A_2} \propto \left| |e| \sin 2\theta \right|^2} \tag{S22}$$

$$I_{A_g(\text{orA}')} \propto \left| (\cos\theta \sin\theta\ 0) \begin{pmatrix} |a|e^{i\phi_a} & 0 & |d|e^{i\phi_d} \\ 0 & |b|e^{i\phi_b} & 0 \\ |d|e^{i\phi_d} & 0 & |c|e^{i\phi_c} \end{pmatrix} \begin{pmatrix} \cos\theta \\ \sin\theta \\ 0 \end{pmatrix} \right|^2$$

$$\boldsymbol{I_{A_g(\text{orA}')} \propto |a|^2 \cos^4 \theta + |b|^2 \sin^4 \theta + 2|a||b| \cos^2 \theta \sin^2 \theta \cos \phi_{ab}} \tag{S23}$$

$$I_{B_g(\text{orA}'')} \propto \left| (\cos\theta \sin\theta\ 0) \begin{pmatrix} 0 & |e|e^{i\phi_e} & 0 \\ |e|e^{i\phi_e} & 0 & |f|e^{i\phi_f} \\ 0 & |f|e^{i\phi_f} & 0 \end{pmatrix} \begin{pmatrix} \cos\theta \\ \sin\theta \\ 0 \end{pmatrix} \right|^2$$

$$\boldsymbol{I_{B_g(\text{orA}'')} \propto \left| |e| \sin 2\theta \right|^2} \tag{S24}$$

Here, the intensity of the $A_g(\text{orA}')$ Raman modes have the same angular dependence as $A_1$, and the $B_g(\text{orA}'')$ Raman modes have the same angular dependence as $A_2$.

When considering **birefringence effect**, the Jones matrix $J(t)$ including phase shift $\delta(\text{t})$ between the light with polarization parallel to the chain direction and cross-chain direction is:

$$J(t) = \begin{pmatrix} 1 & 0 & 0 \\ 0 & e^{i\cdot\delta(\text{t})} & 0 \\ 0 & 0 & 1 \end{pmatrix} \tag{S25}$$

$$I_{A_1} \propto \left| (\cos\theta \sin\theta\ 0) \begin{pmatrix} 1 & 0 & 0 \\ 0 & e^{i\cdot\delta(\text{t})} & 0 \\ 0 & 0 & 1 \end{pmatrix} \begin{pmatrix} a & 0 & 0 \\ 0 & b & 0 \\ 0 & 0 & c \end{pmatrix} \begin{pmatrix} 1 & 0 & 0 \\ 0 & e^{i\cdot\delta(\text{t})} & 0 \\ 0 & 0 & 1 \end{pmatrix} \begin{pmatrix} \cos\theta \\ \sin\theta \\ 0 \end{pmatrix} \right|^2$$

$$\boldsymbol{I_{A_1} \propto a^2 \cos^4 \theta + b^2 \sin^4 \theta + 2ab \cos^2 \theta \sin^2 \theta \cos 2\delta(t)} \tag{S26}$$

**Equation S26 is the same as Equation S21** with having $\cos 2\delta(\text{t})$ instead of $\cos \phi_{ab}$.

$$I_{A_2} \propto \left| (\cos\theta \sin\theta\ 0) \begin{pmatrix} 1 & 0 & 0 \\ 0 & e^{i\cdot\delta(\text{t})} & 0 \\ 0 & 0 & 1 \end{pmatrix} \begin{pmatrix} 0 & e & 0 \\ e & 0 & 0 \\ 0 & 0 & 0 \end{pmatrix} \begin{pmatrix} 1 & 0 & 0 \\ 0 & e^{i\cdot\delta(\text{t})} & 0 \\ 0 & 0 & 1 \end{pmatrix} \begin{pmatrix} \cos\theta \\ \sin\theta \\ 0 \end{pmatrix} \right|^2$$

$$\boldsymbol{I_{A_2} \propto |2\, e\, (\sin\theta * \cos\theta)|^2 \propto |e \sin 2\theta|^2} \tag{S27}$$

**Equation S27 is the same as Equation S22.**

$$I_{A_g(\text{orA}')} \propto \left| (\cos\theta \sin\theta\ 0) \begin{pmatrix} 1 & 0 & 0 \\ 0 & e^{i\cdot\delta(\text{t})} & 0 \\ 0 & 0 & 1 \end{pmatrix} \begin{pmatrix} a & 0 & d \\ 0 & b & 0 \\ d & 0 & c \end{pmatrix} \begin{pmatrix} 1 & 0 & 0 \\ 0 & e^{i\cdot\delta(\text{t})} & 0 \\ 0 & 0 & 1 \end{pmatrix} \begin{pmatrix} \cos\theta \\ \sin\theta \\ 0 \end{pmatrix} \right|^2$$

$$\boldsymbol{I_{A_g(\text{orA}')} \propto a^2 \cos^4\theta + b^2 \sin^4\theta + 2ab \cos^2\theta \sin^2\theta \cos 2\delta(t)} \tag{S28}$$

**Equation S28 is the same as Equation S23** with having $\cos 2\delta(\text{t})$ instead of $\cos \phi_{ab}$.

$$I_{B_g(\text{orA}'')} \propto \left| (\cos\theta \sin\theta\ 0) \begin{pmatrix} 1 & 0 & 0 \\ 0 & e^{i\cdot\delta(\text{t})} & 0 \\ 0 & 0 & 1 \end{pmatrix} \begin{pmatrix} 0 & e & 0 \\ e & 0 & f \\ 0 & f & 0 \end{pmatrix} \begin{pmatrix} 1 & 0 & 0 \\ 0 & e^{i\cdot\delta(\text{t})} & 0 \\ 0 & 0 & 1 \end{pmatrix} \begin{pmatrix} \cos\theta \\ \sin\theta \\ 0 \end{pmatrix} \right|^2$$

$$\boldsymbol{I_{B_g(\text{orA}'')} \propto |e \sin 2\theta|^2} \tag{S29}$$

**Equation S29 is the same as Equation S24**.

*S3.4.2. 1T′ and $T_d$ Molybdenum Ditelluride ($MoTe_2$)*

$MoTe_2$ exists in several phases with distinct physical properties. At room temperature, $MoTe_2$ can be converted between semiconducting hexagonal phase (2H or α-$MoTe_2$) and semi-metallic monoclinic phase (1T′ or β-$MoTe_2$), which exhibits in-plane anisotropy. Moreover, when the temperature is below 250 K, the 1T′ $MoTe_2$ converts to the orthorhombic phase known as $T_d$-$MoTe_2$, which is isostructural to the noncentrosymmetric $T_d$ phase of the previously mentioned $WTe_2$.[5]

The high-temperature monoclinic 1T′ phase of bulk $MoTe_2$ shows an inclined stacking angle of ~93.9° with centrosymmetric space group $P2_1/m$ and point group $C_{2h}^2$, which is the same as monolayer (1L) $T_d$-$WTe_2$. The same symmetry holds for any odd layers 1T′ $MoTe_2$, monolayer includes. However, the inversion symmetry is broken for 1T′ $MoTe_2$ with even layers belonging to space group *Pm* and point group $C_S^1$, which is the same as NL Td-$WTe_2$ ($N$≥2) mentioned above.[5, 12-14] The temperature-induced structural transition converts the monoclinic 1T′ $MoTe_2$ to orthorhombic $T_d$-$MoTe_2$ with a noncentrosymmetric space group $Pmn2_1$ and point group $C_{2v}^7$, which is the same as bulk $T_d$-$WTe_2$. The crystal symmetry of thicker few-layer $T_d$-$MoTe_2$ is independent of the number of layers, and only the mirror symmetry exists with space group *Pm* and point group $C_S^1$, which makes them different from the bulk but the same as NL $T_d$-$WTe_2$ ($N$≥2). Since the monoclinic 1T′ $MoTe_2$ and orthorhombic $T_d$-$MoTe_2$ differ only in the layer-to-layer stacking, the structure of the individual monolayer $MoTe_2$ is identical in both 1T′ and Td-phase, which has the same space group $P2_1/m$ and point group $C_{2h}^2$ as bulk 1T′$MoTe_2$ and 1L $T_d$-$MoTe_2$.

The crystal symmetry analysis of 1T′ and $T_d$-$MoTe_2$ indicates that the derivation of the light-polarization dependent equations for the corresponding Raman modes with the same symmetry will be the same as discussed in the $T_d$-$WTe_2$ cases, and the conclusions for the cases in $T_d$-$WTe_2$ also applied to the cases 1T′ and $T_d$-$MoTe_2$.

*S3.4.3. 1T′ Rhenium Disulfide ($ReS_2$) and Rhenium Diselenide ($ReSe_2$)*

$ReS_2$ and $ReSe_2$ (hereafter denoted as $ReX_2$) exhibit a distorted octahedral phase (denoted 1T′) with much lower symmetry and significant in-plane anisotropy.[7] Compared with the other anisotropic materials, such as orthorhombic black phosphorus and monoclinic 1T′ $MoTe_2$, the 1T′ $ReX_2$ crystallizes in the triclinic structure with space group $P\bar{1}$ and point group $C_i$, showing a reduced symmetry with only inversion symmetry operation.[7, 9, 15, 16] In common with other TMDs, $ReX_2$ is layered material in which Re atomic layer sandwiched between the chalcogen atomic layers, with layers stacking along the *c*-axis. However, in each single layer, the Re atoms are moved away from the metal sites of the ideal octahedral 1T structure into “lozenge” or “diamond” shapes in the plane with four Re atoms, defined as $Re_4$. The driving force for this distortion has been discussed in terms of Peierls distortion.[17, 18] This arrangement of the linked $Re_4$ groups leads to highly anisotropic properties in the layer plane with forming one-dimensional $Re_4$ chains along the *b*-axis of the crystal.

According to the group theory, both monolayer and few-layer $ReX_2$ belong to the $P\bar{1}$ space group and $C_i$ point group, and the monolayer unit cell of $ReX_2$ has 12 atoms giving rise to 36 zone-center phonon modes including 18 $A_g$ modes and 18 $A_u$ modes. All the $A_g$ modes are Raman-active, and the 18 $A_u$ modes include 15 infrared-active modes and 3 acoustic $A_u$ modes. The Raman tensor of the $A_g$ mode is:[16, 18]

$$\hat{R}_{A_g} = \begin{pmatrix} a & d & e \\ d & b & f \\ e & f & c \end{pmatrix}$$

In a back-scattering set-up with parallel configuration where the sample is rotated while keeping the polarizer and analyzer fixed, the polarization vectors lie in the *ab*-plane with the initial incident laser polarized along the *b*-axis (the Re chain direction), and the unitary vectors can be written as a function of angle $\theta$ with respect to the *b*-axis as follows: for the incident beam, $\hat{e}_i = (sin\,\theta\; cos\,\theta\; 0)$, while for the scattered light $\hat{e}_s = (sin\,\theta\; cos\,\theta\; 0)$. Thus, the angular dependent intensity of the $A_g$ ($I_{Ag}$) mode is:

$$\boldsymbol{I_{A_g}} \propto \left|(sin\,\theta\; cos\,\theta\; 0)\begin{pmatrix} a & d & e \\ d & b & f \\ e & f & c \end{pmatrix}\begin{pmatrix} sin\,\theta \\ cos\,\theta \\ 0 \end{pmatrix}\right|^2$$

$$\propto \left|\boldsymbol{b * cos^2\,\theta + a * sin^2\,\theta + d * sin\,2\theta}\right|^2 \quad \text{(S30)}$$

When considering **light absorption** in the material, the Raman tensor elements also present complex values, with real and imaginary parts. The Raman tensor elements relevant to this work (*a*, *b*, *c*, *d*, *e* and *f*) can thus be written as:

$$a = |a|e^{i\phi_a};\ b = |b|e^{i\phi_b};\ c = |c|e^{i\phi_c};\ d = |d|e^{i\phi_d};\ e = |e|e^{i\phi_e};\ f = |f|e^{i\phi_f};$$

$$I_{A_g} \propto \left|(sin\,\theta\; cos\,\theta\; 0)\begin{pmatrix} |a|e^{i\phi_a} & |d|e^{i\phi_d} & |e|e^{i\phi_e} \\ |d|e^{i\phi_d} & |b|e^{i\phi_b} & |f|e^{i\phi_f} \\ |e|e^{i\phi_e} & |f|e^{i\phi_f} & |c|e^{i\phi_c} \end{pmatrix}\begin{pmatrix} sin\,\theta \\ cos\,\theta \\ 0 \end{pmatrix}\right|^2$$

$$\boldsymbol{I_{A_g} \propto |b|^2\,cos^4\,\theta + |a|^2\,sin^4\,\theta + |d|^2\,sin^2\,2\theta + 4|a||d|\,sin^3\,\theta\,cos\,\theta\,cos\,\phi_{ad} +}$$

$$\boldsymbol{4|b||d|\,cos^3\,\theta\,sin\,\theta\,cos\,\phi_{bd} + 2|a||b|\,sin^2\,\theta\,cos^2\,\theta\,cos\,\phi_{ab}} \quad \text{(S31)}$$

Here, $\phi_{ab}$ is the phase difference of $(\phi_a - \phi_b)$; $\phi_{ad}$ is the phase difference of $(\phi_a - \phi_d)$; $\phi_{bd}$ is the phase difference of $(\phi_b - \phi_d)$.

When considering **birefringence effect**, the Jones matrix $J(t)$ including phase shift $\delta(\mathrm{t})$ between the light with polarization parallel to the chain direction and cross-chain direction is:

$$J(t) = \begin{pmatrix} 1 & 0 & 0 \\ 0 & e^{i\cdot\delta(\mathrm{t})} & 0 \\ 0 & 0 & 1 \end{pmatrix} \quad \text{(S32)}$$

$$I_{A_g} \propto \left|(sin\,\theta\; cos\,\theta\; 0)\begin{pmatrix} 1 & 0 & 0 \\ 0 & e^{i\cdot\delta(\mathrm{t})} & 0 \\ 0 & 0 & 1 \end{pmatrix}\begin{pmatrix} a & d & e \\ d & b & f \\ e & f & c \end{pmatrix}\begin{pmatrix} 1 & 0 & 0 \\ 0 & e^{i\cdot\delta(\mathrm{t})} & 0 \\ 0 & 0 & 1 \end{pmatrix}\begin{pmatrix} sin\,\theta \\ cos\,\theta \\ 0 \end{pmatrix}\right|^2$$

$$\boldsymbol{I_{A_g} \propto b^2\,cos^4\,\theta + a^2\,sin^4\,\theta + d^2\,sin^2\,2\theta + 4ad\,sin^3\,\theta\,cos\,\theta\,cos\,\delta(t) +}$$

$$\boldsymbol{4bd\,cos^3\,\theta\,sin\,\theta\,cos\,\delta(t) + 2ab\,cos^2\,\theta\,sin^2\,\theta\,cos\,2\delta(t)} \quad \text{(S33)}$$

**S4. Anisotropic electron-photon and electron-phonon interactions in bulk $Ta_2Ni_3Te_5$**

The first-principles calculations were performed with the Vienna Ab initio Simulation Package (VASP)[19] with projector augmented wave (PAW) pseudopotentials[20] describing the interactions between valence electrons and ion cores. The exchange-correlation energy in the electronic system was described with the Perdew-Burke-Ernzerhof (PBE) functional.[21]

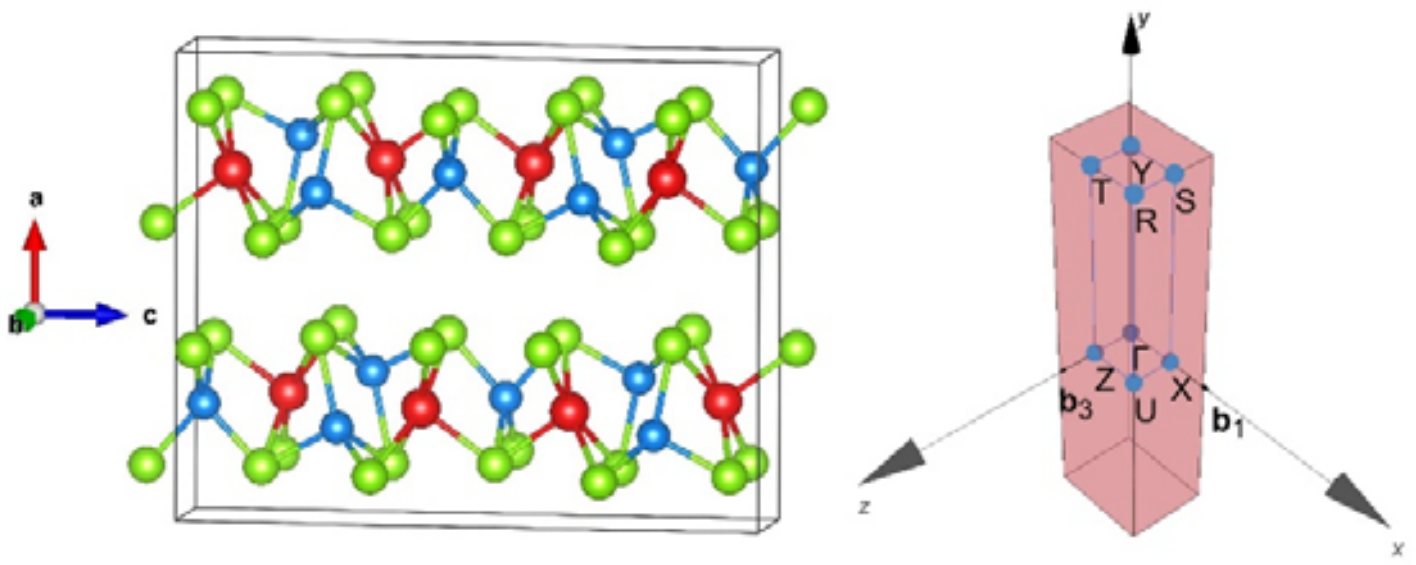

**Figure S12**. Left: the atomic structure of bulk $Ta_2Ni_3Te_5$ used in the band structure and Raman mode calculations. The super cell vectors a, b and c are along the x, y and z axis of the Cartesian coordinate system, respectively. The Ta, Ni and Te atoms are represented with red, blue and green color. Right: the corresponding 3D Brillouin zone with high symmetry points marked.

**Table S3**. The character table of point group $D_{2h}$. The directions of the x, y and z axes are described in Figure S12.

| | E | C2(z) | C2(y) | C2(x) | i | $\sigma$(xy) | $\sigma$(xz) | $\sigma$(yz) | Linear, rotations | Quadratic |
|---|---|---|---|---|---|---|---|---|---|---|
| $A_g$ | 1 | 1 | 1 | 1 | 1 | 1 | 1 | 1 | | $x^2$,$y^2$,$z^2$ |
| $B_{1g}$ | 1 | 1 | -1 | -1 | 1 | 1 | -1 | -1 | Rz | xy |
| $B_{2g}$ | 1 | -1 | 1 | -1 | 1 | -1 | 1 | -1 | Ry | xz |
| $B_{3g}$ | 1 | -1 | -1 | 1 | 1 | -1 | -1 | 1 | Rx | yz |
| $A_u$ | 1 | 1 | 1 | 1 | -1 | -1 | -1 | -1 | | |
| $B_{1u}$ | 1 | 1 | -1 | -1 | -1 | -1 | 1 | 1 | z | |
| $B_{2u}$ | 1 | -1 | 1 | -1 | -1 | 1 | -1 | 1 | y | |
| $B_{3u}$ | 1 | -1 | -1 | 1 | -1 | 1 | 1 | -1 | x | |

**Table S4**. Selection rules of optical transitions for the $D_{2h}$ point group. These selection rules correspond to the **electron-photon matrix element** $\langle f|H_{op}|i\rangle$.

| yy polarized ($B_{2u}$) | | zz polarized ($B_{1u}$) | |
|---|---|---|---|
| $\vert i\rangle$ | $\vert f\rangle$ | $\vert i\rangle$ | $\vert f\rangle$ |
| $A_g$ | $B_{2u}$ | $A_g$ | $B_{1u}$ |
| $B_{1g}$ | $B_{3u}$ | $B_{1g}$ | $A_u$ |
| $B_{2g}$ | $A_u$ | $B_{2g}$ | $B_{3u}$ |
| $B_{3g}$ | $B_{1u}$ | $B_{3g}$ | $B_{2u}$ |
| $A_u$ | $B_{2g}$ | $A_u$ | $B_{1g}$ |
| $B_{1u}$ | $B_{3g}$ | $B_{1u}$ | $A_g$ |
| $B_{2u}$ | $A_g$ | $B_{2u}$ | $B_{3g}$ |
| $B_{3u}$ | $B_{1g}$ | $B_{3u}$ | $B_{2g}$ |

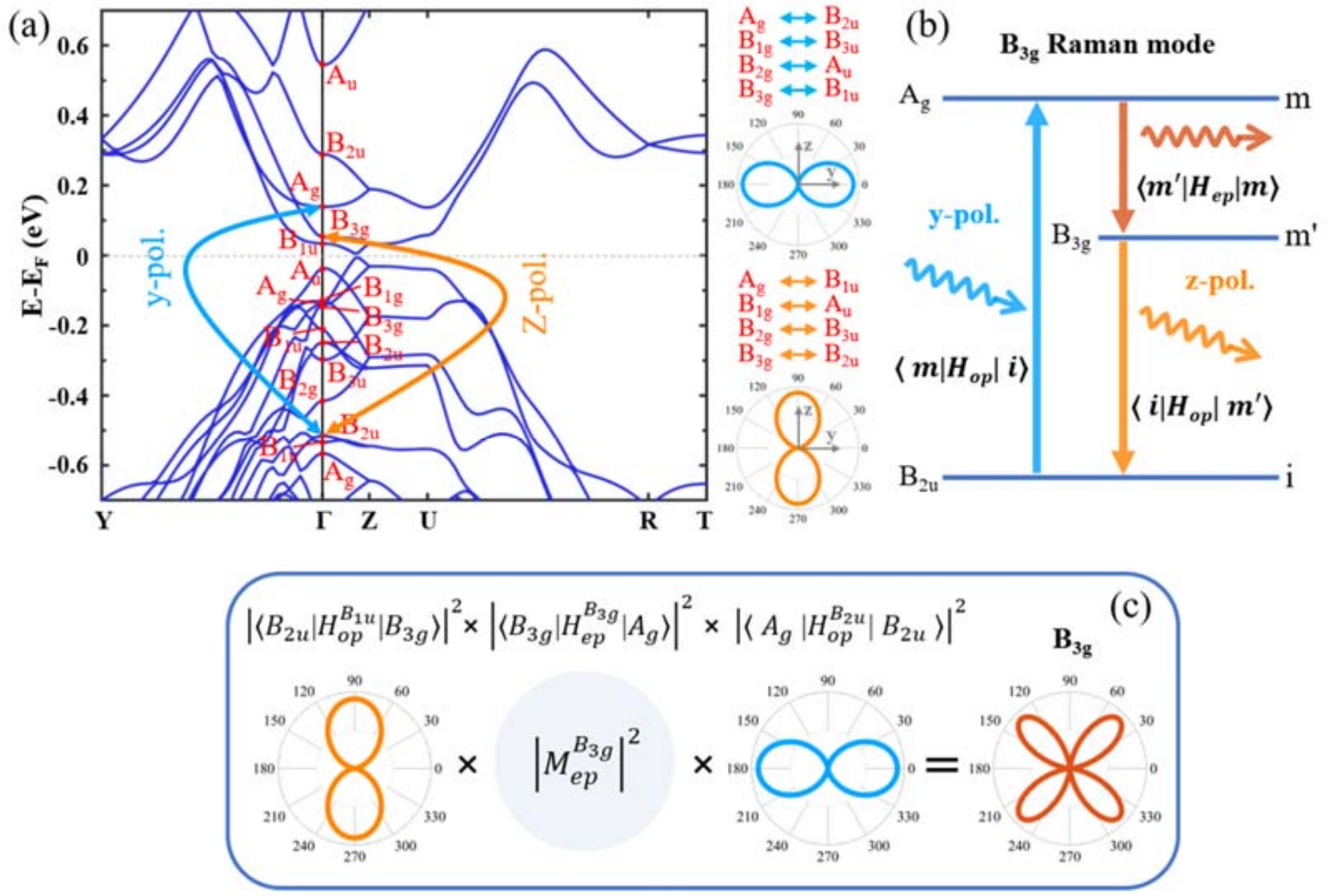

**Figure S13.** (a) The calculated band structures of bulk $Ta_2Ni_3Te_5$ with the PBE function shown in a k point path from Y to T. The irreducible representations of the bands at Γ point have been labeled. The thick blue (orange) line indicates the transition from a $B_{2u}$ ($B_{2u}$) valence band to $A_g$ ($B_{3g}$) conduction band for the y-axis (z-axis) direction polarized light, and its anisotropic absorption as a function of the angle is shown on the right side (the blue and orange polar plots). (b) Schematic representation of the Raman scattering process for the $B_{3g}$ Raman modes with considering the symmetry allowed

selection rules in the full quantum theory; (c) Calculated polar plots of the electron-photon and electron-phonon interactions and corresponding $B_{3g}$ Raman modes in the Raman scattering, where the anisotropy of electron-phonon interaction is neglected.

For the $B_{3g}$ modes, $|m\rangle$ and $|m'\rangle$ have different symmetries. The selection rules of the $B_{3g}$ mode are given in Table S6. Figure S13b shows a symmetry allowed transition for the $B_{3g}$ mode, where the incident and scattered light have opposite polarization dependence, resulting in the 90° period of the $B_{3g}$ polarization profile (Figure S13c).

**Table S5**. Selection rules of Raman scattering for the $A_g$ mode phonon in bulk $Ta_2Ni_3Te_5$. $|i\rangle$, $|f\rangle$, $|m\rangle$ and $|m'\rangle$ are the initial state, final state, and two intermediate states, respectively. yy (zz) means that both the incident and scattered lights are y (z) polarized. These selection rules correspond to the product of the matrix elements: $\langle f|H_{opy}|m'\rangle\langle m'|H_{ep}(A_g)|m\rangle\langle m|H_{opy}|i\rangle$ and $\langle f|H_{opz}|m'\rangle\langle m'|H_{ep}(A_g)|m\rangle\langle m|H_{opz}|i\rangle$, where $H_{opy}$ ($H_{opz}$) is the electron-photon interaction Hamiltonian matrix for the y (z) polarized light, and $H_{ep}$ is the electron-phonon interaction matrix.

| yy polarized ($B_{2u}$) | | zz polarized ($B_{1u}$) | |
|---|---|---|---|
| $\|i\rangle=\|f\rangle$ | $\|m\rangle = \|m'\rangle$ | $\|i\rangle=\|f\rangle$ | $\|m\rangle = \|m'\rangle$ |
| $A_g$ | $B_{2u}$ | $A_g$ | $B_{1u}$ |
| $B_{1g}$ | $B_{3u}$ | $B_{1g}$ | $A_u$ |
| $B_{2g}$ | $A_u$ | $B_{2g}$ | $B_{3u}$ |
| $B_{3g}$ | $B_{1u}$ | $B_{3g}$ | $B_{2u}$ |
| $A_u$ | $B_{2g}$ | $A_u$ | $B_{1g}$ |
| $B_{1u}$ | $B_{3g}$ | $B_{1u}$ | $A_g$ |
| $B_{2u}$ | $A_g$ | $B_{2u}$ | $B_{3g}$ |
| $B_{3u}$ | $B_{1g}$ | $B_{3u}$ | $B_{2g}$ |

**Table S6**. Selection rules of Raman scattering for the $B_{3g}$ mode phonon in bulk $Ta_2Ni_3Te_5$. The polarization of the incident light is different from that of the scattered light for the $B_{3g}$ mode. zy means that the incident light is y-polarized, while the scattered light is z-polarized, and analogous definition is for yz. These selection rules correspond to the product of the matrix elements: $\langle f|H_{opz}|m'\rangle\langle m'|H_{ep}(B_{3g})|m\rangle\langle m|H_{opy}|i\rangle$ and $\langle f|H_{opy}|m'\rangle\langle m'|H_{ep}(B_{3g})|m\rangle\langle m|H_{opz}|i\rangle$, where $H_{opz}$ ($H_{opy}$) is the electron-photon interaction Hamiltonian matrix for the z (y) polarized light, and $H_{ep}$ is the electron-phonon interaction matrix.

| zy polarized | | | yz polarized | | |
|---|---|---|---|---|---|
| $\|i\rangle=\|f\rangle$ | $\|m\rangle$ | $\|m'\rangle$ | $\|i\rangle=\|f\rangle$ | $\|m\rangle$ | $\|m'\rangle$ |
| $A_g$ | $B_{2u}$ | $B_{1u}$ | $A_g$ | $B_{1u}$ | $B_{2u}$ |
| $B_{1g}$ | $B_{3u}$ | $A_u$ | $B_{1g}$ | $A_u$ | $B_{3u}$ |
| $B_{2g}$ | $A_u$ | $B_{3u}$ | $B_{2g}$ | $B_{3u}$ | $A_u$ |
| $B_{3g}$ | $B_{1u}$ | $B_{2u}$ | $B_{3g}$ | $B_{2u}$ | $B_{1u}$ |
| $A_u$ | $B_{2g}$ | $B_{1g}$ | $A_u$ | $B_{1g}$ | $B_{2g}$ |
| $B_{1u}$ | $B_{3g}$ | $A_g$ | $B_{1u}$ | $A_g$ | $B_{3g}$ |
| $B_{2u}$ | $A_g$ | $B_{3g}$ | $B_{2u}$ | $B_{3g}$ | $A_g$ |
| $B_{3u}$ | $B_{1g}$ | $B_{2g}$ | $B_{3u}$ | $B_{2g}$ | $B_{1g}$ |

**S5.1: Anisotropic electron-photon and electron-phonon interactions in monolayer $Ta_2Ni_3Te_5$**

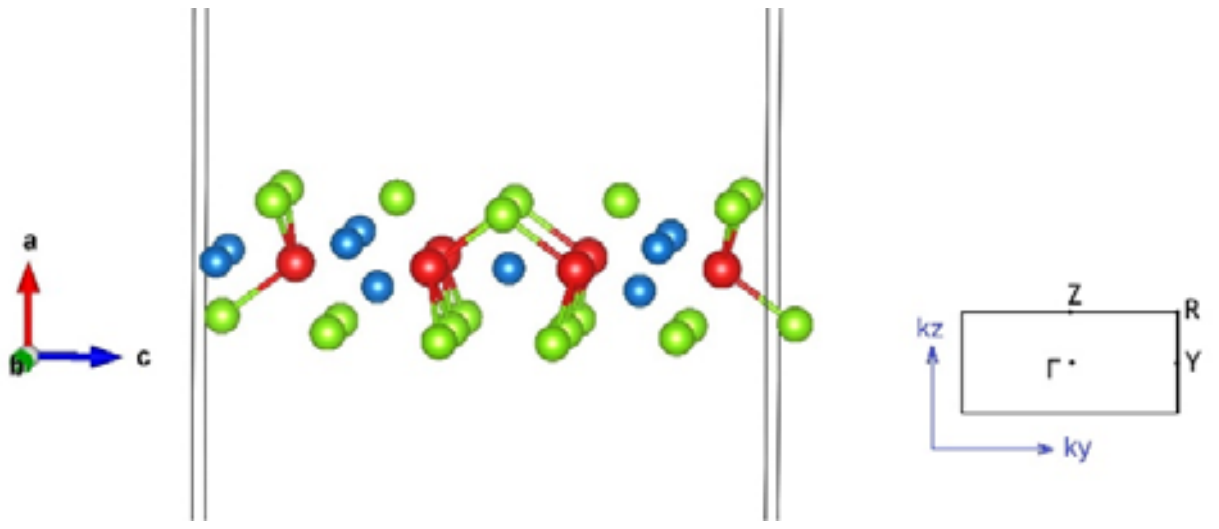

**Figure S14**. The atomic structure of the $Ta_2Ni_3Te_5$ monolayer used in the band structure and Raman mode calculations. The super cell vectors a, b and c are along the x, y and z axis of the Cartesian coordinate system, respectively. Red balls represent Ta atoms, blue balls for Ni atoms, and green ones for Te atoms. The right inset shows the 2D (the yz plane) Brillouin zone.

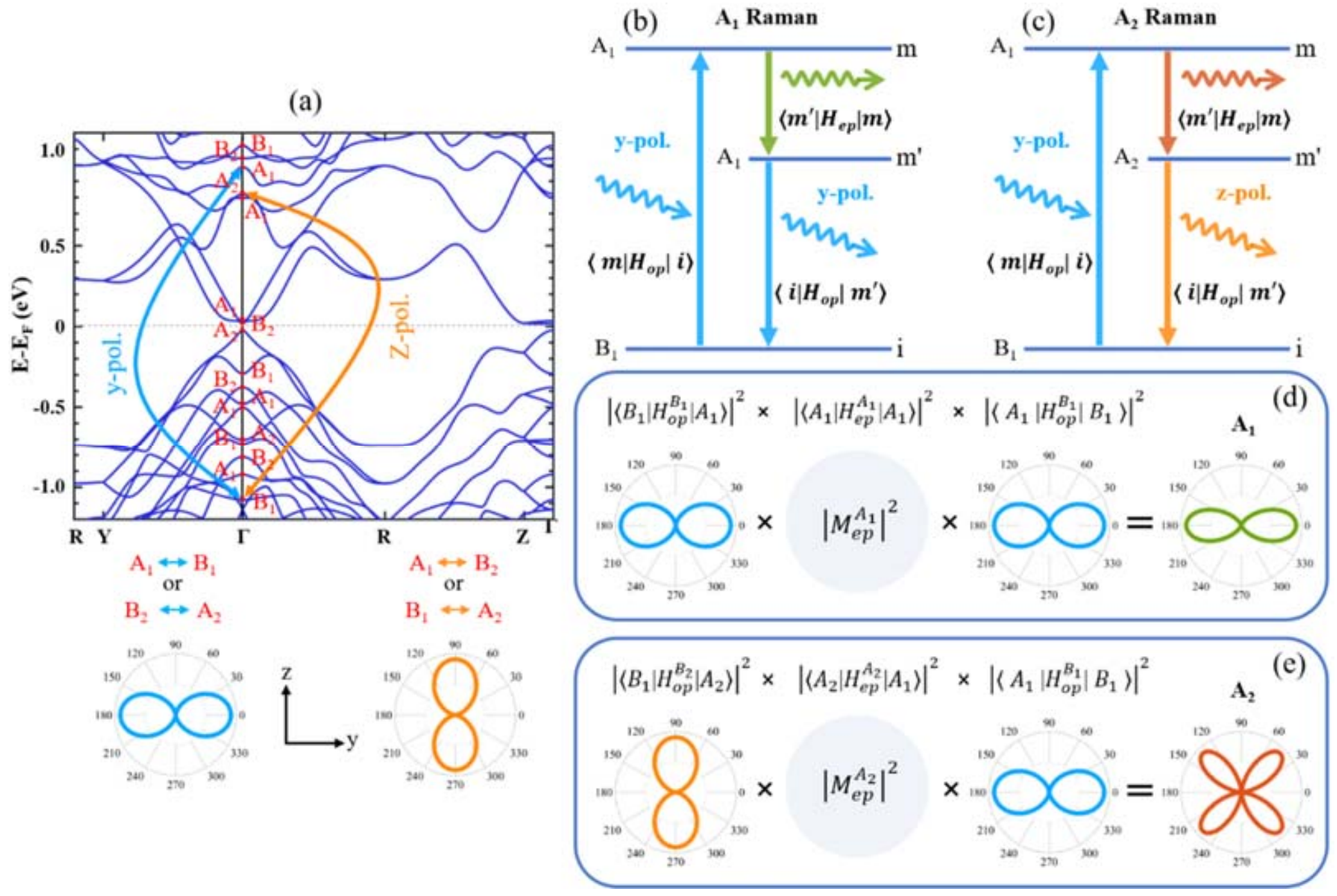

**Figure S15**. Symmetry allowed optical transitions and Raman modes for monolayer $Ta_2Ni_3Te_5$. (a) Symmetry allowed optical transitions are shown in the band structure with irreducible representations marked for bands at the Γ point. The thick blue (orange) line indicates the transition for a $B_1$ ($B_1$) valence band to $A_1$ ($A_2$) conduction band for the y-axis (z-axis) direction polarized light, and its anisotropic absorption as a function of the angle is shown in the lower part (the blue and orange polar plots). Schematic representation of the Raman scattering process for the (b) $A_1$ and (c) $A_2$ Raman modes

with considering the symmetry allowed selection rules in the full quantum theory. Illustration of the polar plots of the electron-photon and electron-phonon interactions and resulting (d) $A_1$ and (e) $A_2$ Raman modes in the Raman scattering, where the anisotropy of electron-phonon interaction is neglected in the calculations.

The atomic structure of $Ta_2Ni_3Te_5$ monolayer is shown in Figure S14, where the 2D monolayer is in the yz plane, the atomic chains are along the y direction. The calculated band structure of monolayer $Ta_2Ni_3Te_5$ is shown in Figure S15a. The gap at the Γ point is about 50 meV. The irreducible representations of the bands at the Γ point are marked. In the calculation, monolayer $Ta_2Ni_3Te_5$ has the $C_{2v}$ (or mm2) point group symmetry. The character table and the base functions of irreducible representations for $C_{2v}$ are shown in Table S7. Notably, although the point group symmetry reduces from $D_{2h}$ (bulk) to $C_{2v}$ (monolayer), the monolayer $A_1$ mode retains the same Raman tensor form as the bulk $A_g$ mode. Furthermore, the calculation results (Figure S15) demonstrate that the key aspects of above analysis persist in the monolayer and are consistent with the bulk findings.

**Table S7**. The character table of point group $C_{2v}$. The directions of the x, y and z axes are described in Figure S14.

| | E | C2(x) | s_v(yx) | s_v'(zx) | Linear rotations | Quadratic functions |
|---|---|---|---|---|---|---|
| $A_1$ | 1 | 1 | 1 | 1 | x | $x^2$, $y^2$, $z^2$ |
| $A_2$ | 1 | 1 | -1 | -1 | Rx | yz |
| $B_1$ | 1 | -1 | 1 | -1 | y, Rz | yx |
| $B_2$ | 1 | -1 | -1 | 1 | z, Ry | zx |

The selection rules of optical transitions between valence bands and conduction bands for the $C_{2v}$ point group are shown in Table S8. Since the irreducible representation of the parallel component of $\vec{D}$ on the $\vec{P}$ direction can be $B_1$ (corresponding to the y-axis polarized light) and $B_2$ (corresponding to the z-axis polarized light), the irreducible representation of the final state $|f\rangle$ can be determined by the group theory of symmetry, as shown in Table S8.

**Table S8.** Selection rules of optical transitions for the $C_{2v}$ point group. These selection rules correspond to the **electron-photon matrix element $\langle f|H_{op}|i\rangle$.**

| yy polarized light (B1) | | zz polarized light (B2) | |
|---|---|---|---|
| $|i\rangle$ | $|f\rangle$ | $|i\rangle$ | $|f\rangle$ |
| $A_1$ | $B_1$ | $A_1$ | $B_2$ |
| $A_2$ | $B_2$ | $A_2$ | $B_1$ |
| $B_1$ | $A_1$ | $B_1$ | $A_2$ |
| $B_2$ | $A_2$ | $B_2$ | $A_1$ |

**Table S9**. Selection rules of Raman scattering for the $A_1$ mode phonon. $|i\rangle$, $|m\rangle$ and $|m'\rangle$ are the initial state and two intermediate states, respectively. yy (zz) are polarization vectors for the incident and scattered light: both are y- (z-) polarized. These selection rules correspond to the product of the matrix elements: $\langle f|H_{opy}|m'\rangle\langle m'|H_{ep}(A_1)|m\rangle\langle m|H_{opy}|i\rangle$ and $\langle f|H_{opz}|m'\rangle\langle m'|H_{ep}(A_1)|m\rangle\langle m|H_{opz}|i\rangle$.

| yy B1 | | zz B2 | |
|---|---|---|---|
| $|i\rangle = |f\rangle$ | $|m\rangle = |m'\rangle$ | $|i\rangle = |f\rangle$ | $|m\rangle = |m'\rangle$ |
| $A_1$ | $B_1$ | $A_1$ | $B_2$ |
| $A_2$ | $B_2$ | $A_2$ | $B_1$ |
| $B_1$ | $A_1$ | $B_1$ | $A_2$ |
| $B_2$ | $A_2$ | $B_2$ | $A_1$ |

**Table S10**. Selection rules of Raman scattering for the $A_2$ mode phonon. The polarization of the incident light is different from that of the scattered light for the $A_2$ mode. yz means that the incident light is z-polarized, and the scattered light is y-polarized. These selection rules correspond to the product of the matrix elements: $\langle f|H_{opy}|m'\rangle\langle m'|H_{ep}(A_2)|m\rangle\langle m|H_{opz}|i\rangle$and $\langle f|H_{opz}|m'\rangle\langle m'|H_{ep}(A_2)|m\rangle\langle m|H_{opy}|i\rangle$.

| yz | | | zy | | |
|---|---|---|---|---|---|
| $|i\rangle = |f\rangle$ | $|m\rangle$ | $|m'\rangle$ | $|i\rangle = |f\rangle$ | $|m\rangle$ | $|m'\rangle$ |
| $A_1$ | $B_2$ | $B_1$ | $A_1$ | $B_1$ | $B_2$ |
| $A_2$ | $B_1$ | $B_2$ | $A_2$ | $B_2$ | $B_1$ |
| $B_1$ | $A_2$ | $A_1$ | $B_1$ | $A_1$ | $A_2$ |
| $B_2$ | $A_1$ | $A_2$ | $B_2$ | $A_2$ | $A_1$ |

**S5.2: Complex Raman tensors calculation in monolayer $Ta_2Ni_3Te_5$ with considering anisotropic electron-phonon interaction:**

The first-principles calculations were performed with the Vienna Ab initio Software Package (VASP)[19] with projector augmented wave (PAW) pseudopotentials[20] describing the interactions between valence electrons and ion cores. The exchange-correlation energy in the electronic system was described with the Perdew-Burke-Ernzerhof (PBE) functional.[21] A supercell of 1x4x1 was used for phonon calculations with a Gamma centered k point mesh of 4x4x4. An energy cutoff of 450 eV was used for the plane-wave basis set expansion. The forces on each atom were relaxed to less than 0.001 eV/Å and the total energy was converged to less than $10^{-6}$ eV. The Phonopy code[22] was used for calculations of force sets and phonon frequencies. The complex Raman tensor calculation was performed with Quantum ESPRESSO[23] and QERaman,[24] where the Optimized Norm-Conserving Vanderbilt Pseudopotentials[25] were used. The energy cutoff of 90 Ry (1224 eV) for wavefunctions and the k mesh of 1x20x6 were used for monolayer calculations.

The back parallel scattering is assumed. The angle is induced by the rotation of the monolayer in the yz plane. The plots are based on the complex Raman tensors calculated in DFPT in QERaman code. It shows the anisotropic Raman spectra with the angle between the y axis and the polarization direction of the incident light.

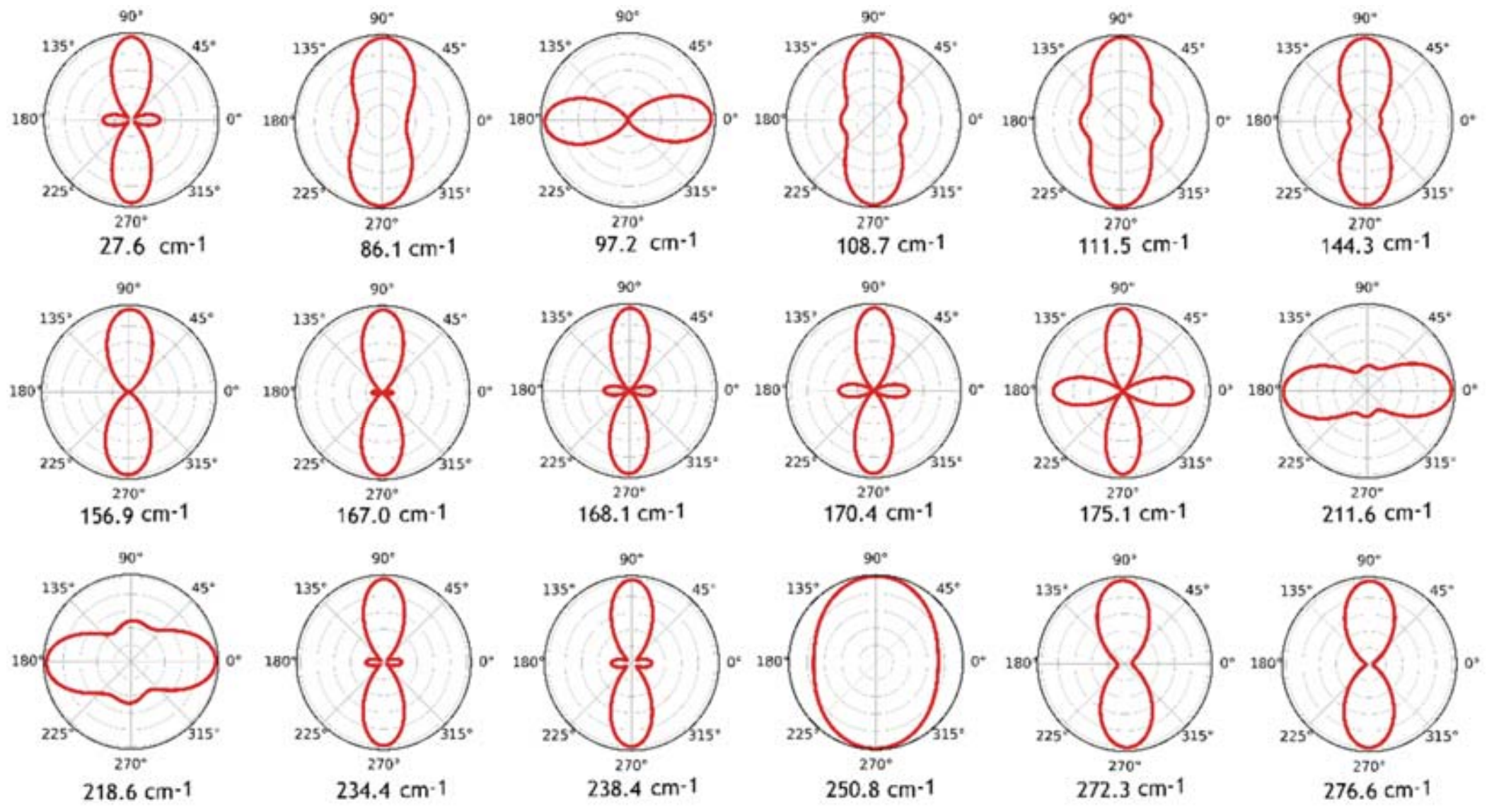

**Figure S16**. The polar plots of the monolayer 18 $A_1$ Raman modes. The angle $\theta$ is between the y axis and the polarization direction of the incident light.

**Table S11**. Lists of the complex Raman tensors of the monolayer $A_1$ Raman modes. For each phonon mode block, the 1st and 2nd columns represent the real and imaginary parts of the first complex number, respectively. The 3rd and 4th columns for the real and imaginary parts of the second complex number, respectively. The 5th and 6th columns for the real and imaginary parts of the third complex number, respectively.

| | | | | | | | | | | | |
|---|---|---|---|---|---|---|---|---|---|---|---|
| Phonon mode: 27.6 $cm^{-1}$ | | | | | | Phonon mode: 86.1 $cm^{-1}$ | | | | | |
| -0.1 | 0.4 | 0.0 | 0.0 | 0.1 | 0.0 | 0.0 | -0.2 | 0.0 | 0.0 | 0.0 | 0.0 |
| 0.0 | 0.0 | -0.6 | -0.2 | 0.0 | 0.0 | 0.0 | 0.0 | 0.3 | 0.5 | 0.0 | 0.0 |
| 0.1 | 0.0 | 0.0 | 0.0 | 1.0 | -0.3 | 0.0 | 0.0 | 0.0 | 0.0 | -0.3 | 1.0 |
| Phonon mode: 97.2 $cm^{-1}$ | | | | | | Phonon mode: 108.7 $cm^{-1}$ | | | | | |
| -0.2 | -0.5 | 0.0 | 0.0 | -0.1 | 0.0 | 0.2 | 0.1 | 0.0 | 0.0 | 0.0 | 0.0 |
| 0.0 | 0.0 | 0.1 | -1.0 | 0.0 | 0.0 | 0.0 | 0.0 | 0.3 | 0.5 | 0.0 | 0.0 |
| -0.1 | 0.0 | 0.0 | 0.0 | -0.1 | 0.1 | 0.0 | 0.0 | 0.0 | 0.0 | -0.6 | 0.8 |
| Phonon mode: 111.5 $cm^{-1}$ | | | | | | Phonon mode: 144.3 $cm^{-1}$ | | | | | |
| 0.2 | 0.1 | 0.0 | 0.0 | 0.0 | 0.0 | -0.5 | 0.0 | 0.0 | 0.0 | 0.0 | 0.0 |
| 0.0 | 0.0 | 0.3 | 0.6 | 0.0 | 0.0 | 0.0 | 0.0 | 0.1 | 0.4 | 0.0 | 0.0 |
| 0.0 | 0.0 | 0.0 | 0.0 | -0.7 | 0.8 | 0.0 | 0.0 | 0.0 | 0.0 | -0.9 | 0.4 |
| Phonon mode: 156.9 $cm^{-1}$ | | | | | | Phonon mode: 167.0 $cm^{-1}$ | | | | | |
| 0.2 | -0.2 | 0.0 | 0.0 | 0.4 | -0.6 | -0.1 | 0.0 | 0.0 | 0.0 | 0.0 | 0.0 |
| 0.0 | 0.0 | -0.1 | 0.1 | 0.0 | 0.0 | 0.0 | 0.0 | -0.3 | 0.1 | 0.0 | 0.0 |
| 0.4 | -0.6 | 0.0 | 0.0 | 0.6 | -0.8 | 0.0 | 0.0 | 0.0 | 0.0 | 0.8 | -0.5 |
| Phonon mode: 168.1 $cm^{-1}$ | | | | | | Phonon mode: 170.4 $cm^{-1}$ | | | | | |
| 0.2 | -0.3 | 0.0 | 0.0 | 0.0 | 0.1 | -0.2 | 0.3 | 0.0 | 0.0 | 0.0 | 0.0 |
| 0.0 | 0.0 | 0.1 | 0.5 | 0.0 | 0.0 | 0.0 | 0.0 | -0.1 | -0.6 | 0.0 | 0.0 |
| 0.0 | 0.1 | 0.0 | 0.0 | 0.2 | -1.0 | 0.0 | 0.0 | 0.0 | 0.0 | -0.1 | 1.0 |
| Phonon mode: 175.1 $cm^{-1}$ | | | | | | Phonon mode: 211.6 $cm^{-1}$ | | | | | |
| 0.2 | -0.3 | 0.0 | 0.0 | 0.1 | 0.0 | -0.1 | 0.2 | 0.0 | 0.0 | -0.2 | 0.2 |
| 0.0 | 0.0 | 0.2 | 0.9 | 0.0 | 0.0 | 0.0 | 0.0 | -0.5 | -0.9 | 0.0 | 0.0 |
| 0.1 | 0.0 | 0.0 | 0.0 | 0.0 | -1.0 | -0.1 | 0.1 | 0.0 | 0.0 | 0.4 | -0.4 |
| Phonon mode: 218.6 $cm^{-1}$ | | | | | | Phonon mode: 234.4 $cm^{-1}$ | | | | | |
| -0.1 | 0.1 | 0.0 | 0.0 | 0.0 | -0.1 | -0.3 | 0.1 | 0.0 | 0.0 | 0.0 | 0.0 |
| 0.0 | 0.0 | -0.6 | -0.8 | 0.0 | 0.0 | 0.0 | 0.0 | -0.4 | -0.2 | 0.0 | 0.0 |
| 0.0 | -0.1 | 0.0 | 0.0 | 0.4 | -0.6 | 0.0 | 0.0 | 0.0 | 0.0 | 0.3 | 0.9 |
| Phonon mode: 238.4 $cm^{-1}$ | | | | | | Phonon mode: 250.8 $cm^{-1}$ | | | | | |
| 0.3 | -0.1 | 0.0 | 0.0 | -0.1 | 0.0 | 0.1 | 0.3 | 0.0 | 0.0 | -0.3 | 0.0 |
| 0.0 | 0.0 | 0.4 | 0.2 | 0.0 | 0.0 | 0.0 | 0.0 | 0.7 | 0.5 | 0.0 | 0.0 |
| -0.1 | 0.0 | 0.0 | 0.0 | -0.2 | -1.0 | -0.2 | 0.1 | 0.0 | 0.0 | 0.9 | 0.5 |
| Phonon mode:272.3 $cm^{-1}$ | | | | | | Phonon mode:276.6 $cm^{-1}$ | | | | | |
| -0.1 | -0.2 | 0.0 | 0.0 | -0.1 | 0.0 | 0.1 | 0.2 | 0.0 | 0.0 | -0.2 | 0.1 |
| 0.0 | 0.0 | -0.2 | -0.2 | 0.0 | 0.0 | 0.0 | 0.0 | 0.1 | 0.2 | 0.0 | 0.0 |
| -0.1 | 0.0 | 0.0 | 0.0 | -1.0 | 0.3 | -0.1 | 0.1 | 0.0 | 0.0 | 0.9 | -0.3 |

The complex Raman tensors calculation was performed with QERaman code, which didn't include the symmetry constraint currently. In the present calculation workflow, the computed tensors could not explicitly project onto the ideal symmetry-constrained Raman-tensor form after the numerical calculation. This may lead to some uncertainty in the unconstrained numerical tensor output. The overall calculated angular responses and, in particular, the systematic mode dependence of the Raman patterns remain consistent with the symmetry-allowed framework and capture the qualitative characteristics and diversity of the anomalous angular responses observed experimentally, indicating the reasonable reliability of the results.

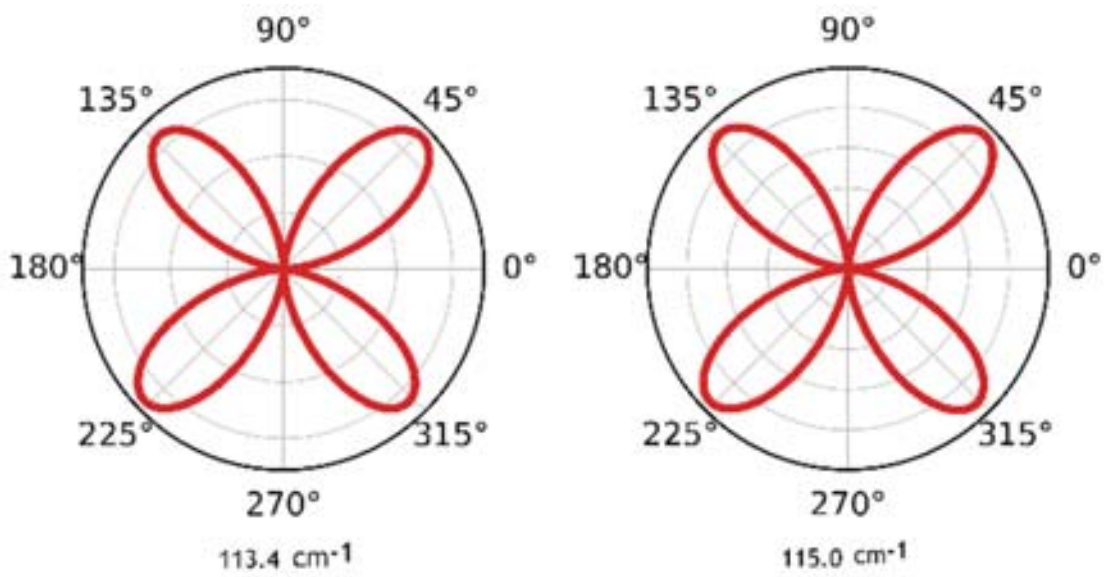

**Figure S17**. The polar plots of two $A_2$ Raman modes.

**Table S12**. Lists of the complex Raman tensors of the two $A_2$ Raman modes.

| Phonon mode: 113.4 $cm^{-1}$ | | | | | | Phonon mode: 115.0 $cm^{-1}$ | | | | | |
|---|---|---|---|---|---|---|---|---|---|---|---|
| -0.1 | 0.0 | 1.0 | -0.1 | 0.0 | 0.0 | 0.0 | 0.0 | -0.7 | 0.1 | 0.0 | 0.0 |
| 1.0 | -0.1 | 0.0 | -0.1 | -0.8 | -0.3 | -0.6 | 0.1 | 0.0 | -0.1 | -0.9 | -0.4 |
| 0.0 | 0.0 | -0.7 | -0.3 | 0.1 | -0.1 | 0.0 | 0.0 | -0.9 | -0.3 | 0.1 | -0.1 |

**S5.3: Anisotropic electron-phonon interaction maps in the first Brillouin zone.**

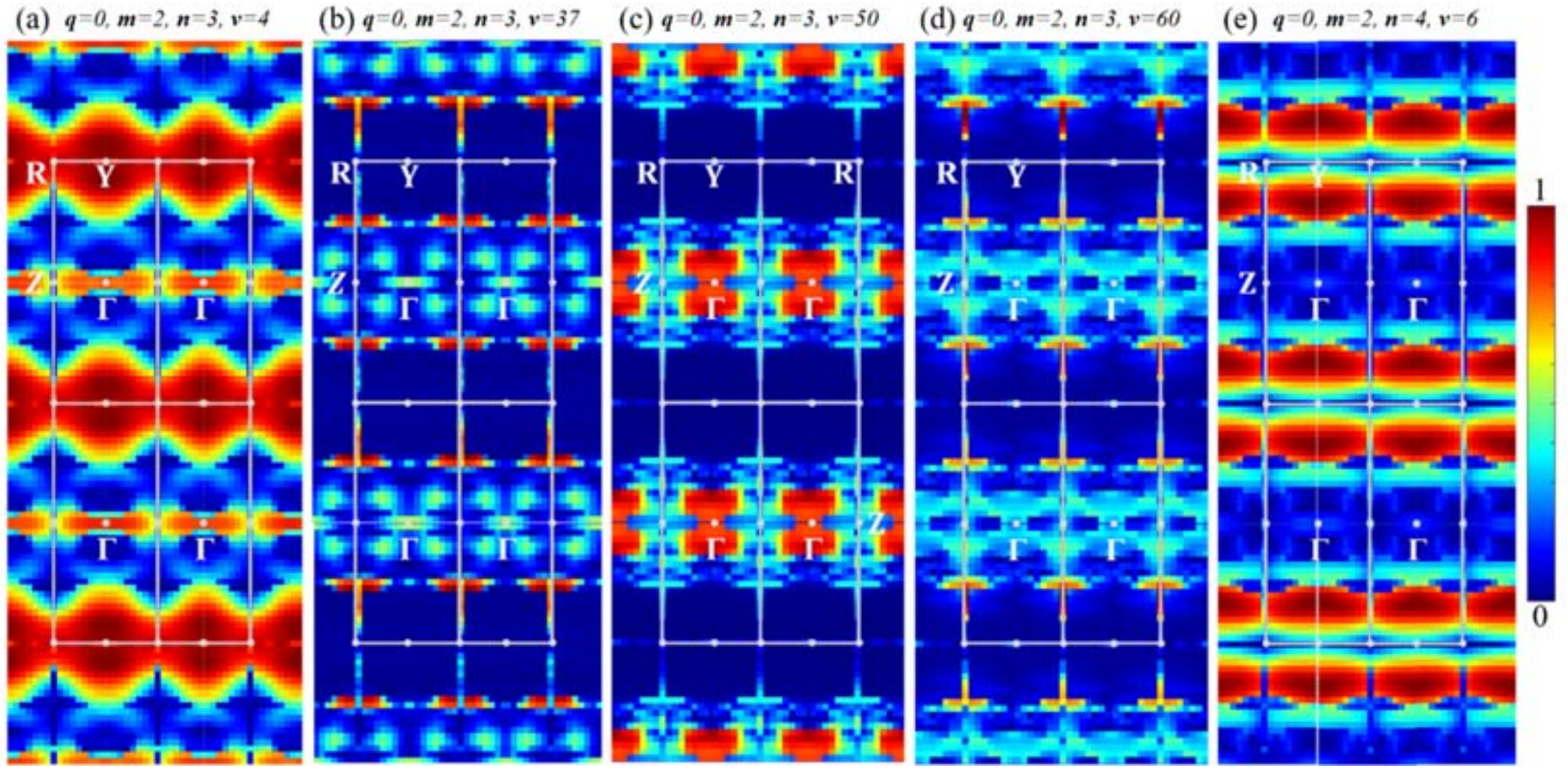

**Fig. S18.** Each plot corresponds to a map of the EPIs matrix as a function of the electronic wavevector $\boldsymbol{k}$ over the entire first Brillouin zone (BZ) for a phonon mode $v$.

The electron-phonon interaction matrix calculation was conducted with Quantum ESPRESSO (QE)[23] and EPW (Electro-Phonon coupling using Wannier functions) codes.[26] The PBE exchange-correlation functional[21] and the Optimized Norm-Conserving Vanderbilt Pseudopotentials[25] were used. The EPIs matrix is defined as $g_{m,n,v}(\boldsymbol{k},\boldsymbol{q}) = \langle\psi_{n,\boldsymbol{k}+\boldsymbol{q}}|\partial_{\boldsymbol{q}v}V|\psi_{m,\boldsymbol{k}}\rangle$, where $\psi_{m,\boldsymbol{k}}$ is the electronic wavefunction for band $m$ with wavevector $\boldsymbol{k}$, and $\partial_{\boldsymbol{q}v}V$ is the derivative of the self-consistent potential associated with a phonon mode $\omega_{\boldsymbol{q}v}$ of wavevector $\boldsymbol{q}$ and branch index $v$. Phonon mode $\omega_{\boldsymbol{q}v}$ and $\partial_{\boldsymbol{q}v}V$ were calculated in QE with the density-functional perturbation theory (DFPT) self-consistently. With a maximally localized Wannierization and interpolation scheme in EPW, the $g_{m,n,v}(\boldsymbol{k},\boldsymbol{q})$ matrix on a dense k-grid can be obtained. For monolayer $Ta_2Ni_3Te_5$, the energy cutoff of 90 Rydberg was set for the plane-wave expansion, and the coarse k-grid of 1x20x8 and q-grid of 1x5x2 were used for electron and phonon state calculations, respectively. The $g_{m,n,v}(\boldsymbol{k},\boldsymbol{q})$ matrix with $\boldsymbol{q}$ =0, corresponding to Raman modes with zero momentum transfer, was mapped on the first Brillouin zone (BZ) with a finer uniform grid 1x40x16 with selected two conduction bands and a phonon branch index. $m$ and $n$ represent the conduction band index counted from the Fermi level upwards.

The figure below shows the EPIs matrix $g_{m,n,v}(\boldsymbol{k}, \boldsymbol{q} = \boldsymbol{0})$ of graphene plotted in the first BZ. *m* and *n* represent the conduction band index counted from the Fermi level upwards, i.e., *m=2* means the 2nd conduction band, and *n=1* means the 1st conduction band. *q=0* means the phonon wavevector is zero, or the momentum transfer between the phonon and electronic states is zero, consistent with the regular Raman scattering.

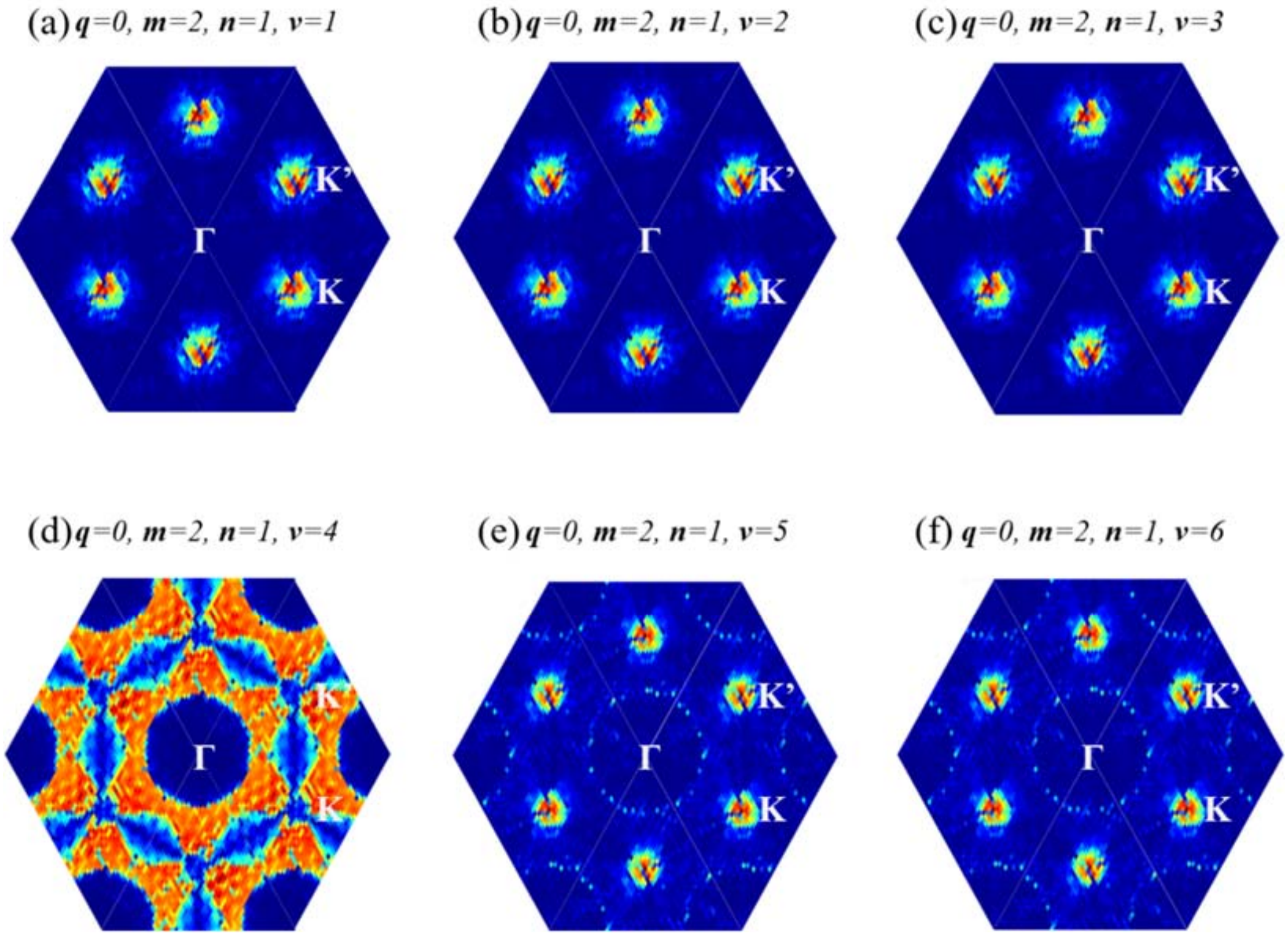

**Fig. S19.** EPIs matrix of the six phonon modes in Graphene. Each plot corresponds to a map of the EPIs matrix as a function of the electronic wavevector $\boldsymbol{k}$ over the entire first Brillouin zone (BZ) for a phonon mode $v$.

## S5.4: Derivation of complex Raman tensor from the quantum Raman scattering

This section provides the explicit derivation showing how the quantum-mechanical Raman scattering, containing the anisotropic electron–photon and electron–phonon interactions, naturally leads to complex Raman tensor.

The observed Raman scattering patterns can be understood from Equation 1 in the main text. The key point is that Equation 1 contains both electron-phonon interaction term, $\langle m' \mid H_{ep}^{\nu} \mid m\rangle$, and the electron-photon interaction terms, $\langle i \mid H_{op} \mid m'\rangle$and $\langle m \mid H_{op} \mid i\rangle$. These contributions are summed over all possible electronic band transitions and weighted by the corresponding energy denominators. The angular dependence enters through the electron-phonon and electron-photon interaction matrix elements, which themselves depend on the relative orientation of the crystal and the incident/scattered light polarizations.

The complete expression for Equation 1 is:

$$I_{\nu}(E_L, E_{RS}) = \left| \sum_{\mathbf{k},i,m,m'} \frac{\langle i|H_{op}|m'\rangle\langle m'|H_{ep}^{\nu}|m\rangle\langle m|H_{op}|i\rangle}{(E_L - \Delta E_{mi}(\mathbf{k}))(E_L - \hbar\omega_{\nu} - \Delta E_{m'i}(\mathbf{k}))} \right|^2 \times \delta(E_{RS} - \hbar\omega_{\nu}) \quad \text{(S34)}$$

where the band states $\mid i\rangle$, $\mid m\rangle$, and $\mid m'\rangle$depend on the wavevector $\boldsymbol{k}$. Here, $E_{RS}$ is the Raman shift, and $\hbar\omega_{\nu}$ is the phonon energy for phonon mode $\nu$. The delta function can be approximated by a Lorentzian broadening function:

$$\delta(E_{RS} - \hbar\omega_{\nu}) \to \frac{1}{\pi}\left[\frac{\Gamma}{(E_{RS} - \hbar\omega_{\nu})^2 + \Gamma^2}\right] \quad \text{(S35)}$$

where $\Gamma$ is proportional to the inverse phonon lifetime. Within the dipole approximation, $\langle i|H_{op}|m'\rangle$ can be expressed as:

$$\langle i|H_{op}|m'\rangle = C_{opt} D^{m'-i}(\mathbf{k}) \cdot \mathbf{P} \quad \text{(S36)}$$

where $C_{opt}$ is the potential associated with the light:

$$C_{opt} = \frac{-ie\hbar}{m}\left(A \cdot expi\left(\mathbf{k}_{opt} \cdot \mathbf{r} - \omega_{opt} t\right) + c.c.\right) \quad \text{(S37)}$$

where $A$ is the magnitude of the vector potential of the light wave, and $\boldsymbol{k}_{\mathbf{opt}}$ and $\omega_{\mathrm{opt}}$ are the wavevector and angular frequency of the light. Here, "c.c." denotes the complex conjugate. $C_{\mathrm{opt}}$(or $C_{\mathrm{opt}}^2$) can be treated as a constant for a given experimental setup. **P** is the polarization vector, which describes the direction and amplitude of the electric field of the light wave. $D^{i-f}(\mathbf{k})$ is the dipole vector of the transition:

$$D^{i\to f}(\mathbf{k}) = \langle f|\nabla|i\rangle \tag{S38}$$

Using the above expressions, the Raman intensity can be written as:

$$I_{\alpha,\beta}(\nu, E_L, E_{RS}) = \left|\mathbf{P}_\alpha^*\mathbf{R}(\nu)\mathbf{P}_\beta\right|^2 \times \delta(E_{RS} - \hbar\omega_\nu) \tag{S39}$$

where $\alpha,$ and $\beta$ represent the polarization direction. $\mathbf{R}(\nu)$ is the Quantum Raman tensor:

$$\mathbf{R}(\nu) = C_{opt}^2 \sum_{\mathbf{k},i,m,m'} \frac{D^{m'\to i}(\mathbf{k}) \left\langle m'\middle|H_{ep}^{\nu}\middle|m\right\rangle D^{i\to m}(\mathbf{k})}{(E_L - \Delta E_{mi}(\mathbf{k}))\left(E_L - \hbar\omega_\nu - \Delta E_{m'i}(\mathbf{k})\right)} \tag{S40}$$

Here, $D^{i-f}(\mathbf{k})$ and $\left\langle m' \mid H_{ep}^{\nu} \mid m\right\rangle$ have complex quantities in this quantum-mechanical derivation. Consequently, the quantum Raman tensor is also complex. Therefore, the complex Raman tensor, which can effectively fit the experimental data, originates from the anisotropic electron-photon and electron-phonon interactions. Moreover, because the electron–phonon matrix elements $\left\langle m' \mid H_{ep}^{\nu} \mid m\right\rangle$ are intrinsically mode-specific (they depend on the phonon mode ν) and momentum-dependent (they vary strongly with k, as directly visualized in Figures 4g–i), the interference among the different scattering channels in the coherent sum differs from mode to mode. This is precisely the microscopic origin of the mode-specific complex phases of the Raman tensor elements: the anisotropic EPI matrix elements reweight the complex-valued contributions of the different inter-band channels, producing the mode-dependent phase differences observed experimentally (Figures 3e–h and S8).

This derivation also establishes the full consistency between the quantum framework and the tensor formalism: the classical (phenomenological) complex Raman tensor is recovered as the coherent sum in Equation (S40), with its complex elements now given explicit microscopic content.

## S6: Temperature-dependent XRD, Raman spectra and phonon decay process in $Ta_2Ni_3Te_5$

To further probe the phonon dynamics and their interactions, complementary temperature-dependent Raman measurements, together with the analysis of three- and four-phonon anharmonic decay processes, are presented in this section.

The temperature-dependent powder X-ray diffraction (XRD) was first performed on $Ta_2Ni_3Te_5$ single crystals. The XRD patterns are strictly consistent between 10 K and 300 K without any splitting or emergence of new diffraction peaks (Figure S20), indicating a stable crystal structure.

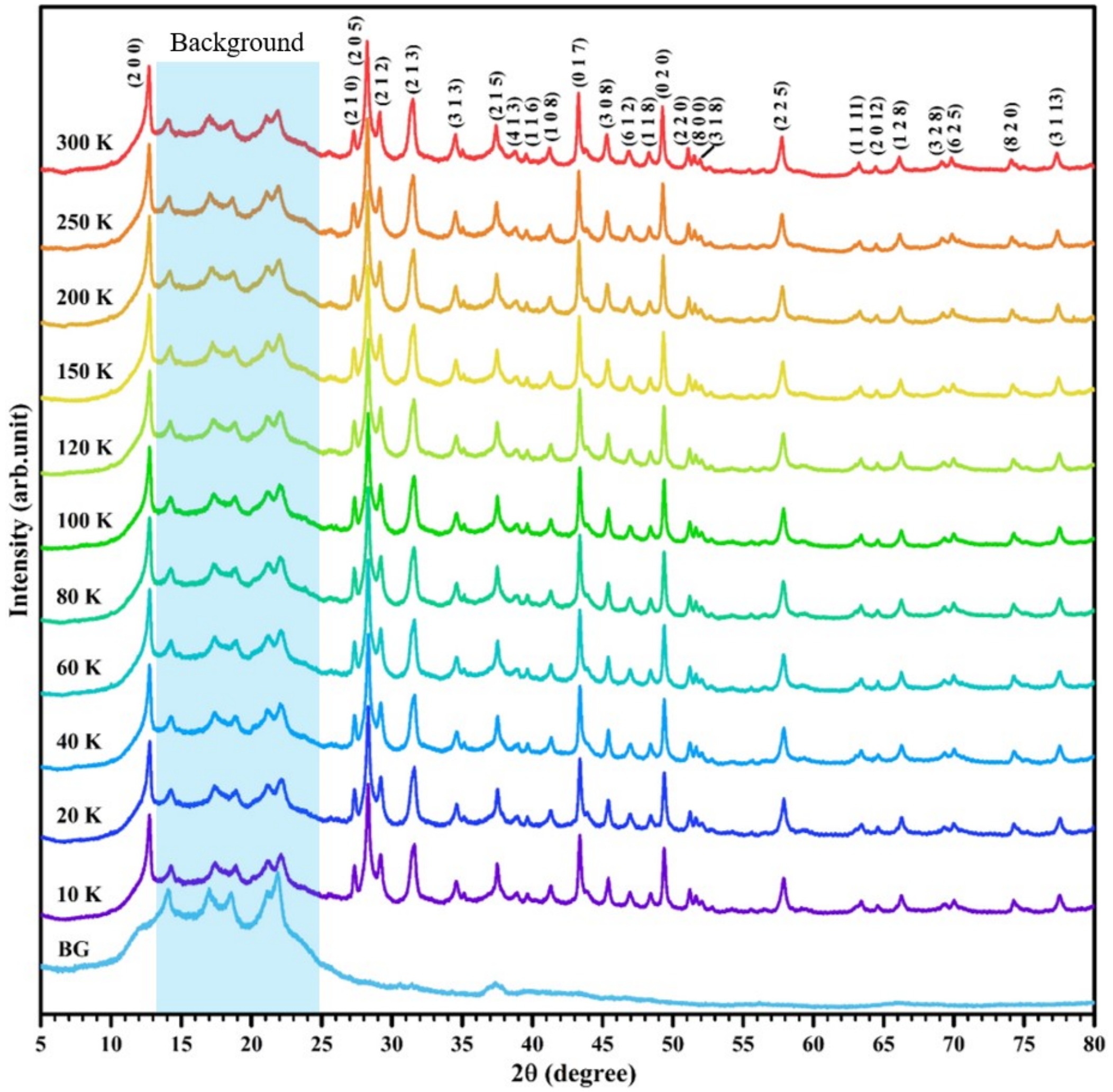

**Figure S20.** Temperature-dependent powder XRD patterns of $Ta_2Ni_3Te_5$ crystals. There is no sign of structural change from the XRD patterns between 10 K and 300 K. The experimental data has a certain background due to the cooling station.

Temperature-dependent Raman measurements were performed in a home-built liquid nitrogen-cooled open-cycle cryostat with temperature range from 80-475 K. Higher-order phonon decay processes have been observed, reflecting the complex interactions and points to strong lattice anharmonicity.

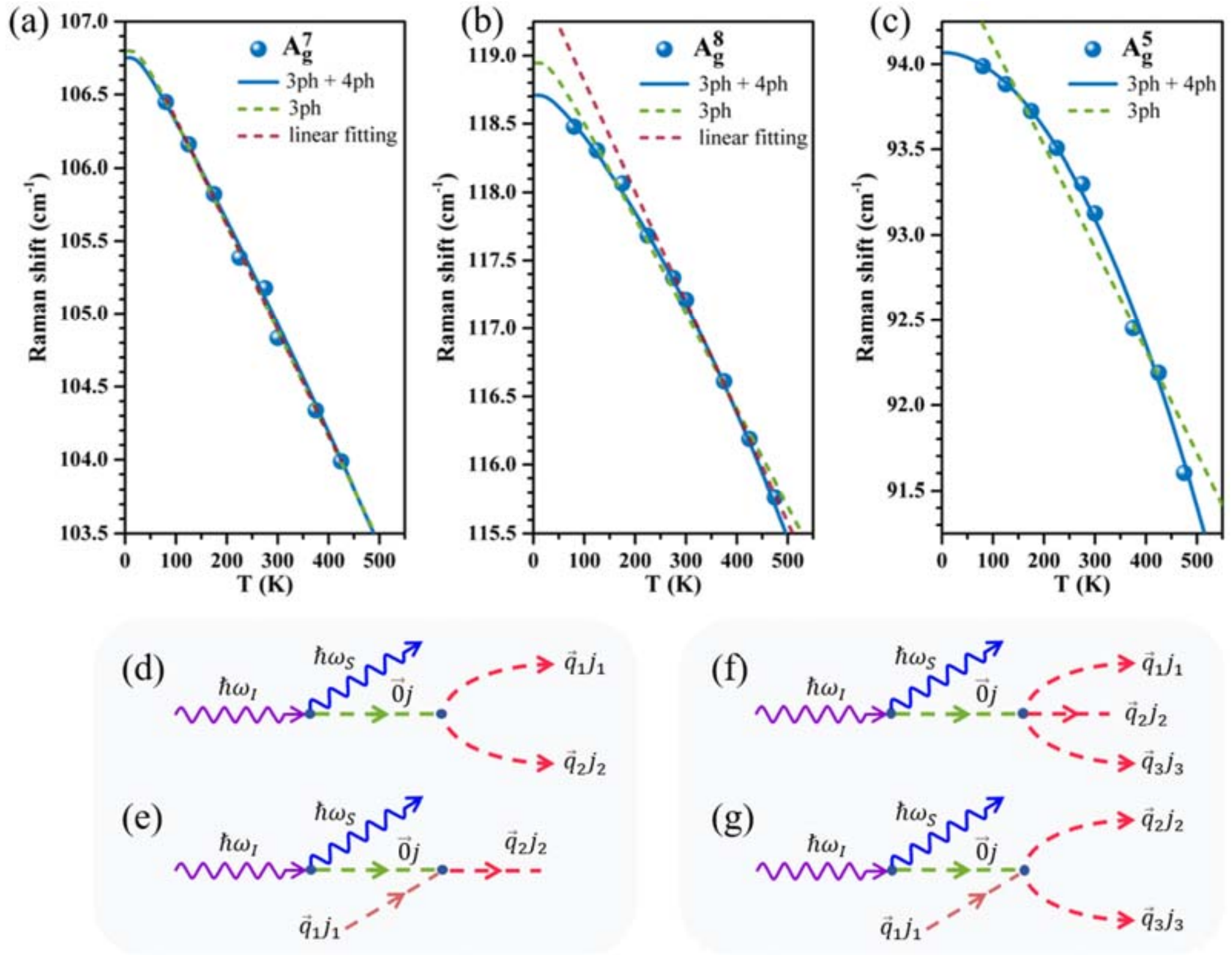

**Figure S21.** Anharmonic effect and phonon decay processes in $Ta_2Ni_3Te_5$. (a-c) Temperature dependence of the Raman modes. Experimental data are represented as blue dots, blue solid lines (green dashed lines) are the corresponding fitting curves with three- and four-phonon (only three-phonon) processes. The red dashed lines in some of the plots are linear fitting of the experimental data in the linear range. (d, e) Three- and (f, g) four- phonon processes contributing to the decay of the Raman active modes.

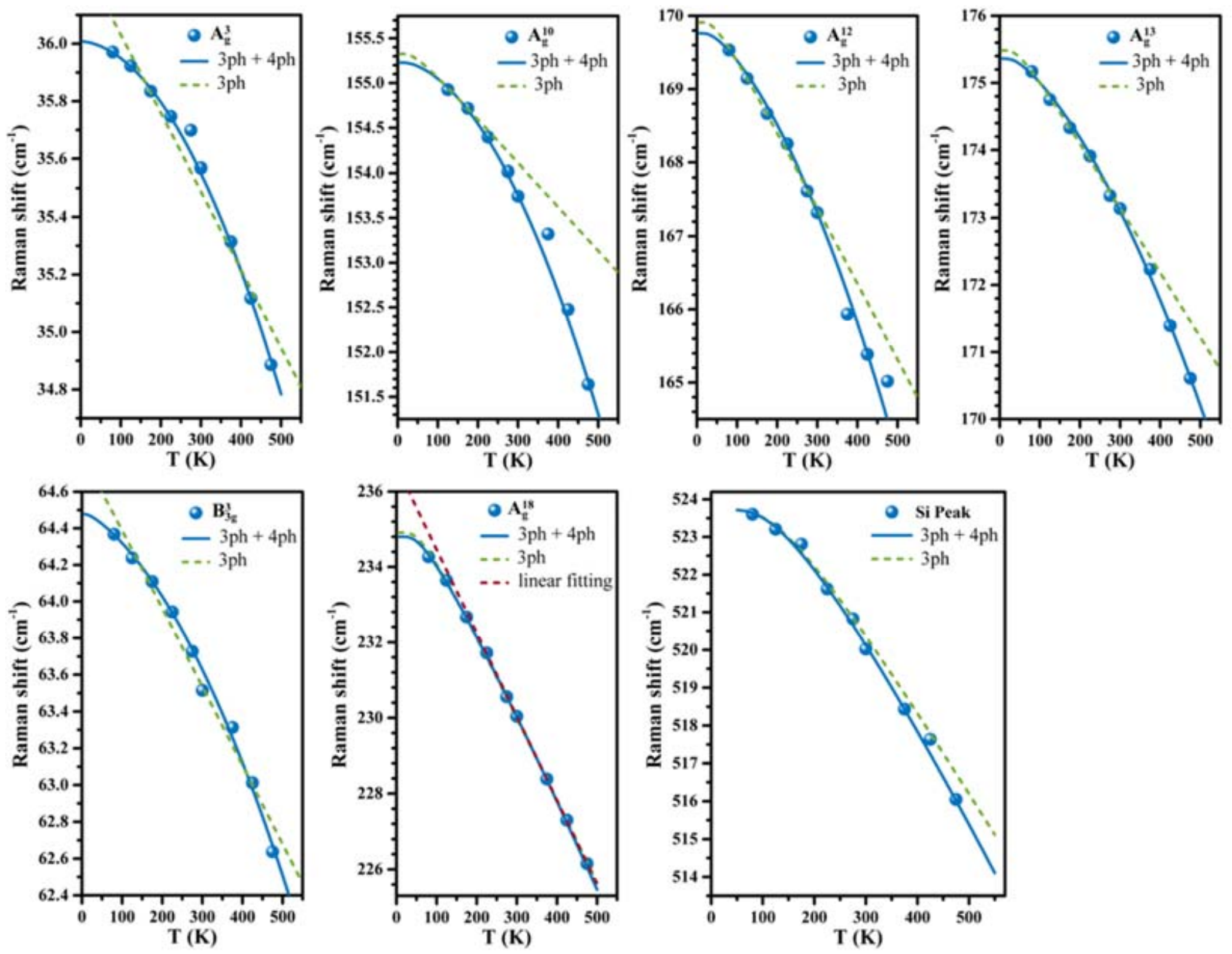

**Figure S22**. Temperature dependence of the frequencies of more Raman modes. Experimental data are represented as blue dots, blue solid lines (green dashed lines) are the corresponding fitting curves with three- and four-phonon (only three-phonon) process. The red dash lines in some of the plots are the linear fitting of the experimental data in the linear range.

Different Raman peaks exhibit distinct temperature dependencies. Notably, the Raman peak at 105.2 cm$^{-1}$ (Figure S21a) displays an almost linear temperature dependence within this range, whereas most other Raman peaks demonstrate strong nonlinear temperature dependence, arising from the anharmonic decay processes. Conventionally, the temperature effects on Raman shifts are primarily attributed to two contributions: the “self-energy” shift arising from anharmonic phonon mode coupling, and the shift due to lattice parameter variations resulting from thermal expansion, which alters the crystal's force constants with volume.[27-30] Consequently, the temperature dependence of Raman mode positions is often approximated using a first-order temperature coefficient according to the equation:

$$\omega(T) = \omega_0 + \left(\frac{\partial \omega}{\partial T}\right)_V \Delta T + \left(\frac{\partial \omega}{\partial V}\right)_T \Delta V = \omega_0 + (\chi_T + \chi_V)\,\Delta T = \omega_0 + \chi\,\Delta T \qquad \text{(S41)}$$

Here, $\omega_0$ is the vibration frequency at 0 $K$. The coefficient $\chi$ comprises two components: the "self-energy" shift term $\chi_T$ and thermal expansion term $\chi_V$. However, the temperature dependencies observed for most Raman mode positions in our study (including the reference Si peak) exhibit apparent nonlinearity, which cannot be adequately explained solely by this first-order temperature coefficient model. A more accurate description of $\omega(T)$ across the entire temperature range can be achieved using the model developed by Balkanski et al.[31] This model was initially proposed to explain the temperature dependence of the silicon Raman mode position and was subsequently applied to GaAs.[32] More recently, the nonlinear temperature-dependent Raman shifts observed in some low-dimensional systems, like $MoS_2$,[33] $WS_2$,[34] $ReSe_2$,[35] $NiPS_3$,[36] black phosphorus,[37] and topological Kagome metal $ScV_6Sn_6$,[38] have been explained with this approach.

To identify the dominant phonon decay mechanism, we analyzed the temperature dependent phonon frequency, $\omega(T)$, using the model established by Balkanski et al.:[31]

$$\omega(T) = \omega_0 + A\left(1 + \frac{2}{e^x - 1}\right) + B\left(1 + \frac{3}{e^y - 1} + \frac{3}{(e^y - 1)^2}\right) \qquad \text{(S42)}$$

where $x = \hbar\omega_0/2k_BT$, $y = \hbar\omega_0/3k_BT$. The coefficients $A$ and $B$ represent the frequency shifts induced by the decay of an optical phonon into two (three-phonon process, second term) or three (four-phonon process, third term) acoustic phonons with equal energy, stemming from cubic and quartic anharmonicity. The physical processes are illustrated in Figure S21d-g.

As a calibration, the Si peak is fitted (shown in Figure S22), yielding parameters consistent with the original work,[31] thereby validating our temperature-dependent measurements for $Ta_2Ni_3Te_5$. As shown in Figure S21 and S22, while three-phonon process adequately fits some data points at low-temperature, they become insufficient to capture the rapid redshift observed in $Ta_2Ni_3Te_5$, particularly at higher temperatures. Excellent agreement is achieved only when four-phonon processes are included or become the dominant term.

**Table S13**. Fitting parameters for the three- and four-phonon decay process, together with the linear fitting parameters in some of the plots.

| Raman modes | $\omega_0$ (cm$^{-1}$) | A (cm$^{-1}$) | B (cm$^{-1}$) | B/A | $\omega_0'$ (cm$^{-1}$) | $\chi$ (cm$^{-1}$*K$^{-1}$) |
|---|---|---|---|---|---|---|
| $A_g^3$ | 36.01 | -0.0018 | -0.00046 | 0.255556 | -- | -- |
| $B_{3g}^3$ | 64.51 | -0.031 | -0.0017 | 0.054839 | -- | -- |
| $A_g^5$ | 94.08 | -0.006 | -0.007 | 1.166667 | -- | -- |
| $A_g^7$ | 107 | -0.247 | -0.0013 | 0.005263 | 107.05 | -0.0072 |
| $A_g^8$ | 118.88 | -0.167 | -0.0063 | 0.037725 | 119.62 | -0.00808 |
| $A_g^{10}$ | 155.3 | -0.045 | -0.0264 | 0.586667 | -- | -- |
| $A_g^{12}$ | 170.07 | -0.275 | -0.0335 | 0.121818 | -- | -- |
| $A_g^{13}$ | 175.7 | -0.31 | -0.0285 | 0.091935 | -- | -- |
| $A_g^{18}$ | 236.5 | -1.68 | -0.018 | 0.010714 | 236.67 | -0.02212 |
| Si | 527.9 | -4.064 | -0.104 | 0.025591 | -- | -- |

## S7: Thickness-dependent angle-resolved polarized Raman measurements

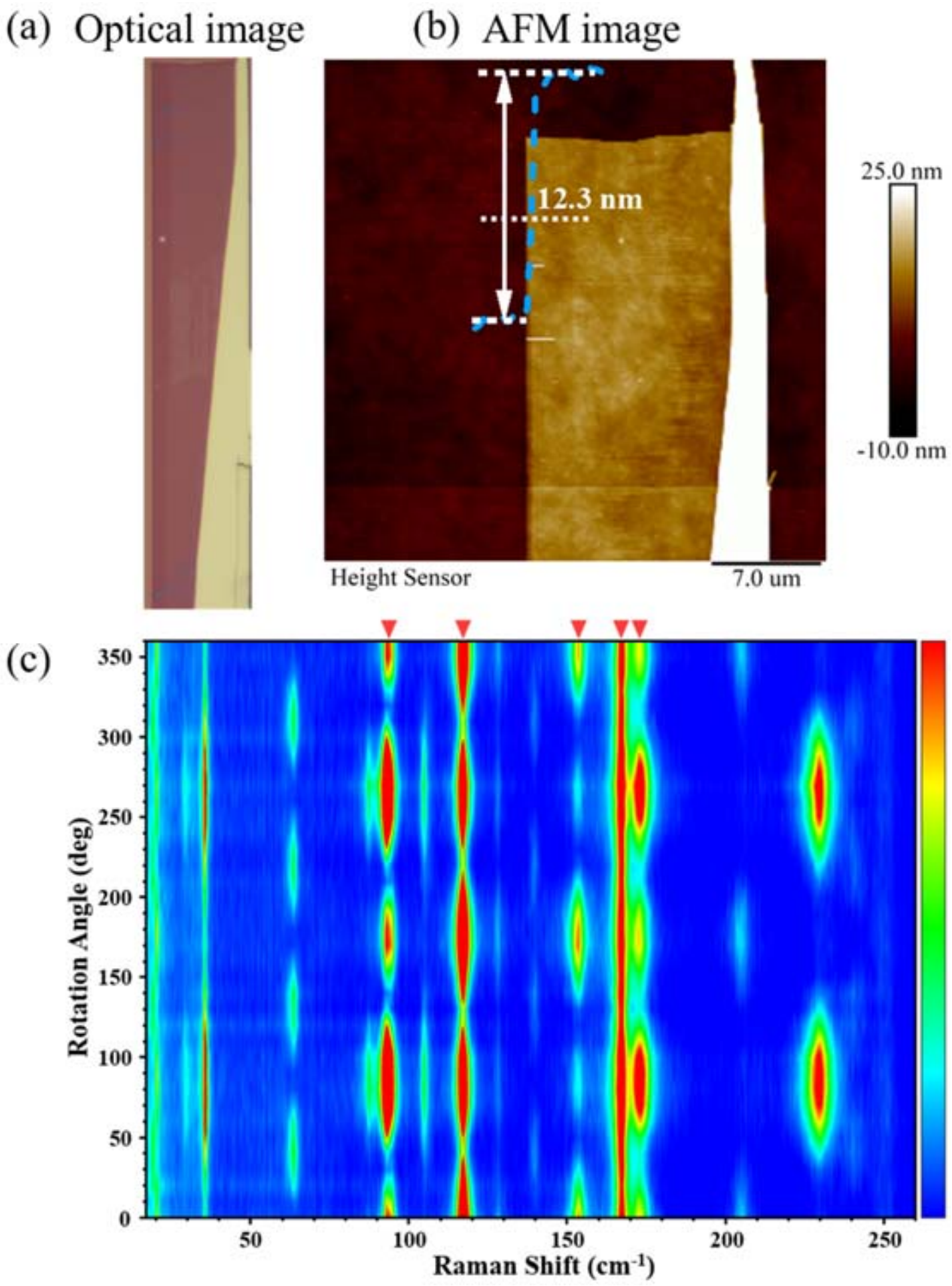

**Figure S23.** Thickness-dependent polarized Raman measurements on a 17-layer $Ta_2Ni_3Te_5$ nanoflake. (a) Optical image showing the ribbon-like morphology with edges aligned along the atomic chain direction. (b) AFM image confirming a thickness of 12.3 nm (17 layers). Scale bar: 7.0 μm. (c) False-color map of angular-dependent Raman intensity under parallel polarization (XX). Red triangles mark the anomalous $A_g$ modes. The polarization profiles are fully consistent with the 12L sample in Figure 3b, confirming thickness-independent intrinsic behavior.

To confirm that the anomalous multi-lobed angular patterns are intrinsic to $Ta_2Ni_3Te_5$, we performed angle-resolved polarized Raman measurements on a 17-layer nanoflake (~12.3 nm), as characterized by the optical image and AFM in Figure S23a,b. This thickness is close to the 12.8 nm sample in Ref. 28 (main text), enabling direct comparison with that study.

Figure S23c shows the false-color angular Raman intensity map measured under identical conditions to Figure 3b. The red triangles mark the anomalous $A_g$ modes. All phonon modes — both standard two-lobed and anomalous multi-lobed $A_g$ modes — show consistent angular patterns between the 12L (~8.7 nm) and 17L (~12.3 nm) samples (Figure S24 and S25). If birefringence were the primary origin, the phase shift $\delta = 2\pi \cdot \Delta n \cdot t / \lambda$ would scale with thickness. The observed thickness-independent behavior supports that the anomalous patterns arise from intrinsic anisotropic EPIs.

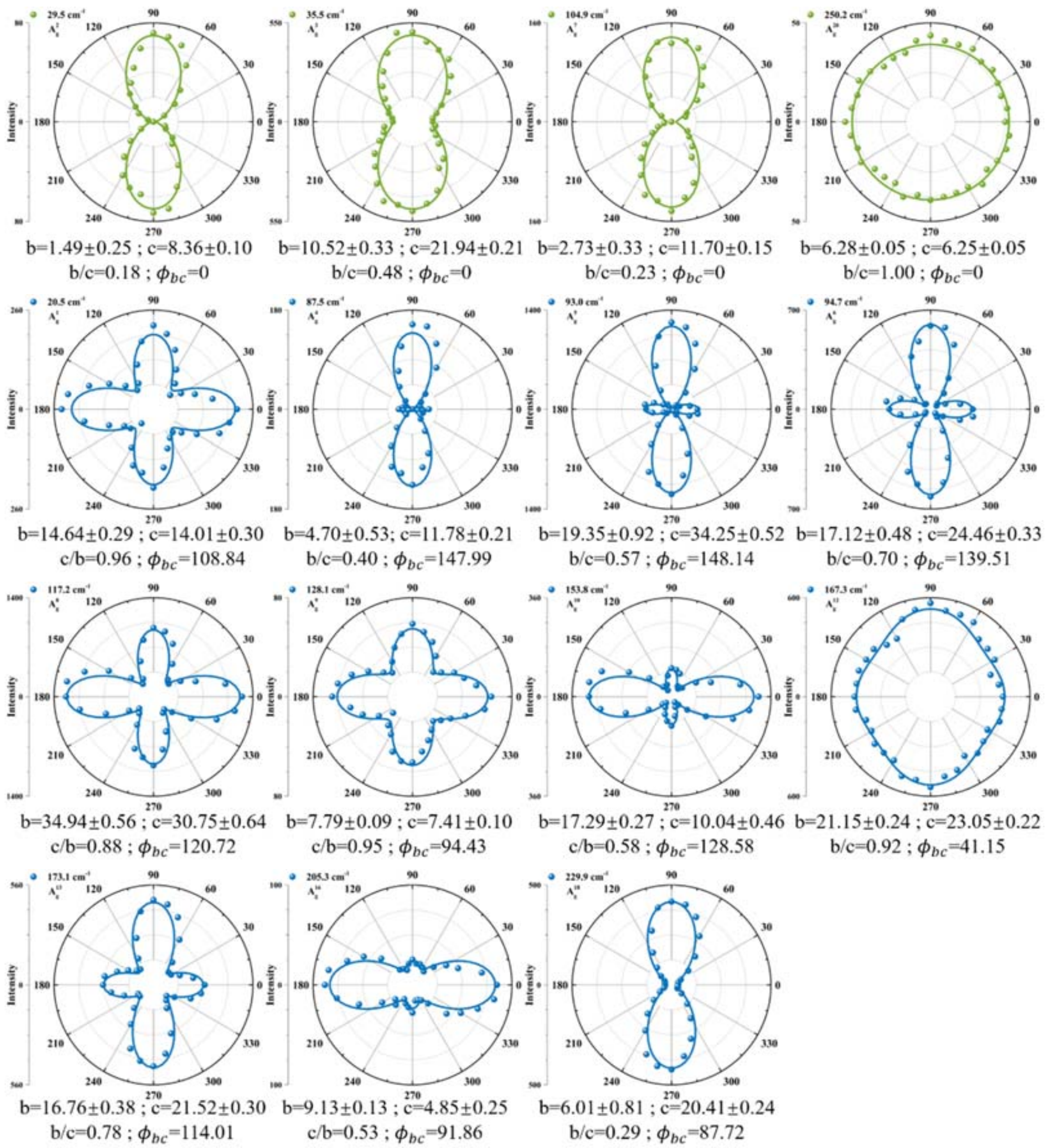

**Figure S24.** The polar plots and corresponding fitted Raman tensor components of each observed $A_g$ Raman modes in the 17-layer sample. Experimental data are represented as dots, while numerical fittings are plotted in solid lines.

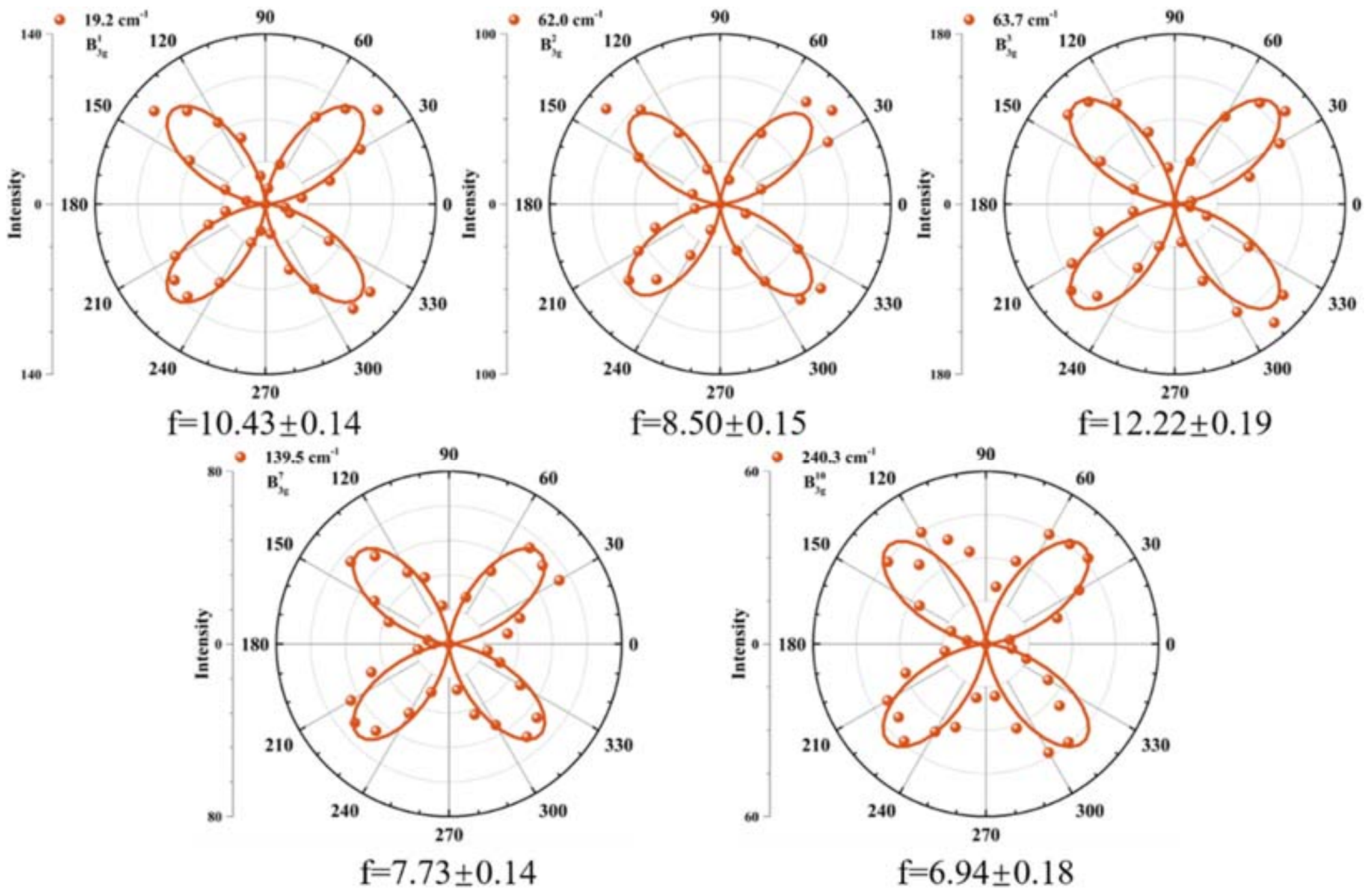

**Figure S25.** The polar plots and corresponding fitted Raman tensor components of each observed $B_{3g}$ Raman modes in the 17-layer sample. Experimental data are represented as dots, while numerical fittings are plotted in solid lines.

**S8. Effect of surface degradation on polarized Raman measurements**

To directly assess the impact of surface degradation on the intrinsic Raman properties of $Ta_2Ni_3Te_5$, we performed a graphite encapsulation control experiment. $Ta_2Ni_3Te_5$ nanoflakes were exfoliated in an Ar atmosphere glove box to minimize initial oxidation. A few layer graphite flake was then transferred onto part of the $Ta_2Ni_3Te_5$ nanoflake using a dry transfer technique, creating two distinct measurement areas on the same flake: a graphite covered area (P1) and an uncovered area (P2), as shown in Figure S26a. Graphite was selected as the encapsulation layer for two reasons: it is highly transparent at the 532 nm excitation wavelength, and its Raman modes do not overlap with those of $Ta_2Ni_3Te_5$, ensuring the measurements remain unambiguous. The sample was then exposed to ambient conditions prior to Raman measurements, allowing the uncovered area P2 to undergo natural surface degradation while P1 remained protected.

Polarized Raman measurements were performed on both P1 and P2 at three representative polarization angles: 0°, 45°, and 90°, as shown in Figure S26b. At all three angles, the Raman spectra measured on the graphite covered area (P1 w/Gr) are identical to those measured on the uncovered area (P2 w/o Gr) in peak frequencies, lineshapes, relative intensities, and polarization dependence. No additional peaks, broadening, or intensity suppression attributable to surface degradation are observed in P2 relative to P1.

These results directly confirm that the surface degradation observed in the HAADF-STEM cross sections (Figure 1f,g in the main text), which is self-limiting and confined to 3 to 4 atomic layers at the top surface, has no observable impact on the intrinsic phonon properties measured by Raman spectroscopy. This is consistent with two additional observations: first, the 532 nm laser penetration depth in $Ta_2Ni_3Te_5$ is substantially greater than the thickness of the degraded surface layer, meaning the Raman signal is dominated by scattering from the pristine crystalline interior; second, the excellent agreement between the experimental Raman frequencies and DFT calculations for a perfect undegraded crystal lattice (Figure 2a) confirms that the measured signal reflects intrinsic bulk properties.

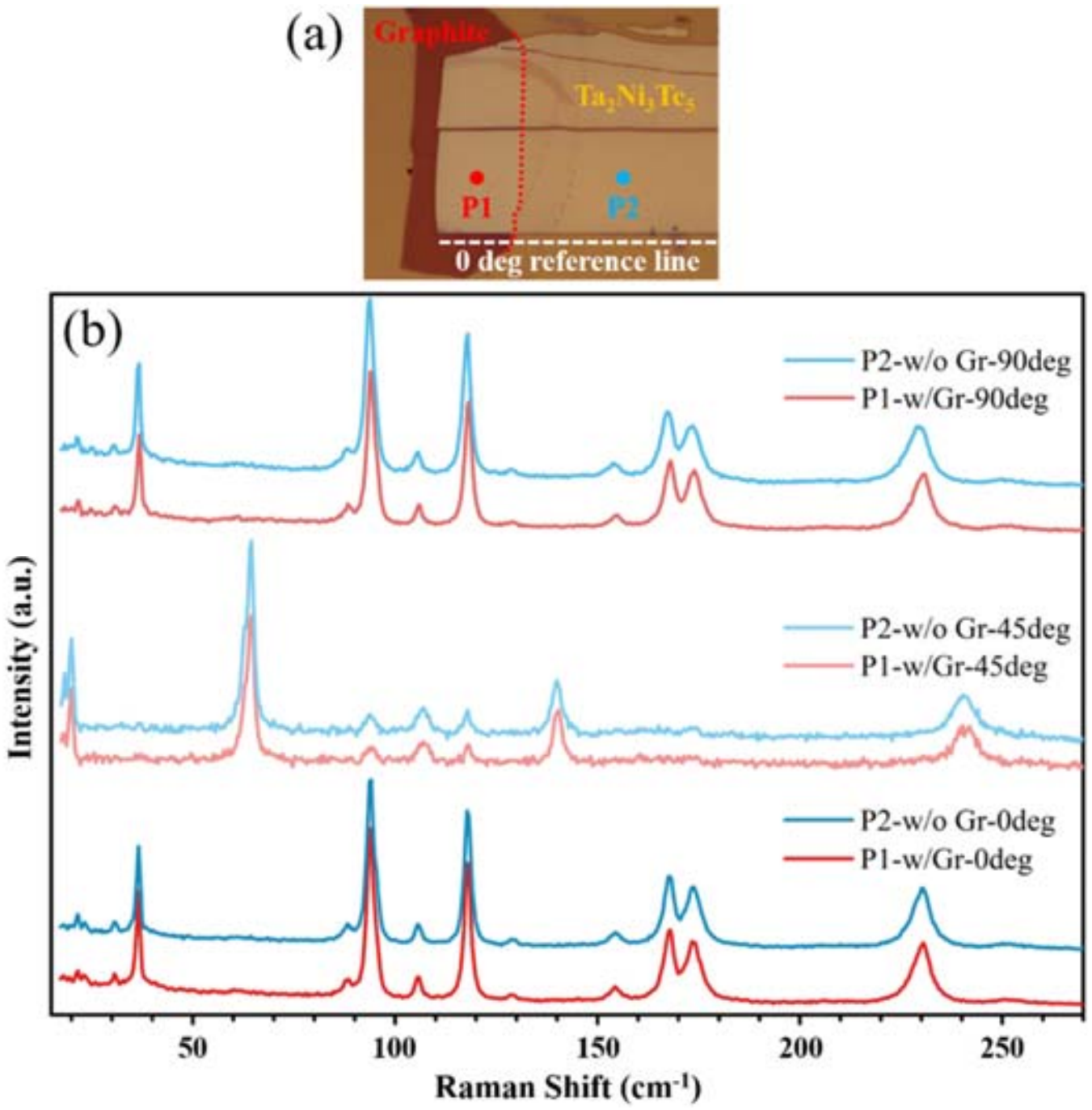

**Figure S26.** Control experiment confirming that surface degradation does not affect polarized Raman measurements. (a) Optical image of the $Ta_2Ni_3Te_5$ nanoflake with a few layer graphite flake transferred onto part of its surface. P1 (red dot) denotes the graphite covered area and P2 (blue dot) denotes the uncovered area. The red dashed line marks the boundary between the two regions. The white dashed line indicates the 0° reference alignment direction. (b) Polarized Raman spectra measured at three polarization angles (0°, 45°, and 90°) on P1 (red and pink curves, w/Gr) and P2 (blue and cyan curves, w/o Gr). The spectra are identical at all polarization angles in peak frequencies, lineshapes, and relative intensities, confirming that surface degradation has no observable effect on the intrinsic Raman properties of $Ta_2Ni_3Te_5$.